\documentclass[twocolumn]{aastex63}
\pdfoutput=1 
\usepackage{amsmath,amstext}
\usepackage[T1]{fontenc}
\usepackage{apjfonts} 
\usepackage[figure,figure*]{hypcap}
\usepackage{ulem}

\renewcommand*{\object}{NGC 3923}

\received{March 27, 2020}
\accepted{August 13, 2020}

\shorttitle{Measuring  stellar populations of \object}
\shortauthors{Feldmeier-Krause et al.}

\begin{document}

\title{Measuring the stellar population parameters of the early-type galaxy NGC 3923 - The challenging measurement of the initial mass function}

\correspondingauthor{Anja Feldmeier-Krause}
\email{afeldmei@astro.uchicago.edu}

\author{A. Feldmeier-Krause}
\affiliation{The Department of Astronomy and Astrophysics, The University of Chicago, 5640 S. Ellis Ave, Chicago, IL 60637, USA}
\author{I. Lonoce}
\affiliation{The Department of Astronomy and Astrophysics, The University of Chicago, 5640 S. Ellis Ave, Chicago, IL 60637, USA}
\author{W. L. Freedman}
\affiliation{The Department of Astronomy and Astrophysics, The University of Chicago, 5640 S. Ellis Ave, Chicago, IL 60637, USA}

\begin{abstract}

Recent studies of early-type galaxies have suggested that the initial mass function (IMF) slope is bottom-heavy, i.e. they contain a larger fraction of low-mass stars than  the Milky Way. However, measurements of the IMF remain challenging in unresolved galaxies because features in their observed spectra are sensitive to a number of factors including the stellar age, metallicity, and elemental abundances, in addition to the IMF. In this paper, we use new high signal-to-noise  IMACS (Magellan) spectra to study the elliptical shell galaxy NGC 3923 at optical (3700--6600\,\AA), and near-infrared (7900--8500\,\AA) wavelengths, as a function of radius.
We have undertaken a number of independent approaches to better understand the uncertainties in our results. 1) We compare two different stellar population model libraries; 2) we undertake spectral index fitting as well as full spectral fitting; 3) we have performed simulations for which we \textit{a priori} know the input IMF, and which closely match our data; 4) we also investigate the effects of including a two-component, rather than a single stellar population. We show that our  results are sensitive to the assumptions we make and to the methods we use. In addition, we evaluate the accuracy and precision of our results based on simulated mock data. 
 We find some indication (although assumption-dependent) 
 for a bottom-heavy  IMF  in the mass-range 0.5--1.0\,$M_\sun$, while the IMF in the mass-range  0.08--0.5\,$M_\sun$ appears to be Milky-Way like and constant. 
 Including  near-infrared data to our analysis  gives consistent results, and  improves the precision.
 
 \end{abstract}



\section{Introduction}
Is the IMF universal (similar to what is observed in the Milky Way), or does it vary as a function of environment? The answer to this simple question remains elusive and important.  For example, this question has serious implications for  photometric and spectroscopic studies of the high-redshift universe. If the IMF varies widely in different environments, interpreting the star formation history with redshift becomes vastly more challenging. 

The star formation histories of galaxies can be complex. Most stellar populations do not have a unique age or metallicity, but rather have unknown {\it a priori} distributions of those parameters. Analogously, there are no individual spectral features that are unique markers of age and metallicity; rather various features reflect those parameters in differing amounts. Spectral modeling is required to disentangle the information in the observed spectra of galaxies.  Accurate spectral modeling requires that the libraries of stellar spectra reflect those of the galaxies being studied. Elliptical galaxies have higher metallicities than stars within the Milky Way. Furthermore, the abundances of the alpha elements are greater than observed in the solar neighborhood.  Ensuring that the libraries are complete remains a serious challenge. 

In the midst of these already challenging degeneracies is the unknown IMF. A further challenge comes from the fact that much of the integrated light in an elliptical galaxy results from the contribution of its brighter giant stars, while the fainter main sequence stars contribute little. There are only a few (generally weak) features that provide clues about the ratio of giant to dwarf stars, and whether the IMF is varying. It is not surprising therefore, that there remain conflicting conclusions in the  literature concerning whether the IMF varies.

In this paper, the first in a series, we have studied the stellar population in the elliptical galaxy, \object. In later papers, we will explore a larger sample of elliptical galaxies, our own Milky Way bulge, and the integrated light of star clusters. Given the inherent difficulties in modeling stellar populations, in this study we have applied several independent approaches to the analysis of the same galaxy:  we have used different models, undertaken simulations, and incorporated different types of spectral analysis. We wish to ask how model-dependent, or how spectral-analysis dependent any resulting conclusions might be. 

We have explored both individual spectral-index fitting \citep[following the procedure outlined in ][]{2013MNRAS.433.3017L,2015MNRAS.447.1033M,2016MNRAS.457.1468L}, as well as full-spectral fitting \citep[following][and using their code]{2018MNRAS.475.1073V,2018MNRAS.479.2443V}. For a more comprehensive overview of relevant literature in this field we refer the reader to Section \ref{sec:context}. In this work, we incorporate  two different models: the MILES/E-MILES models of \cite{2010MNRAS.404.1639V,2015MNRAS.449.1177V,2016MNRAS.463.3409V}, 
and the Conroy models \citep{2012ApJ...747...69C,2018ApJ...854..139C}. We have undertaken simulations (Lonoce et al. in preparation),  in which we know {\it a priori} the underlying stellar population, and which we fit in the same way as the data for \object. Our goal has been to ascertain whether, despite modeling uncertainties and underlying degeneracies, we can find robust evidence for an IMF that differs from that observed in the Milky Way. 
We test if optical spectra  alone can be used to constrain the IMF, or if near-infrared data are required. 
We compare the different methods, wavelength regions,  and SPS models  to understand the effects of different assumptions on the measurements.

This paper is organized as follows: We discuss previous studies of the IMF  in Section \ref{sec:context}.  We present our data and measure spectral indices in Section \ref{sec:data}. 
We use those indices for a stellar population analysis in Section \ref{sec:spindex}. In Section \ref{sec:pystaff} we apply full-spectral fitting as second approach. We compare our different approaches with those in the literature in Section \ref{sec:disc}, and summarize our results in Section \ref{sec:sum}.

\section{Previous Work and Context}
\label{sec:context}

Our understanding of  galaxies and galaxy evolution depends heavily on assumptions regarding the stellar initial mass function (IMF), which  parameterizes  the shape of the initial stellar mass distribution  for a  population of stars. 
A galaxy consisting of a stellar population with a  \textit{top-heavy} IMF,  i.e. a larger fraction of high-mass stars, evolves faster, has  more stellar feedback,   a higher [$\alpha$/Fe] abundance, larger amounts of dust and metals, and more   stellar remnants.  
 On the other hand,  a \textit{bottom-heavy} IMF, 
 leads to a slower chemical enrichment in the galaxy, and  a larger fraction of M dwarf stars. The IMF determines several observables of a galaxy, such as the color, total  luminosity, mass-to-light ratio, 
 half-light radius, and star-formation rates
\citep{2013MNRAS.436.2254B,2018MNRAS.479.5448B}.

IMF studies have had a long history, but to date there is still debate about whether the IMF is universal. 
The first measurements of the IMF were made using field stars of the Milky Way.  \cite{1955ApJ...121..161S} parameterized the IMF in the form of a single power-law, with $dN/d\text{log}m \propto m^{-\Gamma}$, where $m$ is the initial mass of a star as it reaches the main sequence, $N$ is the number of stars in a logarithmic mass bin. The  IMF slope $\Gamma$ was measured to be  1.35. Later studies found that a single power-law   overestimates the number of stars  with lower masses than the Sun in the Milky Way,  and the IMF flattens and turns over at stellar masses  0.5--1\,$M_\sun$ \citep{2001MNRAS.322..231K,2003PASP..115..763C}. Recent studies of resolved stellar populations in the Milky Way  find similar IMF slopes for  field stars,  several star  clusters and associations \citep[e.g.][and references therein]{2010ARA&A..48..339B,2014PhR...539...49K}, with only few possible exceptions, which are still under debate. In addition,  the high-mass IMF slopes in the   LMC and  SMC are compatible with the Milky Way's IMF \citep{2003ARA&A..41...15M,2008AJ....135..173S,2009ApJ...696..528D}. 

Theoretical considerations, however, suggest a dependence of the IMF on various physical processes and conditions during star formation, such as fragmentation, turbulence, accretion, magnetic fields, stellar interactions,  feedback, or metallicity \citep[e.g.][]{2007prpl.conf..149B,2010ApJ...720L..26L,2013MNRAS.433..170H,2013MNRAS.436.2254B,2014ApJ...796...75C}. Until now,  no   star formation theory has  been able to implement all effects and predict their combined influence on the IMF. 

It is important to measure the IMF in different galaxies, to get a better understanding of the processes that shape it, and how the IMF influences the evolution of a galaxy. The IMF may be non-universal  and change for  different environments, and  stellar populations formed at different times.

In the past decade,  several studies  have measured the  IMF beyond the Local Group, using integrated light observations and  stellar population synthesis (SPS) models.  Two basic approaches have been used. In the first approach one  computes  the mass-to-light ratio (M/L) of SPS models with different IMFs and compares them to the measured M/L of a galaxy. The stellar M/L can be derived using stellar kinematics and dynamical modeling \citep[e.g.][]{2012Natur.484..485C,2013MNRAS.432.1862C,2016MNRAS.463.3220L},
gravitational lensing \citep[e.g.][]{2008MNRAS.383..857F,2010MNRAS.409L..30F,2012ApJ...752..163S,2016MNRAS.459.3677L}
or a combination of both \citep[e.g.][]{2010ApJ...709.1195T,2010ApJ...721L.163A,2011MNRAS.415..545T,2011MNRAS.417.3000S,2017ApJ...845..157N}. 
However, the derived stellar M/L is dependent on the assumed dark matter distribution of a galaxy. 
Most studies assume a constant stellar M/L, but this is not necessarily true, as the stellar populations of galaxies change with radius. Constraining both the dark matter distribution and a possible stellar  M/L gradient is a degenerate problem. 
Further, this method only allows one to exclude  certain IMF shapes, but is unable to distinguish if a high M/L  is caused by a bottom-heavy IMF, with many low-mass stars, or a top-heavy IMF, with many high-mass stars that evolved to  dark remnants. 

The second approach to measuring the IMF  is to compare the integrated light spectra  with SPS model spectra. Several spectral features in a galaxy spectrum are sensitive to the surface gravity of a star.  The integrated light of a galaxy with an old stellar population is  a mixture of the light from   massive and bright giant stars, and less massive and faint, but more abundant dwarf stars. A galaxy spectrum changes depending on the ratio of dwarf stars to giant stars.  
This technique was used in  several studies \citep[e.g.][and many others]{2003MNRAS.339L..12C,2010Natur.468..940V,2012ApJ...760...70V,2012ApJ...760...71C,2013MNRAS.433.3017L,2014MNRAS.438.1483S,2015MNRAS.451.1081M,2015ApJ...806L..31M,2019A&A...626A.124M}, 
and these authors found    early-type galaxies  with a higher fraction of low-mass stars than the Milky Way. Some galaxies have indications of  an IMF gradient, with a higher fraction of low-mass stars in the galaxy center  \citep{2015MNRAS.447.1033M,2017ApJ...841...68V,2018MNRAS.478.4084S}, 
while  other studies have found galaxies with a constant IMF \citep{2017MNRAS.468.1594A,2018MNRAS.478.4464A,2018MNRAS.475.1073V}. 
However, in addition to the IMF,  the  stellar age, metallicity,  elemental abundances, and star formation history influence spectral lines. 
These parameters can have a stronger influence on  certain spectral features than the IMF has, and there are correlations and degenerate solutions. 
If the surface-gravity sensitive effects can be disentangled from abundance and star-formation history effects, then  it may be possible to constrain the low-mass end of the IMF  in early-type galaxies. For these  reasons,  it remains a challenging task to measure the  IMF slope.

Previous studies \citep[e.g.][]{2016MNRAS.457.1468L,2017MNRAS.464.3597L,2018MNRAS.478.4084S,2018MNRAS.479.2443V} have generally used  spectra in the optical ($\sim$4000-6500\,\AA) to near-infrared wavelength range (\textgreater8000\,\AA). While longer wavelength spectra are more sensitive to low-mass stars, they are also more expensive in terms of observing time.


\begin{table}[h!]
\renewcommand{\thetable}{\arabic{table}}
\centering
\caption{Summary of observations} \label{tab:observations}
\begin{tabular}{ccccc }
\tablewidth{0pt}
\hline
\hline
Date & Grating & Slit & Exposure&Position angle\\
 & &width& time&wrt major axis\\
\hline
\decimals
2015-05-19 & 600$\ell$/9\fdg78 & 2\farcs5 & 1200\,s $\times$ 3 &0\degr\\
2015-05-20 & 600$\ell$/9\fdg82 & 2\farcs5 & 1200\,s $\times$ 4&0\degr\\
2015-05-19 & 600$\ell$/16\fdg6 & 2\farcs5 & 1200\,s  $\times$ 3&0\degr\\
2018-05-11 & 600$\ell$/10\fdg46 & 2\farcs5 & 1200\,s $\times$ 2&48\degr\\
2018-05-11 & 600$\ell$/10\fdg46 & 2\farcs5 & 770\,s $\times$  1&48\degr\\
\hline
\end{tabular}
\end{table}

\section{Data}
\label{sec:data}

In this section we give an overview of our observations, our data reduction techniques, and describe the quality of our data.
Our target is \object, an E4-5 galaxy  with more than twenty symmetric shells \citep{1988ApJ...326..596P}, located in the constellation of Hydra. \object\space is at a distance of 30-33\,Mpc,  it belongs to the \object\space group at redshift $z$=0.0046 \citep{1993A&AS..100...47G,2003ApJS..145...39M}.

\subsection{Observations}
We observed \object\space on three nights: May 19, 2015, May 20, 2015, and  May 11, 2018 with IMACS \citep{2006SPIE.6269E..0FD}, on the Magellan Baade 6.5-meter telescope. The observations were obtained with the $f$/4 camera, which provides a slit length of 15\,arcmin, in slow read-out mode. In 2015, the slit was placed along the  galaxy's major  axis. For the observation on May 11, 2018, the position angle was at 48\degr\space with respect to the major axis, along the Galactic North-South. The slit length is larger than  \object, which has an effective radius of  $R_{h}$=86\farcs4\space \citep{2011ApJS..197...21H}, 
 and thus allows  simultaneous sky observations. In May 2015, we additionally observed blank fields on the sky. The IMACS detector consists of eight chips with 2048 wavelength  pixels$\times$4096 spatial pixels each. Combined they give a 8192 pixel $\times$ 8192 pixel mosaic with one spatial gap and three wavelength gaps. We placed the slit such that the galaxy  falls entirely on the upper part of the slit, and only on four of the  eight detector chips.  
We used a slit width of  2\farcs5, and different gratings to provide greater wavelength coverage. For the two nights in 2015, we had the  600$\ell$/9\fdg78 and 600$\ell$/9\fdg82 grating to cover the spectrum at $\lambda$=3380--6725\,\AA, and the 600$\ell$/16\fdg6 grating to cover 7800--8600\,\AA; for the night in May 2018 we used the  600$\ell$/10\fdg46 grating  and covered $\lambda$=3900--7120\,\AA. Exposure times were 1200\,s, however, one exposure taken in 2018-05-11 was shorter (770\,s), because the telescope was approaching zenith.  We summarize the observations in Table \ref{tab:observations}.


\subsection{Data reduction}

\subsubsection{Instrumental calibrations}
Each detector chip  was reduced separately, including bias subtraction, cosmic ray removal, distortion correction, wavelength calibration,  sky subtraction, flat fielding, and flux calibration
using  \textsc{idl} and \textsc{iraf} scripts.

We estimated the bias using the overscan region of each chip, and subtracted it. With  the \textsc{idl} routine \textsc{l.a.cosmic} \citep{2001PASP..113.1420V}  we  identified cosmic rays and created  bad pixel masks. 
We rectified the distortion along chip rows  by tracing several emission lines on the He/Ne/Ar  arc lamp exposures.  For each night, we took pinhole exposures, and we used those to trace the distortion along chip columns. We used the \textsc{iraf} tasks \textsc{identify}, \textsc{reidentify}, and \textsc{fitcoords} for simultaneous  wavelength calibration  to air wavelengths. The wavelength gaps between individual chips are 10--20\,\AA\space wide.

We traced the position of the galaxy along the slit as a function of wavelength by fitting a Cauchy function to the galaxy light  profile. We fit various arc lines close to the galaxy trace 
to estimate the spectral resolution $R=\lambda/\Delta\lambda$. 
 $R$ is increasing with wavelength,  from about $R$=675 at $\lambda$=3900\,\AA\space to about $R$=1150 at $\lambda$=6700\,\AA, and $R$=1500 at $\lambda$=8000\,\AA\space for a slit width of 2\farcs5. 

\subsubsection{Spectral extraction and telluric correction}

We extracted the one-dimensional spectra from each exposure in several radial bins. The central spectrum was extracted in a   1\farcs5 wide region, further at  distances of 0\farcs75--3\arcsec\space from the center, 3\arcsec--$\frac{1}{8}$\,$R_{h}$, $\frac{1}{8}$\,$R_{h}$--\onequarter\,$R_{h}$, \onequarter\,$R_{h}$--\onehalf\,$R_{h}$, \onehalf\,$R_{h}$--$\frac{3}{4}$\,$R_{h}$, \onehalf\,$R_{h}$--1\,$R_{h}$, and  
 $\frac{3}{4}$\,$R_{h}$--1\,$R_{h}$. 
 We adopted 1\,$R_{h}$=86\farcs4 and summed the  respective regions  in the upper and lower part of the slit together.  For the spectra observed on 2018-05-11, we corrected for  the different P.A. as follows: We assumed an ellipticity of $\epsilon$=0.271, and modified the extraction regions such that they contain  the same isophote regions as the observations along the major axis. 	

Further, we extracted  the sky from a region \textgreater~3\,$R_h$ from the galaxy and subtracted it from the spectra, see Appendix~\ref{sec:skyap} for details. We applied  flat fielding and  flux calibration, derived from standard star observations (Feige 67 and Hip 59167) taken on the same nights.

The spectra  have telluric absorption lines from H$_2$O and O$_2$  molecules in the  Earth's atmosphere. We used the ESO tool \textsc{molecfit} \citep{2015A&A...576A..77S,2015A&A...576A..78K} to correct the atmospheric absorption. \textsc{molecfit}  creates a synthetic atmosphere spectrum and derives a correction function, taking the spectral resolution into account. Molecular absorption lines vary with time. The advantage of \textsc{molecfit} is that it uses the science observations directly and  not observations of a telluric standard star, taken at a different time than the target.  We derived  the atmospheric absorption correction 
for each exposure using the central  spectrum, and applied the same correction to all  spectra.

\begin{figure}
\gridline{\fig{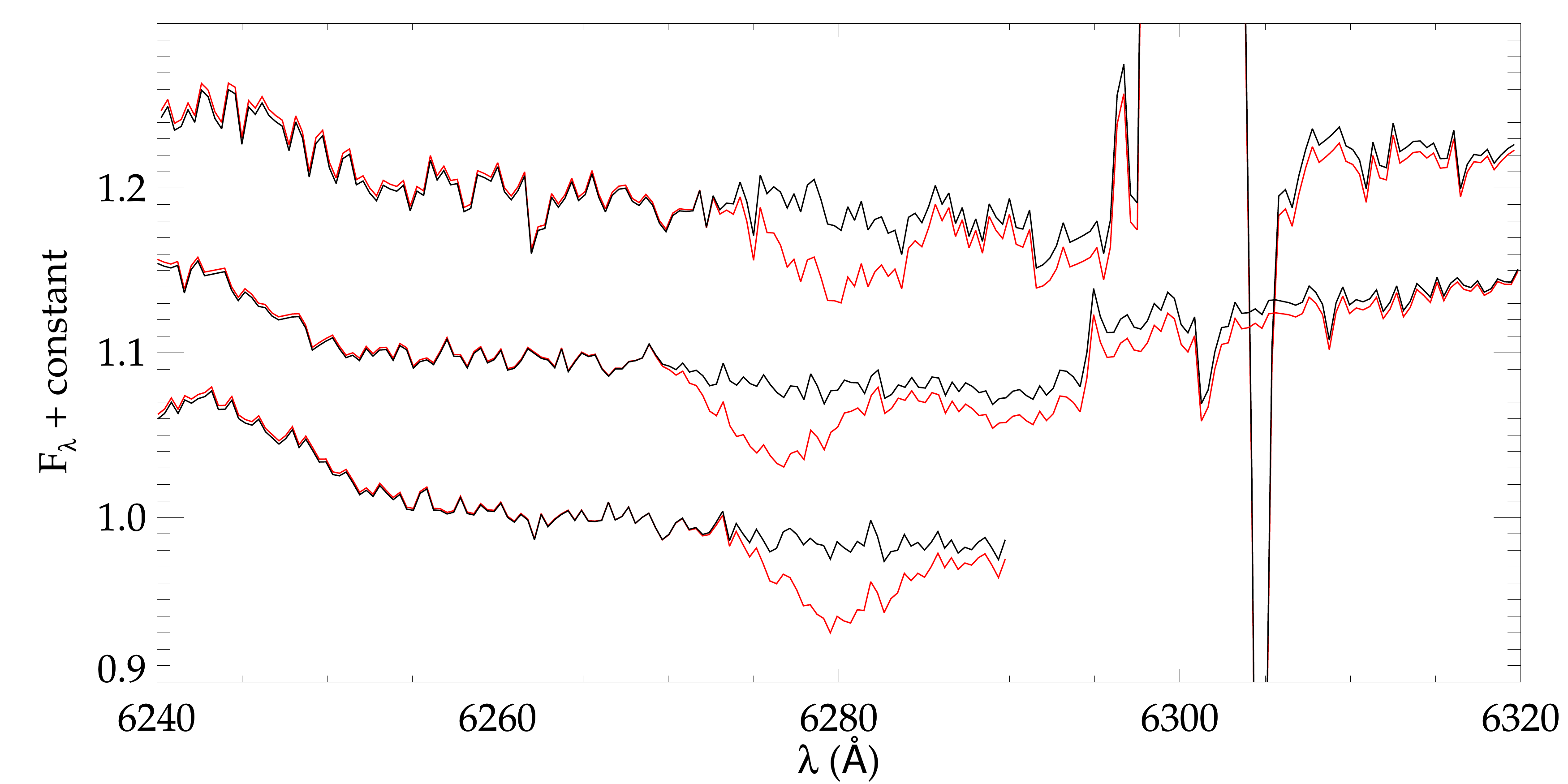}{0.45\textwidth}{}}
\gridline{\fig{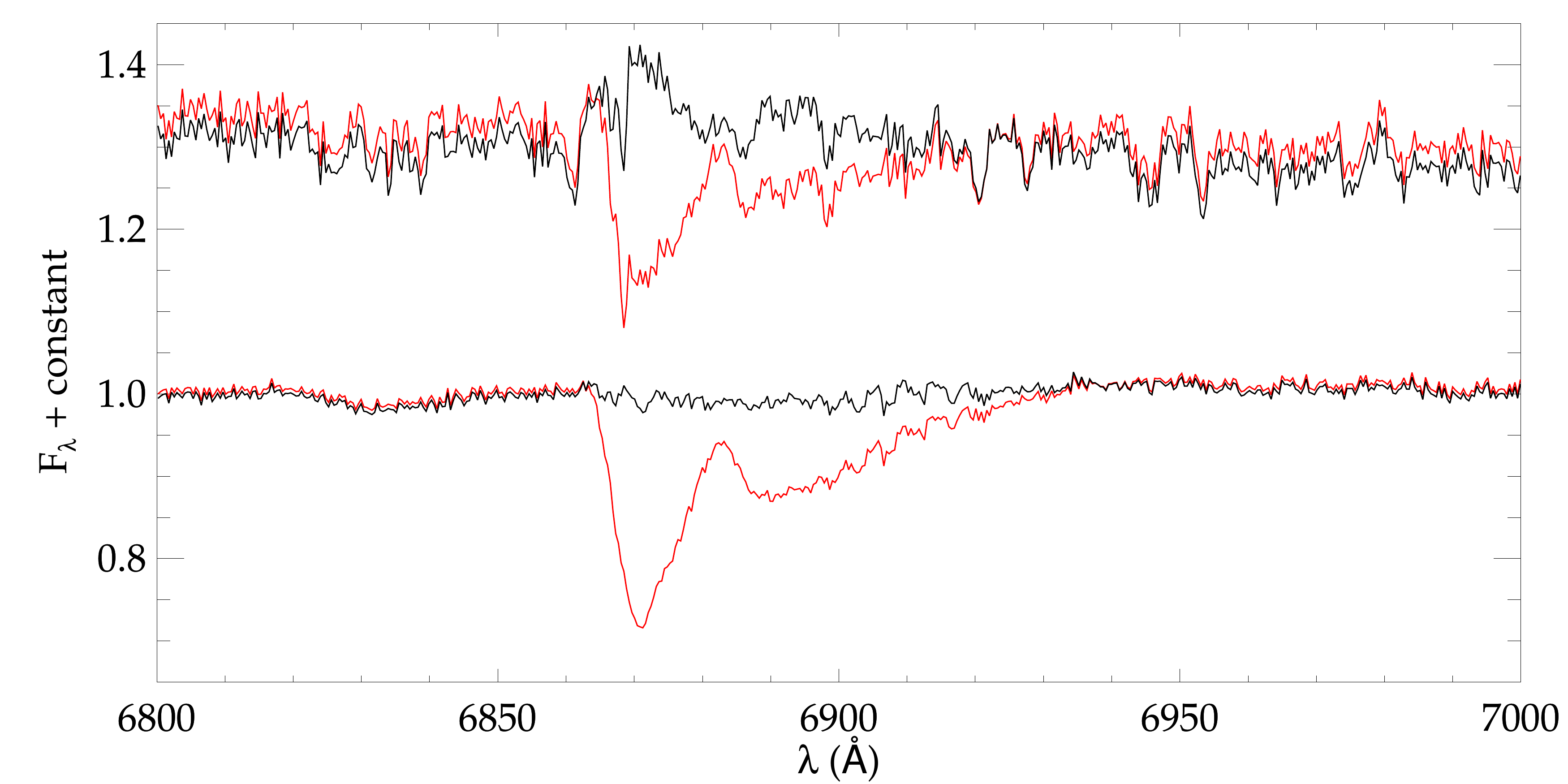}{0.45\textwidth}{}}
\caption{Spectra of \object\space before (red) and after (black) telluric absorption correction. The upper panel shows three exposures observed on three different nights in the wavelength region 6250--6320\,\AA. The three different spectra were extracted at different regions of the galaxy; some of them have sky residuals around 6300\,\AA.  For one of the three exposures, observed 2018-05-11, the chip gap starts at 6290\,\AA. The lower panel shows two out of three exposures observed 2018-05-11, in the longer wavelength region 6800--7000\,\AA, also at different distances from the galaxy's center.  \label{fig:pyramid}}
\end{figure}

We fit the atmospheric absorption in the wavelength regions 6253--6300\,\AA,  6820--6976\,\AA, and 8170--8300\,\AA\space 
and made sure that these regions are free of prominent emission or absorption lines. 
Example spectra with the applied \textsc{molecfit} correction are shown in  Fig.\ref{fig:pyramid}. The spectra before telluric correction are in red, and spectra after telluric correction are in black.  We show spectra observed on different nights, and extracted for different regions, to cover a wide range of S/N. Telluric correction in the region 6270\,\AA\space is satisfactory, but there are sky residuals at 6295--6305\,\AA\space (two upper spectra), which led us to perform a second order sky subtraction (see Appendix~\ref{sec:secsky}). There are some residuals from telluric correction at 6870\,\AA\space (see also Fig.~\ref{fig:spec2}) for spectra with lower S/N. We excluded this spectral region from our analysis.

After applying  these corrections, we summed the different exposures taken at the same nights to one spectrum per night and IMACS chip. 
Finally, we performed a  second order sky subtraction (see Appendix~\ref{sec:secsky}) and  used the   velocity measurements to shift each spectrum to  rest wavelength.

\subsubsection{Data quality}
We combined the spectra of the two 2015 nights  to obtain the final spectra as a function of distance from the center of \object. 
As the 2018 observations were taken at a different P.A. and grating angle, we did not sum the 2018 spectra together with the 2015 spectra. As result, we have a set of spectra in the optical wavelength region 3500--6640\,\AA, observed along the major axis and as a function of galactocentric radius (shown in Fig.~\ref{fig:spec2}), and another set of spectra in the optical wavelength region 3900--7050\,\AA, at a P.A. 48\degr\space offset from the galaxy major axis (Fig.~\ref{fig:spec}). In addition, we have spectra in the near-infrared wavelength region 7800--8600\,\AA\space along the major axis (Fig.~\ref{fig:pystafffit}). 
 Due to the longer total exposure time (2h 20 min), the optical spectra along the major axis have a higher S/N than the spectra along P.A.=48\degr\space (53 min), and the near-infrared spectra along the major axis (60 min). 
 
\begin{figure}
\gridline{\fig{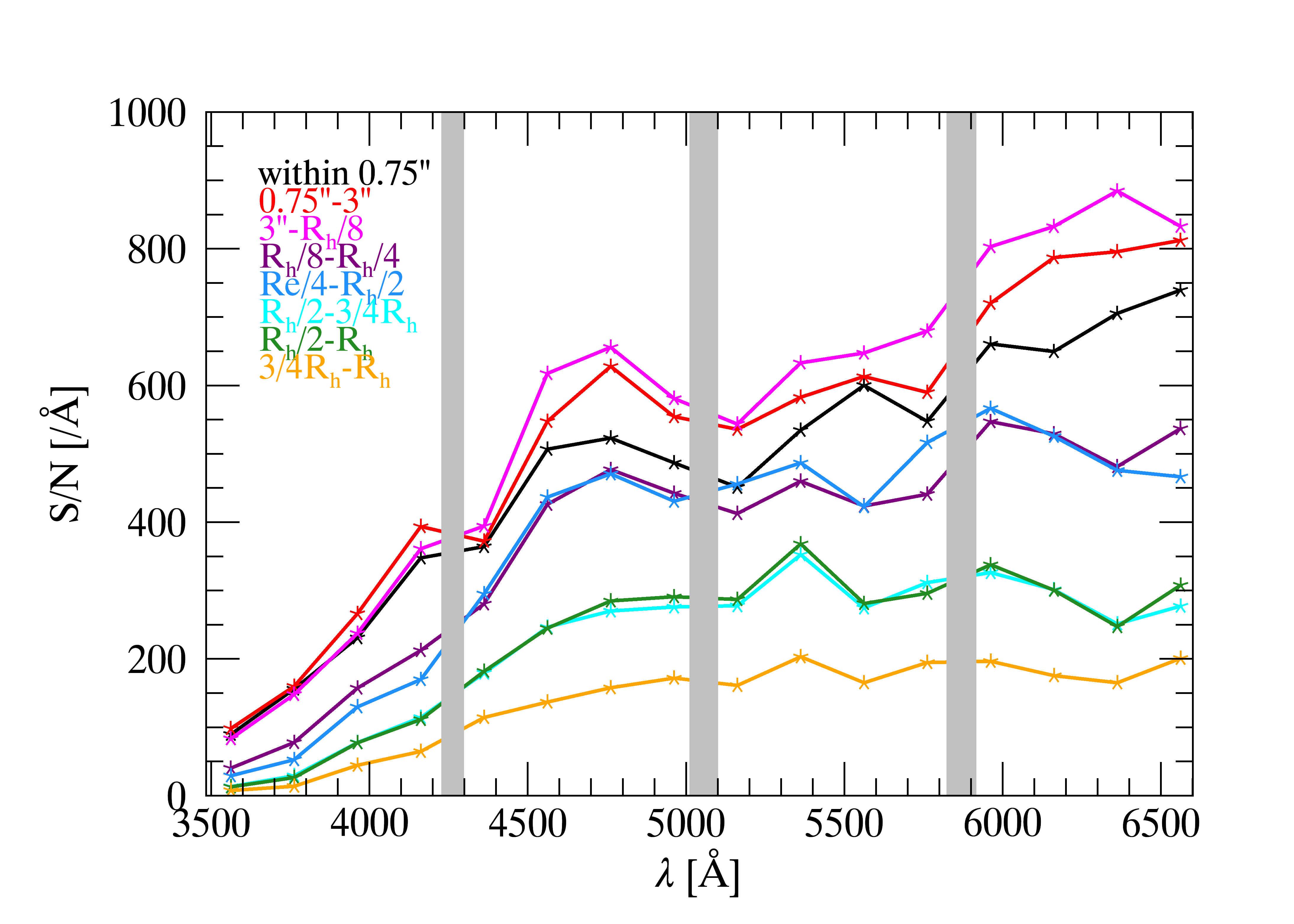}{0.45\textwidth}{}}
\gridline{\fig{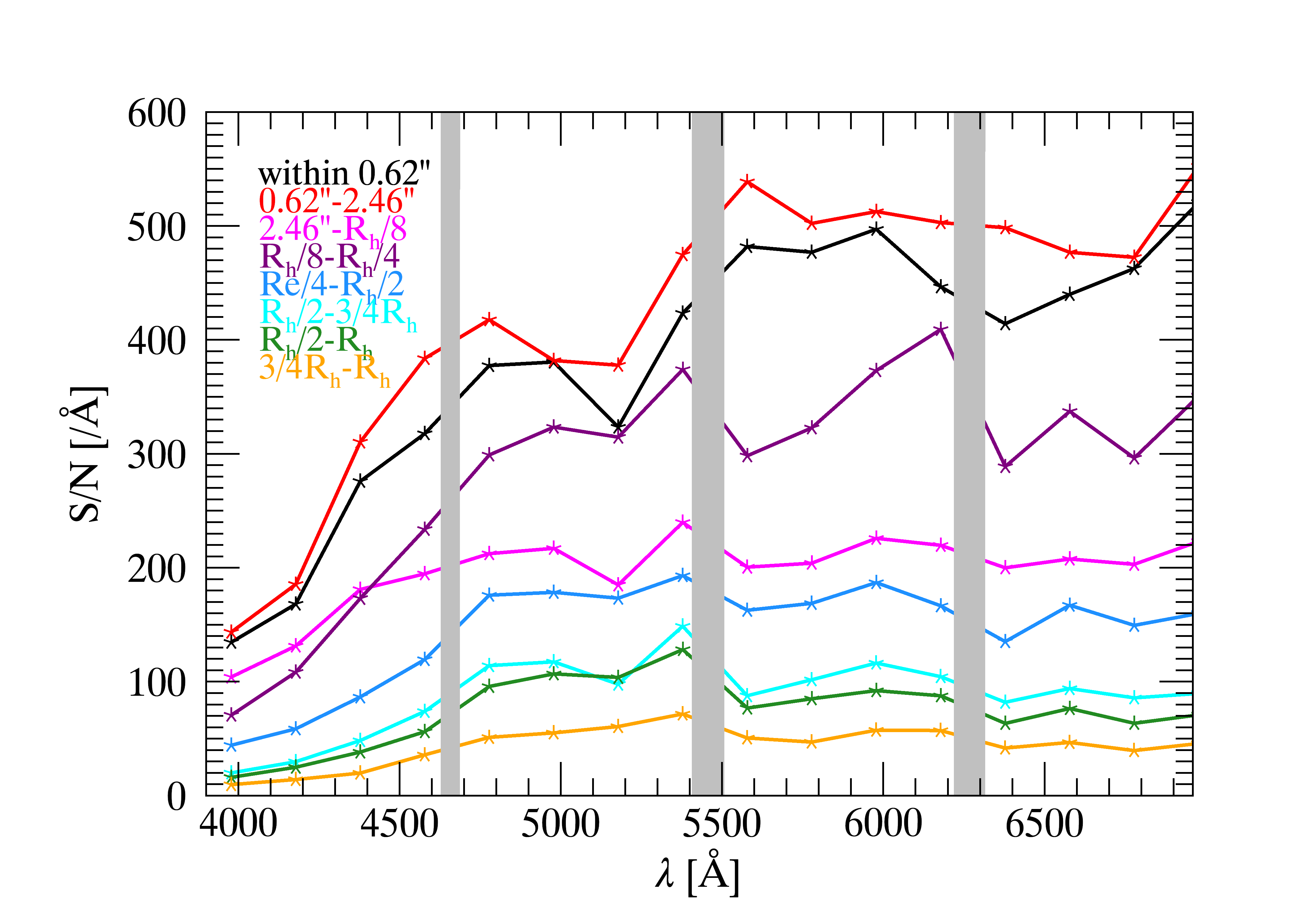}{0.45\textwidth}{}}
\caption{Signal-to-noise ratio of the spectra  as a function of  wavelength for different extraction radii,  for  spectra with P.A.=0\degr\space on the upper panel, and    P.A. = 48\degr\space on the lower panel. Grey shaded regions mark  chip gaps. \label{fig:snr}}
\end{figure}

  We show the signal-to-noise ratio (S/N) of the optical spectra as a function of wavelength for each spectrum in Fig.~\ref{fig:snr}; the different colors denote the extraction regions. 
 The S/N is lowest in the blue end of the spectrum, $\lambda$\textless\,4000\AA.  Except for the outermost spectrum, we have a S/N$\gtrsim$200\,\AA$^{-1}$\space  for  the P.A.=0\degr\space spectra at $\lambda$\textgreater 4500\,\AA. Our near-infrared spectra (7800--8600\,\AA) have S/N $\approx$ 400 in the center, decreasing to $\approx$ 100 in the outer three bins. For the P.A.=48\degr\space spectra, the S/N does not exceed 200\,\AA$^{-1}$\space  for the outer four spectra. Recent studies used spectra with S/N$\gtrsim$100\,\AA$^{-1}$\space \citep[e.g.][]{2015MNRAS.447.1033M,2018MNRAS.478.4084S} for spectral index fitting. Some of our spectra also  exceed the S/N of the spectra used by \cite{2012ApJ...760...70V} and \cite{2012ApJ...760...71C}, which go up to 400\,\AA$^{-1}$\space at 5000\,\AA.   Thus, our data have S/N that is comparable or even higher than the S/N of data used in other studies.

\subsection{Measurements}
In the following section we describe our procedure for measuring   stellar kinematics, gas emission, and spectral indices on our optical spectra. For the spectral indices we account for the  effects of resolution, and non-Gaussian line-of-sight velocity distribution (LOSVD). 

\subsubsection{Stellar kinematics and emission line correction}
\label{sec:kin}

We   fit the  stellar LOSVD of the final combined spectra with the program \textsc{pPXF} \citep{2004PASP..116..138C, 2017MNRAS.466..798C} and SSP templates. The SSP models are convolved with a LOSVD, to obtain the four best-fit  kinematic moments (V, $\sigma$, $h_3$, $h_4$). Together with the kinematics, we  fit an optimal template, which is a linear combination of the input SSP templates. 

To obtain the correct kinematic solution, it is important that the template library contains spectra that match the stellar populations of the data. For this reason, we used a wide range of stellar parameters for the template models. In particular, we used the  E-MILES SSP  models \citep{2016MNRAS.463.3409V}  with a  bimodal IMF parameterization, with 14 different IMF slopes (see Sect. \ref{sec:ageppxf} for details), 15 ages (1--17.8 Gyr), and 6 metallicities (-1.7 to +0.2\,dex). See Appendix~\ref{sec:modelsum} for an overview of SSP models.

We measured the stellar kinematics in the spectral ranges 3760--6640\,\AA\space (P.A.=0\degr) and 3920--6700\,\AA\space (P.A.=48\degr), with  additive polynomials (10th and 12th order, respectively) to correct the model continuum shape. We masked bad pixels for the fit.
To account for the possibility of gas emission, we used the code \textsc{GandALF} (Gas AND Absorption Line Fitting) by \cite{2006MNRAS.366.1151S} to fit  stellar populations and  gas emission simultaneously. We fixed the stellar kinematics and fit the  spectra  with multiplicative polynomials (10th and 12th order, respectively). The fit emission lines were subtracted from the spectra if their amplitudes were three times higher than the noise in the respective spectral region. This was  fulfilled for the Balmer lines in the outer four spectra along the major axis.  

Given that   the spectra have been shifted to rest wavelength (Appendix \ref{sec:secsky}), the resulting velocities are close to zero. The velocity dispersion in the center reaches 275\,km\,s$^{-1}$, but decreases down to 170\,km\,s$^{-1}$ at 54\arcsec\space and beyond. 
The  Gauss-Hermite moments h$_3$ and h$_4$ are  within -0.01\textless h$_3$\textless0.02 and 0.07\textless h$_4$\textless0.14. This means that the LOSVD is nearly symmetric, as the skewness $h_3$ is close to zero. The positive kurtosis $h_4$ indicates that the  LOSVD is more heavy-tailed than a Gaussian, possibly due to   a radial  velocity anisotropy.   We list our kinematic results in Table \ref{tab:decimal}. We derived the  uncertainties  from the dispersion  of 500 Monte Carlo runs, in which we added random noise to the spectra before measuring the LOSVD.

\begin{table}[h!]
\renewcommand{\thetable}{\arabic{table}}
\centering
\caption{LOSVD results for P.A.=0\degr\space data} \label{tab:decimal}
\begin{tabular}{cc c c }
\tablewidth{0pt}
\hline
\hline
radius & $\sigma$ & $ h_3$ & $h_4$\\
(arcsec) & (km$\cdot$ s$^{-1}$) & &\\
\hline
\decimals
  0.00 &  276.0  $\pm$     2.7 &  -0.01  $\pm$    0.01 &   0.08  $\pm$    0.01\\
  1.85 &  270.1  $\pm$     2.6 &  -0.01  $\pm$    0.01 &   0.08  $\pm$    0.01\\
  6.90 &  244.3  $\pm$     2.6 &  -0.01  $\pm$    0.01 &   0.10  $\pm$    0.01\\
 16.20 &  210.3  $\pm$     3.1 &  -0.00  $\pm$    0.01 &   0.11  $\pm$    0.01\\
 32.40 &  196.3  $\pm$     3.9 &   0.00  $\pm$    0.01 &   0.11  $\pm$    0.01\\
 54.00 &  174.0  $\pm$     6.9 &   0.00  $\pm$    0.01 &   0.13  $\pm$    0.03\\
 64.80 &  169.7  $\pm$     7.3 &   0.01  $\pm$    0.01 &   0.13  $\pm$    0.03\\
  75.60 &  170.3  $\pm$    12.5 &   0.02  $\pm$    0.02 &   0.14  $\pm$    0.06\\

\hline
\end{tabular}
\tablecomments{Kinematic uncertainties are  statistical only}
\end{table}

\subsubsection{Spectral index measurements}

\begin{figure*}[ht!]
\plotone{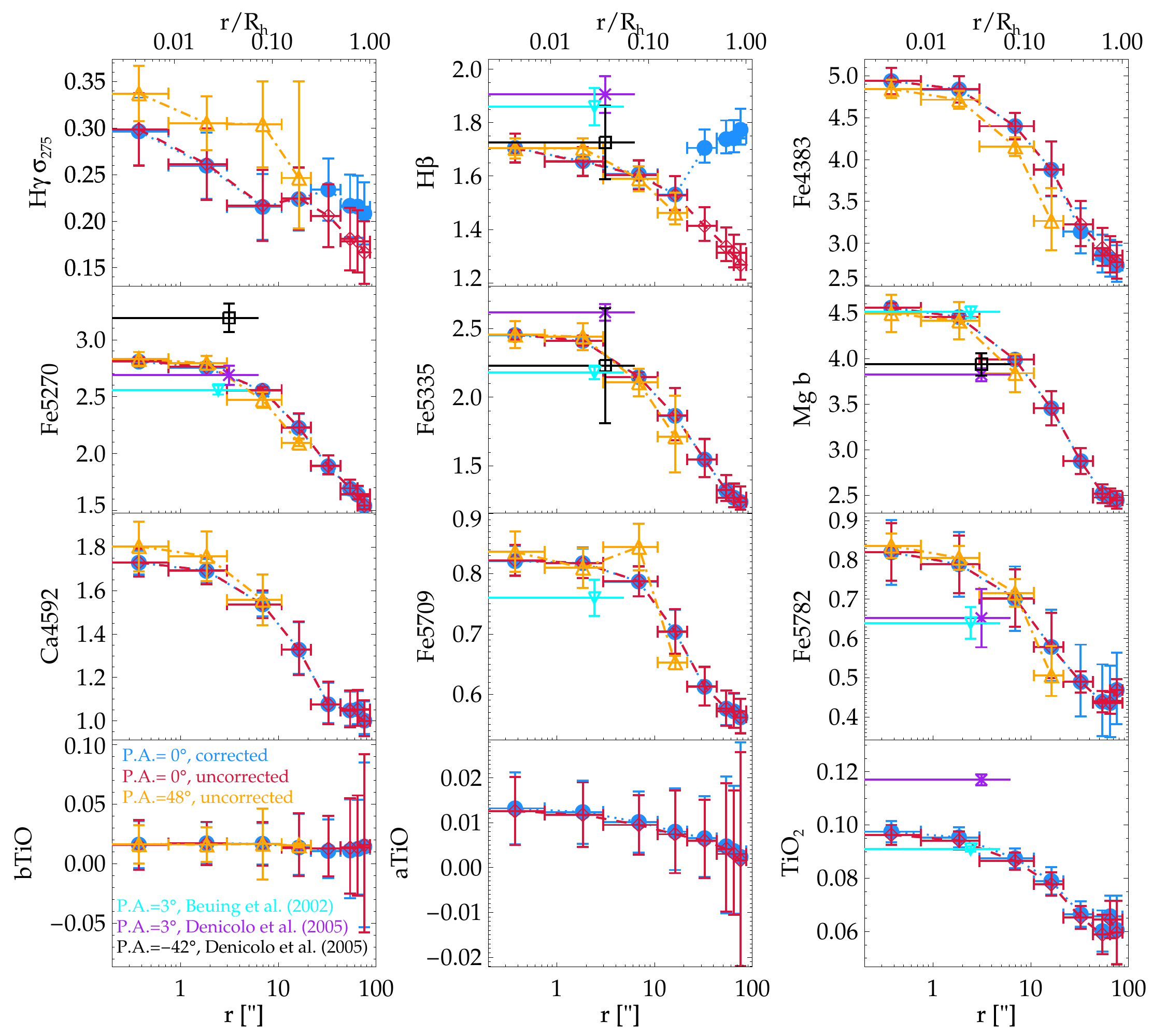}
\caption{Absorption line index gradients measured for  \object. Open red diamonds denote index measurements without emission line correction, blue closed circles with correction, both for P.A.=0\degr. Orange  open triangles denote P.A.=48\degr\space data, without  correction. The data are plotted at constant r/$R_{h}$ ($R_{h}$=86\farcs4), the P.A.=48\degr\space data are actually at smaller values of r ["]. Error bars are the quadratic sum of systematic and  statistical uncertainties (see Appendix~\ref{sec:spindunc}). Purple x-symbols denote the measurements of \cite{2005MNRAS.356.1440D} within r$\leq$6\farcs25, at  P.A.=3\degr; open black square symbols at P.A.=--42\degr. Cyan downward-facing triangles denote the measurements of \cite{2002A&A...395..431B}, also at P.A.=3\degr,  r$\leq$4\farcs85. We use all  of the shown indices  for spectral index fitting, except H$\gamma_{\sigma275}$ and H$\beta$.
\label{fig:gradient}}
\end{figure*}

We interpolated over bad pixels and  measured spectral indices (as listed in Table \ref{tab:spindexdef}) on both the final combined spectra and, if needed, on the emission-line subtracted spectra. If we omit the subtraction, the value of several Balmer line indices changes (see red diamonds and blue circle symbols in Fig. \ref{fig:gradient}), leading to a different age estimate by several Gyr. By subtracting the gas emission,   we ensure that the spectral index measures  stellar population rather than gas. Our procedure to derive spectral index uncertainties is described in Appendix~\ref{sec:spindunc}.

\begin{deluxetable*}{l |c |c| c |l l }
\tablecaption{Spectral index definitions \label{tab:spindexdef}}
\tablecolumns{6}
\tablewidth{0pt}
\tablehead{
\colhead{Index} &
\colhead{blue continuum [\AA]} &
\colhead{feature} [\AA]&
\colhead{red continuum} [\AA]&
\colhead{type\tablenotemark{a}}&
\colhead{reference\tablenotemark{b}}
}
\startdata
H$\gamma \sigma_{275}$ &4331.500 -- 4341.00&  4331.500 --  4351.875&  4359.250 -- 4368.750&A  &   1\\
H$\beta$& 4827.875 --  4847.875&  4847.875  -- 4876.625&  4876.625 --  4891.625& A  &       2     \\
Fe4383&4359.125 -- 4370.375&  4369.125 --  4420.375&  4442.875 --  4455.375&A  &    2     \\
Fe5270&5233.150 -- 5248.150&  5245.650 --  5285.650&  5285.650 --  5318.150& A  &         2   \\  
Fe5335&5304.625 -- 5315.875&  5312.125  -- 5352.125&  5353.375 --  5363.375& A  &         2    \\ 
Mg\textit{b}&5142.625 -- 5161.375&  5160.125 --  5192.625&  5191.375 --  5206.375& A  &            2 \\    
Ca4592&4502.500 -- 4512.000&  4578.000 --  4603.000&  4611.000 --  4628.000& A  &       3\\
Fe5709&5672.875 -- 5696.625&  5696.625 --  5720.375&  5722.875 --  5736.625& A  &       2  \\   
Fe5782&5765.375 -- 5775.375&  5776.625 --  5796.625&  5797.875 --  5811.625& A  &      2  \\   
bTiO&4742.750 -- 4756.500&  4758.500 --  4800.000&  4827.875 --  4847.875& M  &         4\\
aTiO&5420.000 -- 5442.000&  5445.000 --  5600.000&  5630.000 --  5655.000& M  &        4\\
TiO$_2$ &6066.625 -- 6141.625&  6189.625 --  6272.125&  6372.625 --  6415.125& M  &         2\\
\enddata
\tablenotetext{a}{A denotes atomic, M molecular index definition}
\tablenotetext{b}{1: \cite{1999ApJ...525..144V}, 2: \cite{1998ApJS..116....1T}, 3: \cite{1994AJ....108.2164G},  4: \cite{2014MNRAS.438.1483S}}
\end{deluxetable*}

\subsubsection{Resolution and LOSVD correction}
To constrain the stellar populations of \object, we compared our measured spectral indices  with those indices from  SSP models. For a meaningful comparison, data and model indices must be measured at the same resolution. 
However,  each observed spectrum has a different resolution because of the  varying  velocity dispersion, and a different non-zero h$_3$ and h$_4$ (see Sect.~\ref{sec:kin}). 
Both,  velocity broadening and non-Gaussian LOSVD,  affect the spectral index measurements \citep{2004A&A...426..737K}. 
Before comparing our data with the  model indices, we therefore   corrected for the different resolutions and  non-Gaussian LOSVDs  as follows.

First, we chose a common resolution for our data and model reference indices.
Spectral line indices are often measured in the ``Line Index System"  \citep[LIS,][]{2010MNRAS.404.1639V}  
with  a resolution of FWHM=14\,\AA. Due to velocity broadening, the FWHM of the  three central \object\space  spectra is higher than the reference value of  FWHM=14\,\AA, whereas the outer spectra have a higher resolution and lower FWHM. For the data with higher resolution, we simply convolved the spectra  to  FWHM=14\,\AA\space before measuring spectral indices. 

In order to derive a correction for the lower resolution data and non-Gaussian LOSVD, we measured the line indices not only on the data,  but also  on our best-fit model spectra  obtained with \textsc{pPXF}. We measured the indices on two sets of best-fit model spectra: On the best-fit model spectrum at the LIS reference spectral resolution   (FWHM=14\,\AA, with h$_3$ = h$_4$ = 0), and on the best-fit spectrum that was convolved  with the same LOSVD as measured on the data with \textsc{pPXF}. The difference  of these two measurements is nonzero, which confirms that the broader FWHM caused by the high velocity dispersion, and the non-Gaussian LOSVD of our data should not be ignored. We used the ratio of these two model index measurements as factor to transfer our atomic 
index measurements (measured in units of \AA) to the LIS (i.e. h$_3$ = h$_4$ = 0 and FWHM=14\,\AA).   This factor changes the index measurements of our data  by up to 20\%.  For molecular indices (measured in magnitudes), the correction is not a factor but the difference of the two best-fit model  measurements. Also for indices that can have negative values (e.g. H$\delta_A$, H$\delta_F$, H$\gamma_A$, H$\gamma_F$), the correction is additive \citep{2004A&A...426..737K}. 
After this correction, we can compare the index measurements of our data to indices measured on SSP models.

\subsubsection{Spectral index gradients}
We show a selection of 12 spectral index gradients in Fig.~\ref{fig:gradient}, as measured on the spectra and corrected to FWHM=14\,\AA. The error bars shown in Fig.~\ref{fig:gradient} are  the quadratic sum of statistical and systematic uncertainties, see Appendix~\ref{sec:spindunc} for details.
The index gradients are larger than the statistical uncertainties,  indicating real stellar population gradients.
Most of the absorption line indices  decrease with increasing radius, while bTiO and aTiO are approximately constant to within their uncertainties. 
H$\beta$ is nearly constant for the outer four bins after emission line correction, and roughly at the level of the most central  bin.   The high sensitivity of H$\beta$ and other  Balmer indices to the gas subtraction makes it challenging to estimate the age from  absorption line indices, as the derived ages depend on the adopted emission line correction. For this reason, we refrain from using the indices H$\beta$ and H$\gamma \sigma_{275}$ for spectral index fitting. 
The other indices are  insensitive to our gas emission correction.

When possible, we  measured  indices  on both spectra, the  spectra observed along the major axis, and the spectra with P.A. = 48\degr\space offset.  Some indices (e.g. parts of  Mg\textit{b}, aTiO, TiO$_2$) we can measure only on one spectrum, due to the different gratings and chip gaps. We show  the central four bins of the P.A. = 48\degr\space data in Fig.~\ref{fig:gradient}, as they cover roughly the same region as the P.A. = 0\degr\space data. While we see some differences for the indices measured at different spectra (e.g. at large radii Fe4383, Fe5270), the overall trends agree very well. 

\cite{2002A&A...395..431B} and \cite{2005MNRAS.356.1440D} also measured spectral indices for \object\space at $r\leq$ 4\farcs85 and  $r\leq$ 6\farcs25, though at different P.A. and the resolution of the Lick system.  We transformed their spectral indices  
to  LIS resolution, and compared the measurements. We have good agreement for Fe5270,  Fe5335, Fe5782,   and Mg\textit{b}, with most deviations at  \textless1.5$\sigma$. However, there is  disagreement for some of the Balmer line indices. \cite{2005MNRAS.356.1440D} performed an emission line correction for H$\beta$ by 0.13\,\AA, while  
\cite{2002A&A...395..431B} did not correct their data. The TiO$_2$ measurement of \cite{2005MNRAS.356.1440D} differs by 0.02\,mag from our results, but we have good agreement with \cite{2002A&A...395..431B}. As our S/N in this spectral region is by at least  a factor 12 higher than the S/N obtained by \cite{2005MNRAS.356.1440D}, we consider our measurements more accurate. 


\section{Spectral index stellar population analysis}
\label{sec:spindex}

One approach to measuring stellar population parameters is to use   stellar absorption line indices, or for short, spectral indices. Instead of using the full spectrum, only certain regions are selected.  Spectral indices  respond to several stellar population parameters simultaneously, but with varying sensitivity \citep{1984ApJ...287..586B,1994ApJS...94..687W,1998ApJS..116....1T}.
This method is widely used  for determining age, metallicity, and [$\alpha$/Fe]. 
It has also been applied to measure the IMF  with different combinations of spectral indices  e.g. by \cite{2011ApJ...735L..13V,2012ApJ...760...70V,2013MNRAS.429L..15F, 2013MNRAS.433.3017L, 2014MNRAS.438.1483S, 2015ApJ...803...87S, 2015MNRAS.447.1033M, 2015MNRAS.451.1081M,  2015ApJ...806L..31M, 2016MNRAS.457.1468L, 2018MNRAS.478.4084S}.
Simultaneously estimating the age, IMF, elemental abundances and  metallicity of a stellar population  with spectral indices is extremely challenging. All of these parameters  affect the stellar absorption lines to a certain degree and lead to degenerate results. For this reason, we first   estimate the luminosity-weighted age and derive  [$\alpha$/Fe] of the stellar population, before we evaluate  different IMF slopes.  We give an overview of different spectral indices and their sensitivity to stellar population parameters in Appendix \ref{sec:indexselect}. 

\subsection{Basic stellar population parameters: age, metallicity and [$\alpha$/Fe]}
\label{sec:ages}
In order to test the robustness of our results, we  investigated a number of methods for measuring the age, metallicities and [$\alpha$/Fe] values for a given stellar population. As we shall see, the age results are highly dependent on what is assumed.

\subsubsection{Index-index grids}
\label{sec:alphaest}

\begin{figure*}
\begin{centering}
\includegraphics[width=8cm]{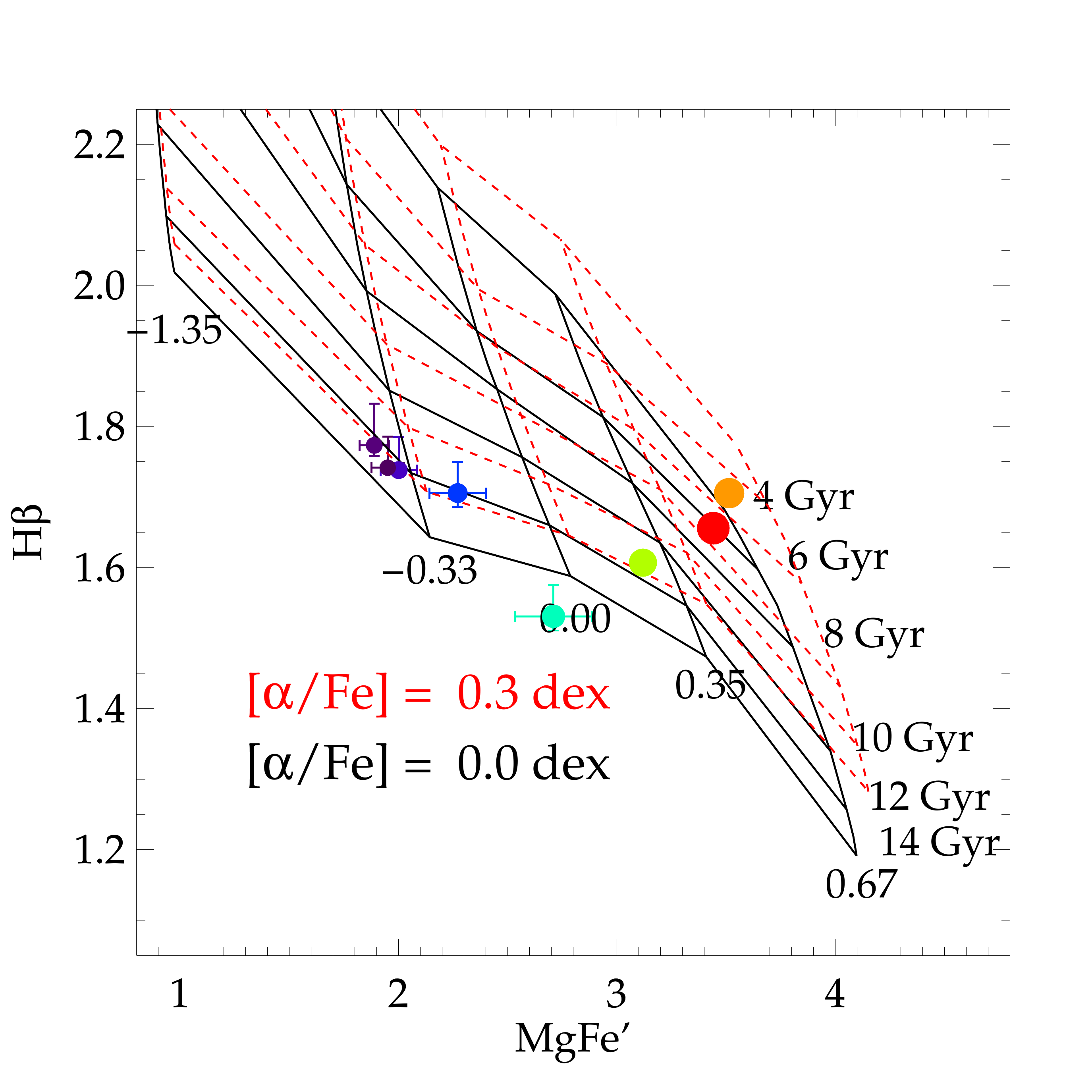}
\includegraphics[width=8cm]{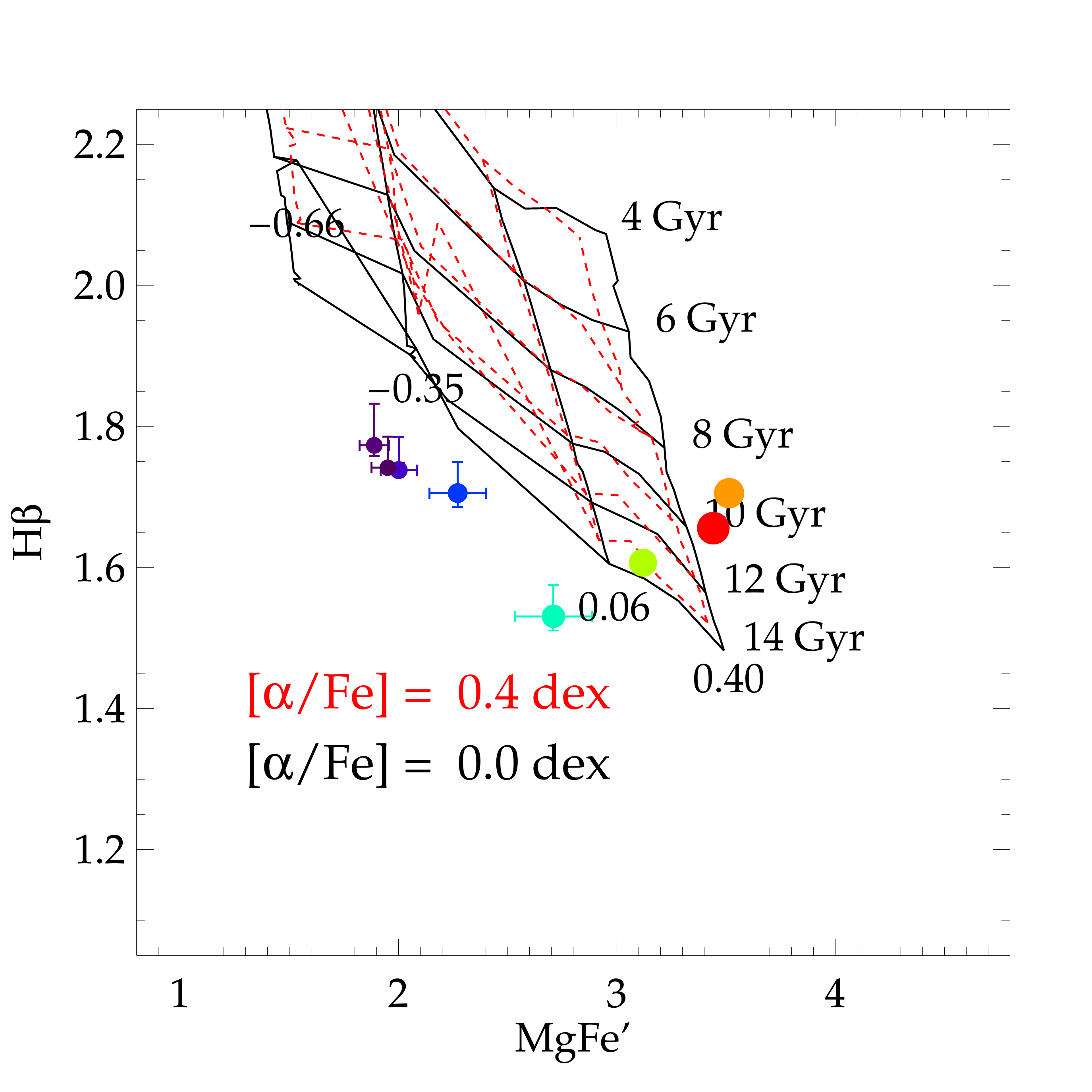}
\caption{Index-index plot of MgFe' and H$\beta$. On the left panels, the lines indicate the \cite{2011MNRAS.412.2183T} (TMJ) SSP model grid with Salpeter IMF, on the right panels the \cite{2015MNRAS.449.1177V} MILES models with Salpeter  IMF slope (unimodal $\Gamma_\text{u}$=1.3). The colored circle points indicate the index measurements, with red to blue color and decreasing symbol size denoting spectra from the center to the outer bins. 
We plot  grids with two different values for  [$\alpha$/Fe] (0.0 and 0.3\,dex; and 0.0 and 0.4\,dex, respectively) indicated by black solid and red dashed lines, which influences the age result. Also  MILES models with different bimodal IMF  result in different ages. For a fixed IMF, the models suggest a younger age for the center of \object\space in the MgFe' and H$\beta$ plane than at  larger radi.}
\label{fig:alphaindexindex}
\end{centering}
\end{figure*}

\begin{figure*}
\begin{centering}
\includegraphics[width=8cm]{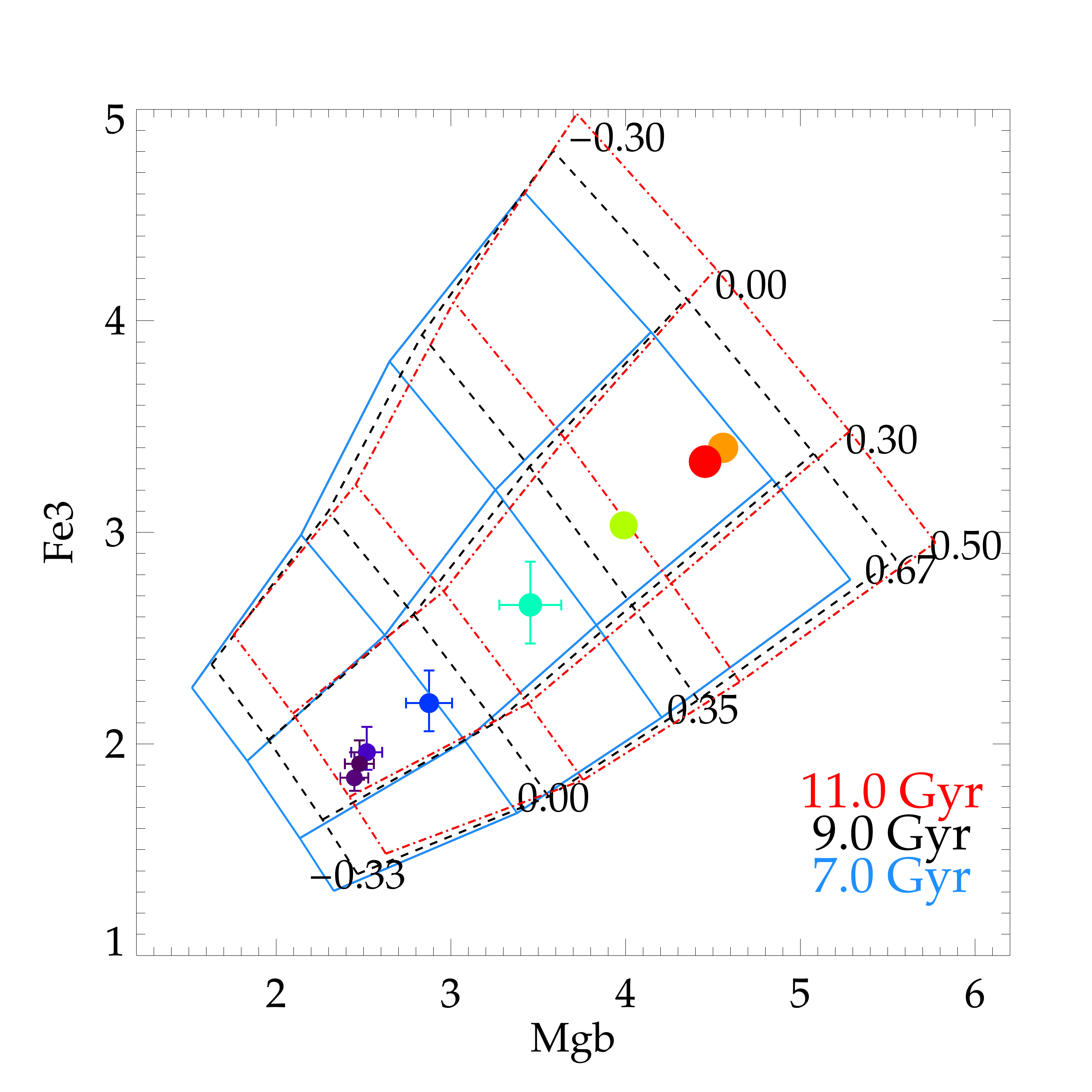}
\includegraphics[width=8cm]{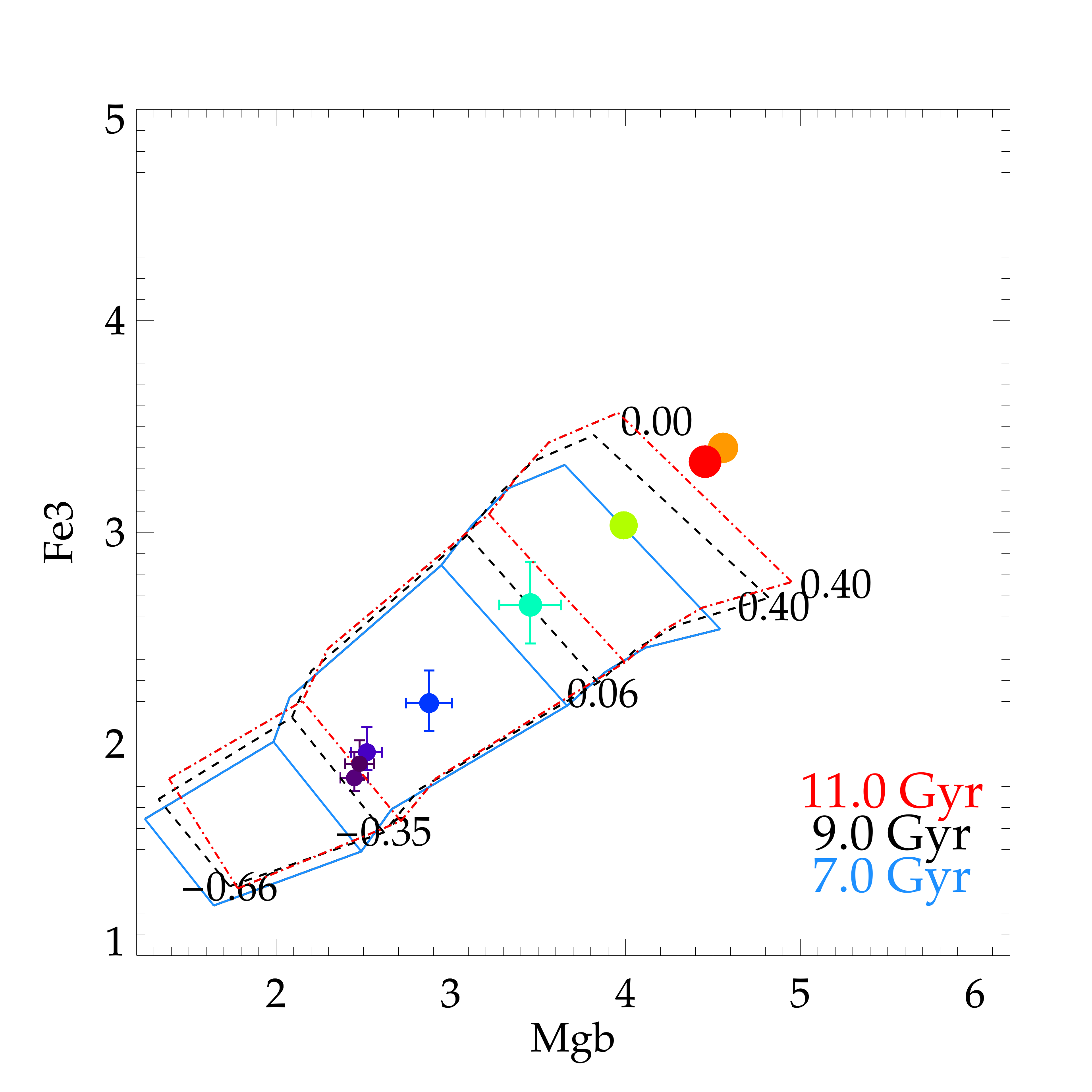}
\caption{Same as Fig. \ref{fig:alphaindexindex}, but for Mg\textit{b} and Fe3. For each model set (TMJ on the left, MILES on the right, both  with Salpeter IMF), we plot  grids with three different ages, denoted by different lines (blue solid lines for 7.0 Gyr, black dashed for 9.0 Gyr, red dot-dashed for 11.0 Gyr).  We  tested also MILES models with different bimodal IMF $\Gamma_\text{b}$=0.3, 1.3, and 3.5 (not shown). All models, ages, and IMF slopes indicate a roughly constant [$\alpha$/Fe]$\sim$0.2\,dex.}
\label{fig:alphaindexindex2}
\end{centering}
\end{figure*}

A common approach for estimating SSP parameters is to compare two  spectral indices with a model grid. We have attempted to estimate the age with the spectral indices H$\beta$ and  MgFe'=$\surd$(Mg\textit{b} $\cdot$ (0.72$\cdot$ Fe5270 + 0.28$\cdot$Fe5335)) (Figure  \ref{fig:alphaindexindex}); and  [$\alpha$/Fe] with Mg\textit{b} and Fe3 = (Fe4383 + Fe5270 + Fe5335)/3 (Figure \ref{fig:alphaindexindex2}). We show the \citet[TMJ, see Appendix~\ref{sec:modelsum} for details]{2011MNRAS.412.2183T}  and MILES \citep{2015MNRAS.449.1177V} SSP model  grids at our resolution of FWHM=14\,\AA. These two grids of SSP models have [$\alpha$/Fe] as parameter (unlike \citealt{2018ApJ...854..139C} SSP models). The TMJ models cover  a larger range of metallicity, and have a finer sampling than the MILES models. 
Large,  red  and orange data points denote the central bins, green, cyan, and blue data points with decreasing size denote indices measured on the outer spectra.

The  H$\beta$ - MgFe'  index-index plot, 
in combination with either model set, suggests a younger age for the center of \object\space than at larger radii. The  data points of the two central bins lie at isochrones for  ages of $\sim$6  Gyr for TMJ models (left panel Fig.~\ref{fig:alphaindexindex});  the outer, emission-line corrected data points  at $\sim$14 Gyr. We obtain a similar trend with the MILES SSP model grid (right panel), but several Gyr older ages. Both  model sets use a \cite{1955ApJ...121..161S} IMF. 
We note that H$\beta$  measurements  that are not emission-corrected  have as small values as H$\beta$=1.3\,\AA, and lie far below the extent of the model grids, at extremely old ages (\textgreater15\,Gyr). This is unphysical and confirms our finding   that  Balmer emission line subtraction is  required for the outer bins.

\cite{2017MNRAS.472.2889C} also studied the stellar populations of \object\space out to 1 R$_h$, to measure age, metallicity, and [$\alpha$/Fe]. 
They used  eight spectral indices (H$\beta$, Fe5015, Mg$_{1}$, Mg$_{2}$, Mg$_{b}$, Fe5270, Fe5335 and Fe5406), five of which overlap with ours, and    the SSP models of \cite{2003MNRAS.339..897T}. 
They also  found younger ages in the center of \object\space with spectral index fitting, but  obtained older ages of about 10\,Gyr  with full-spectral fitting. 
\cite{2017MNRAS.472.2889C} explain this discrepancy with a higher sensitivity of  index fitting to the presence of a young stellar sub-population compared to full-spectral fitting (see also \citealt{2007MNRAS.374..769S}).

To estimate the value of [$\alpha$/Fe], we compared our index measurements of Mg\textit{b} and Fe3 (Fig. \ref{fig:alphaindexindex2}) with the TMJ (left) and MILES (right) models.  All the measurements lie nearly parallel to the lines of constant  [$\alpha$/Fe] (which have positive slope),  between 0.0 and 0.3\,dex. The  SSP grids shown have ages of 7, 9, and 11\,Gyr. We tested different ages (from 6-12 Gyr) and IMF slopes ($\Gamma_\text{b}$=0.3, 1.3, 3.5), and found a  shift of the lines of constant metallicity (which have negative slope), but  the lines of constant [$\alpha$/Fe]   barely moved.   We conclude that the nearly flat [$\alpha$/Fe]-profile derived from this plot is  model- and age-independent. Moreover, the MILES SSP model grids with bottom-heavy to bottom-light  IMF suggest the same constant [$\alpha$/Fe]-profile. 
We note also that the [$\alpha$/Fe] indicated by the Mg\textit{b}-Fe3 plot is in agreement with the measurements of \cite{2017MNRAS.472.2889C}. Their results scatter around [$\alpha$/Fe]=0.27\,dex  from  $\sim$0.15 to 0.4\,dex, with an almost flat [$\alpha$/Fe] profile. 

All index-index plots indicate a metallicity gradient ranging from super-solar ([M/H]\textgreater 0.4\,dex) in the center to sub-solar ([M/H]\textless -0.3\,dex) at 0.8\,R$_\text{h}$. The exact values depend on the assumed age. 

We conclude that our [$\alpha$/Fe] estimate from index-index grids is well-determined: we obtain similar results when we use a different set of  the SSP models, alter the  stellar age by several Gyr, or modify the IMF slope. An alternative approach described in Section \ref{sec:alphaproxy} also gives consistent values for [$\alpha$/Fe]. 
The age estimate from index-index plots is, however, sensitive to changes of [$\alpha$/Fe],  and the H$\beta$ index lies even beyond the range of the  MILES SSP models with a \cite{2001MNRAS.322..231K} IMF. We conclude that the index-index plots are not reliably able to  constrain the stellar age of \object.

\subsubsection{Age constraints  with \textsc{pPXF}}
\label{sec:ageppxf}
In another attempt to constrain the stellar age and metallicity, we did not use spectral indices but rather full spectral fitting with \textsc{pPXF} in the wavelength region $\lambda$= 4000--5600\,\AA, which contains several age- and metallicity-sensitive Balmer and Fe lines.  
As mentioned in Sect.~\ref{sec:kin}, \textsc{pPXF} assigns weights to the template model spectra, and  the optimal template is a linear combination of all input models. This method allows  the determination of  a mass-weighted-mean age and metallicity, consisting of more than one stellar population.  
To determine  age and metallicity,  we used the program in a different manner than in Sect. \ref{sec:kin}, where we fit the LOSVD. The details are given in Appendix \ref{sec:ppxffitapp}. We did not include SSP models with all possible IMF values, but rather fit the spectra for each IMF slope value separately. 
We used the MILES SSP models \citep{2015MNRAS.449.1177V}  
with  seven ages in the range of 2--14\,Gyr, and nine metallicities at
\textit{Z}= $-$1.49, $-$1.26, $-$0.96, $-$0.66, $-$0.35, $-$0.25,  0.06, 0.15, and 0.26\,dex.   We considered  SSP models with [$\alpha$/Fe]=0\,dex and [$\alpha$/Fe]=0.4\,dex in separate fits.

 \begin{figure}
\gridline{\fig{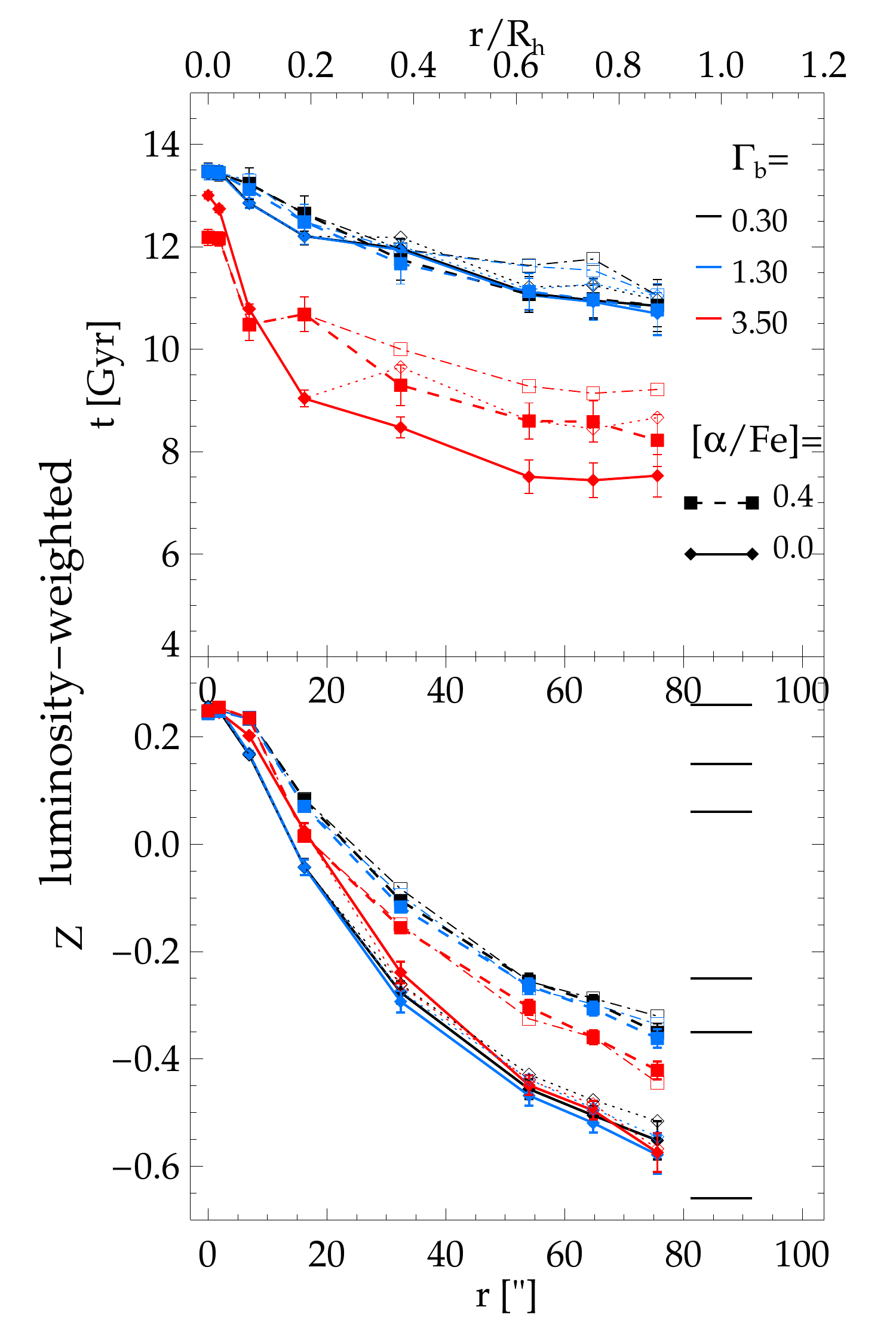}{0.45\textwidth}{}}
\caption{Age and metallicity fit results of \object\space as a function of distance from the galaxy's center, for the data at P.A.=0\degr, obtained with \textsc{pPXF} (Sect.~\ref{sec:ageppxf}) and 
\cite{2015MNRAS.449.1177V} MILES models. 
Colors denote three different fixed IMF slopes $\Gamma_\text{b}$, in particular extremely bottom-light (black, $\Gamma_\text{b}$=0.3), Milky-Way-like (blue, $\Gamma_\text{b}$=1.3) and extremely bottom-heavy (red, $\Gamma_\text{b}$=3.5).
Different symbols and lines denote models with different [$\alpha$/Fe], filled diamond symbols with solid lines are  [$\alpha$/Fe]=0.0\,dex, filled square symbols with dashed lines are [$\alpha$/Fe]=0.4\,dex.
Balmer emission-line correction was only required for the outer four spectra (filled symbols). Omitting the correction results in older ages (open symbols).   Horizontal black lines in the lower panel show the spacing of the SSP models in metallicity space.  While \textit{Z} is rather robust, the age depends on [$\alpha$/Fe], $\Gamma_\text{b}$, and the emission-line correction. 
\label{fig:ppxfagemet}}
\end{figure}

 \begin{figure}
\gridline{\fig{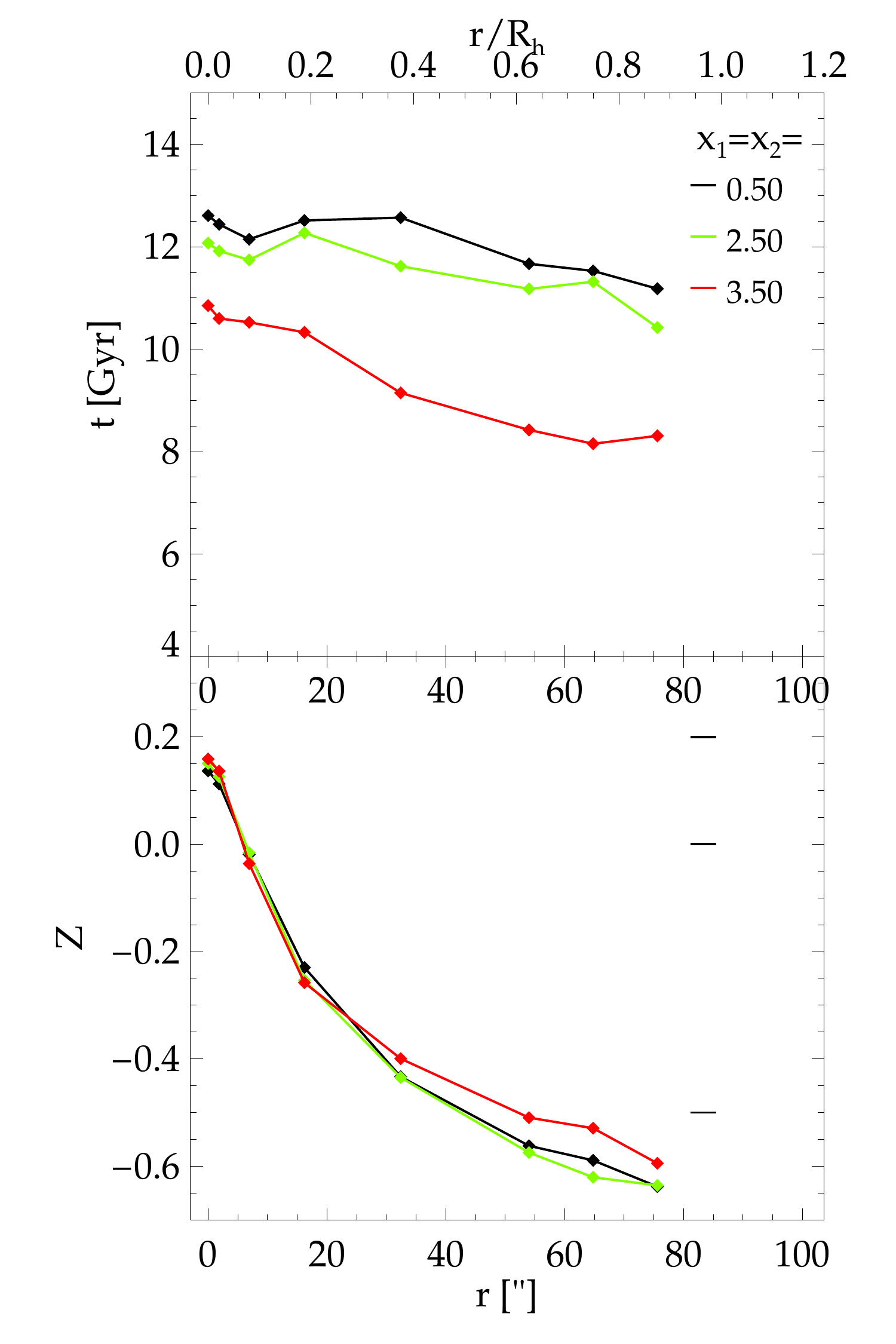}{0.45\textwidth}{}}
\caption{Same as Fig.~\ref{fig:ppxfagemet}, but using the \cite{2018ApJ...854..139C} models with solar abundances. 
Colors denote three different fixed IMF slopes, in particular extremely bottom-light (black, $x_1$=$x_2$=0.5), Salpeter-like (green, $x_1$=$x_2$=2.5) and extremely bottom-heavy (red, $x_1$=$x_2$=3.5).
  Horizontal black lines in the lower panel show the spacing of the SSP models in metallicity space. 
\label{fig:ppxfagemetcvd}}
\end{figure}

We show our luminosity-weighted age and metallicity results as a function of radius in Fig.~\ref{fig:ppxfagemet}, age on the upper panel, metallicity on the lower panel. 
 The results are shown with filled diamond symbols connected by solid lines for [$\alpha$/Fe]=0\,dex, and filled square symbols, connected by dashed lines for [$\alpha$/Fe]=0.4\,dex. Different colors denote different values for the IMF slope $\Gamma_\text{b}$. We show only the results for $\Gamma_\text{b}$=0.3, 1.3, and 3.5, but we fitted MILES models with all 14 possible bimodal  IMF slopes  individually.

The luminosity-weighted metallicity \textit{Z}  of the \textsc{pPXF} fit shows a significant gradient from +0.25\,dex in the center  to about -0.4\,dex at the outermost bin at 1\,R$_h$. 
The total metallicity is not strongly affected by different IMF slopes. The [$\alpha$/Fe]=0.4\,dex SSP models have a higher metallicity \textit{Z}, by 0.1--0.2\,dex compared to the [$\alpha$/Fe]=0.0\,dex models. The emission-line correction changes the metallicity values only slightly, by \textless0.04\,dex.

\begin{deluxetable*}{ll|l|l|l}
\tablecaption{Absorption line index sets \label{tab:indexset2}}
\tablecolumns{5}
\tablewidth{0pt}
\tablehead{
\colhead{Sect.} &
\colhead{Used indices} &
\colhead{Free parameters} &
\colhead{Fixed} &
\colhead{Symbols\tablenotemark{a}}
}
\startdata
\ref{sec:alphaproxy}&Mg\textit{b}&$Z_{\text{Mg}b}$ &$t$, $\Gamma_\text{b}$, [X/H]=0 &\\
\ref{sec:alphaproxy}&Fe4383, Fe5270, Fe5335&$Z_\text{Fe}$&$t$, $\Gamma_\text{b}$, [X/H]=0 &\\
\hline
\ref{sec:indeximf}& MgFe', TiO$_2$, aTiO, bTiO (base set) &$t_\text{p}$, \textit{Z}, $\Gamma_\text{b}$ & [$\alpha$/Fe]=0.2& black x-symbols\\
\ref{sec:indeximf}& MgFe', TiO$_2$, aTiO, bTiO, Ca4592  &$t_\text{p}$,  \textit{Z}, $\Gamma_\text{b}$ & [$\alpha$/Fe]=0.2& blue diamonds\\
\ref{sec:indeximf}\tablenotemark{b}& MgFe', TiO$_2$, aTiO, bTiO, Fe5709&$t_\text{p}$,  \textit{Z}, $\Gamma_\text{b}$ & [$\alpha$/Fe]=0.2& green  triangles\\
\ref{sec:indeximf}& MgFe', TiO$_2$, aTiO, bTiO, Fe5782&$t_\text{p}$,  \textit{Z}, $\Gamma_\text{b}$ & [$\alpha$/Fe]=0.2& orange squares\\
\ref{sec:indeximf}\tablenotemark{b}& MgFe', TiO$_2$, aTiO, bTiO, Ca4592, Fe5709,  Fe5782&$t_\text{p}$,  \textit{Z}, $\Gamma_\text{b}$ & [$\alpha$/Fe]=0.2& red asterisks\\
\ref{sec:indeximf}\tablenotemark{b}& MgFe', TiO$_2$, aTiO, bTiO, Ca4592, Fe5709,  Fe5782, Fe4531, Fe5406&$t_\text{p}$,  \textit{Z}, $\Gamma_\text{b}$ & [$\alpha$/Fe]=0.2& cyan  circles\\
\hline
\ref{sec:indeximf2}& MgFe', TiO$_2$, aTiO, bTiO, Ca4592, Fe5709,  Fe5782&$t_1$, $t_2$, \textit{Z$_1$}, \textit{Z$_2$}, $w_{1/2}$, $\Gamma_\text{b}$ & [$\alpha$/Fe]=0.2& \\
\ref{sec:indeximf2}& MgFe', TiO$_2$, aTiO, bTiO, Ca4592, Fe5709,  Fe5782, Fe4531, Fe5406&$t_1$, $t_2$, \textit{Z$_1$}, \textit{Z$_2$}, $w_{1/2}$, $\Gamma_\text{b}$ & [$\alpha$/Fe]=0.2&cyan  circles \\
\hline
\enddata
\tablenotetext{a}{Figs. \ref{fig:imf1}--\ref{fig:imf2sp}}
\tablenotetext{b}{Most accurate spectral index sets}\tablecomments{age fitting with age prior: $t_\text{p}$ from Sect. \ref{sec:ageppxf}; fixed  [X/H]=0 means only  solar abundances.}
\end{deluxetable*}

Our results for the stellar  age  are more complicated than the metallicity, as the age results depend strongly on our assumptions on  the  IMF slope,  emission line correction, and a lesser degree the [$\alpha$/Fe] abundance. 
While the age difference between  a bottom-light ($\Gamma_\text{b}$=0.3, black) and a Milky Way-like IMF ($\Gamma_\text{b}$=1.3, blue) is rather small, an increasingly bottom-heavy IMF ($\Gamma_\text{b}$=3.5, red)  is several Gyr younger.
The age difference between [$\alpha$/Fe]=0\,dex and [$\alpha$/Fe]=0.4\,dex ranges from +0.8 to -1.6 Gyr, but is negligible except for high values of $\Gamma_b$. 
We note that we applied the Balmer emission line correction derived in Sect.~\ref{sec:kin} for each of our fits. If we instead derived individual Balmer emission line corrections, using each set of model spectra, the age differences derived with [$\alpha$/Fe]=0\,dex and [$\alpha$/Fe]=0.4\,dex SSP models would be $\sim$1.5\,Gyr in most of the outer spectra, irrespective of the assumed IMF slope.

We cannot test the \textsc{pPXF} fitting with the TMJ models, as we do not have the model spectra, only the model indices. 
But we applied  the same fitting approach on a subset of the \cite{2018ApJ...854..139C} models, all with solar abundances. 
The results are shown in Fig.~\ref{fig:ppxfagemetcvd} assuming three different IMF slopes ($x_1$=$x_2$=0.5, 2.5, or 3.5).  Due to the different IMF slope definitions (Appendix~\ref{sec:modelsum}), it is not possible to make a quantitative comparison to the results obtained with MILES models. However, we would like to make a few general comments: (1) The outer four spatial bins indicate that Balmer emission line subtraction is needed, as found with MILES models; (2) for a more bottom heavy IMF-slope, the derived age is lower, as found with MILES models; (3) the age in the center is largest, and decreases with radius; however, the gradient is less significant compared to the MILES models; (4) there is a Z gradient, though it is offset to lower values than with the MILES models.

 Overall, using \textsc{pPXF} we find an age gradient with oldest ages in the center.  This is in contrast to the age estimate  from the index-index grids, where the center is youngest (see Sect.~\ref{sec:alphaest}). 
 \cite{2017MNRAS.472.2889C}  also analyzed data of \object\space with both spectral indices and \textsc{pPXF} fitting.  Like us, they  found  younger ages ($\sim$4\,Gyr)  with spectral indices, but older ages  ($\sim$10\,Gyr) with \textsc{pPXF} in the center of \object. This discrepancy may be caused by a young sub-population, which  biases the Balmer-line-weighted age to the age of the young component, as found by \cite{2007MNRAS.374..769S}. Depending on the age of the young component, this bias is important even though the  mass-fraction of the young sub-population may be far less than 10 per cent. Also \cite{2017MNRAS.472.2889C} found  that spectral indices are more sensitive to a young sub-population than full-spectral fitting.  Full-spectral fitting appears to give  a better estimate of the dominant, older stellar population. For this reason, we consider the \textsc{pPXF} age a better overall age estimate than the spectral index age. 
 
\cite{2017MNRAS.472.2889C} fit their spectra of \object\space with \textsc{pPXF}  using the PEGASE-HR SSP models \citep{2004A&A...425..881L} with  a Salpeter IMF. In this case, they found a flat age profile close to 10\,Gyr along major and minor axes, while we obtained a decreasing age with radius.  A Salpeter IMF  
corresponds roughly to the case $x_1$=$x_2$=2.5 in the \cite{2018ApJ...854..139C} models; and a more bottom-heavy IMF  than $\Gamma_\text{b}$=1.3 in the MILES models (see Figs. \ref{fig:ppxfagemet} and \ref{fig:ppxfagemetcvd}).  \cite{2017MNRAS.472.2889C} did not correct their outer spectra of \object\space for Balmer emission, as we did. 
We  tested omitting  the Balmer emission correction and obtained increased
 best-fit ages by up to $\sim$1.3\,Gyr in the outer bins, and thus a flatter age gradient (see open symbols in Fig. \ref{fig:ppxfagemet}). This is   closer to the results of  \cite{2017MNRAS.472.2889C} and likely accounts for our different age profiles. 

We note that there are several factors that influence  our age estimate: (a) the IMF slope,  (b)  the emission-line correction, and (c) [$\alpha$/Fe]. 
In order to derive the IMF with spectral indices we make two assumptions which must be kept in mind when assessing the significance of any result: First, we  use the emission line correction, because we  found that it  improves the full-spectral fit to SSP models  in the outer bins  (Sect.~\ref{sec:kin} and \ref{sec:alphaest}), and because omission causes  a low  H$\beta$ spectral index, which can only be explained by stellar ages \textgreater15\,Gyr (Sect.~\ref{sec:alphaest}).
Second, as we constrained [$\alpha$/Fe] to be approximately 0.2\,dex at all radii (Sect.~\ref{sec:alphaest} and ~\ref{sec:alphaproxy}), we linearly  interpolated  the \textsc{pPXF} output ages and MILES  SSP model indices to [$\alpha$/Fe]=0.2\,dex.

\subsubsection{Estimation of [$\alpha$/Fe] with spectral-index fitting}
\label{sec:alphaproxy}
We also  applied the method described by \cite{2013MNRAS.433.3017L} to estimate [$\alpha$/Fe]. We  compared the metallicity $Z_{\text{Mg}b}$ derived from Mg\textit{b} with the metallicity $Z_{\text{Fe}}$ derived from Fe3.  To obtain those metallicities, we fixed the IMF slope $\Gamma_\text{b}$ and age to  certain values. 
We minimised $\chi^2$ following the equation
 \begin{equation}
\chi^2(Z,\text{t})=  \sum_i^N\left[\frac{(\text{EW}_i-\text{EW}_{\text{M},i}}{\sigma_{\text{EW}_i}}\right]^2,
\label{eq:chieqage}
\end{equation}
where EW$_i$ denote measurements of the different absorption line indices, $\sigma_\text{EW}$ the respective uncertainties, and EW$_{\text{M},i}$ the SSP model absorption-line indices at a given IMF slope $\Gamma_\text{b}$.    

Using only the Mg\textit{b} index we derived $Z_{\text{Mg}b}$;   using  Fe3 we derived $Z_{\text{Fe}}$, for each bin and spectrum.  \cite{2015MNRAS.449.1177V} showed that  0.59$\cdot(Z_{\text{Mg}b}-Z_{\text{Fe}})$ can be used as solar proxy of [$\alpha$/Fe], with an accuracy  of 0.025\,dex in  [$\alpha$/Fe]. 
We applied the MILES SSP models \citep{2015MNRAS.449.1177V} with [$\alpha$/Fe]=0\,dex, and interpolated the index measurements to  a finer metallicity grid with spacing $\Delta$\textit{Z}=0.02\,dex.  The high central  Mg\textit{b} index of our data made  it necessary  to extrapolate Mg\textit{b} of the models up to  \textit{Z}=0.6\,dex. 

\begin{figure}
\gridline{\fig{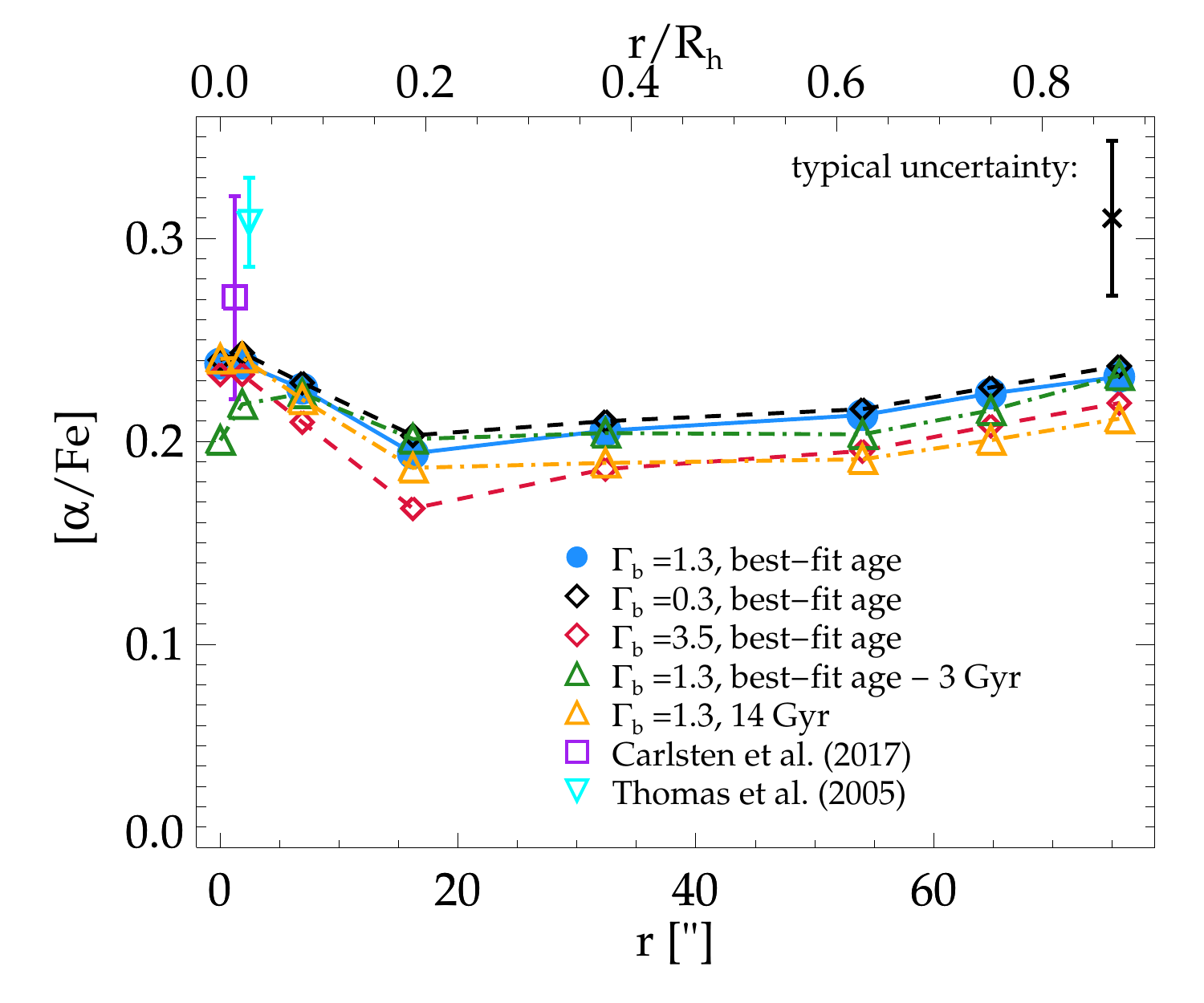}{0.45\textwidth}{}}
\caption{[$\alpha$/Fe] profile, measured from the relation [$\alpha$/Fe]=0.59$\times(Z_{\text{Mg}b}-Z_{\text{Fe}})$ \citep{2015MNRAS.449.1177V}, derived with $\Gamma_\text{b}$=0.3, 1.3, and  3.5 (black, blue, and red colored lines);  for a higher  age of 14\,Gyr at all radii (orange triangles), or a lower age (3\,Gyr less than the best-fit age of Sect.~\ref{sec:ageppxf}, green triangles). The black error bar at the upper right denotes the typical uncertainty of this method.  The central value of \cite{2005ApJ...621..673T}  is a downward-facing cyan triangle,  and \cite{2017MNRAS.472.2889C}  a purple square symbol. 
 \label{fig:alpha}}
\end{figure}

We present our results for the [$\alpha$/Fe] estimate in Fig.~\ref{fig:alpha} as a function of radius. The profile is consistent with being constant  to within the uncertainties. 
We derived [$\alpha$/Fe]  for $\Gamma_\text{b}$=0.3, 1.3, and  3.5  and their respective best-fit ages derived in see Sect.~\ref{sec:ageppxf}  (black, blue, and red colored lines), and the differences are \textless0.04\,dex. This is in agreement with  \cite{2016MNRAS.457.1468L}, who  found that the variation of [$\alpha$/Fe]  for different $\Gamma_\text{b}$ is   at most  0.05\,dex, which corresponds to more than two steps in our metallicity grid.

We compared our results to \cite{2005ApJ...621..673T}, who measured a higher value of  [$\alpha$/Fe] in the central R$_h$/10  of \object\space (0.3\,dex, cyan triangle symbol). They used the indices H$\beta$, Mg\textit{b}, and $\langle$Fe$\rangle$=0.5$\cdot$(Fe5270+
Fe5335) measured by \cite{2002A&A...395..431B} and the models from \cite{2003MNRAS.339..897T}. When we use the \cite{2002A&A...395..431B}   measurements of  Mg\textit{b}, Fe5270, and 
Fe5335 for the central bin, with our Fe4383 measurement and age measurement, we obtain a higher value of [$\alpha$/Fe]=0.28\,dex in the center,  and have  better agreement with \cite{2005ApJ...621..673T}. This suggests that the different value for [$\alpha$/Fe] is mostly caused by  differences of the data rather than the different method or models.

To investigate the effect of age, we derived [$\alpha$/Fe]  with a fixed age of 14\,Gyr (upper bound of our age grid, shown with orange triangles for the case $\Gamma_\text{b}$=1.3), and with an age that is 3\,Gyr less than the best-fit age (green triangles for $\Gamma_\text{b}$=1.3).  [$\alpha$/Fe] changes by no more than 0.04\,dex. However, due to the deep Mg\textit{b}  line in the most central bin, the derived value of $Z_{\text{Mg}b}$ is at the upper bound of the grid (i.e. 0.6\,dex), to  compensate for the younger age (3\,Gyr less than the best-fit age). This leads to a bias in [$\alpha$/Fe] to a lower value. If we force even younger ages, this bias affects also the second and third bins, and can lead to a drop of [$\alpha$/Fe] to $\sim$0.1\,dex in the center. Apart from this bias effect in the center, a different age does not strongly influence [$\alpha$/Fe].  This method is consistent with  our assumption of a constant [$\alpha$/Fe]=0.2, independent of the assumed IMF and age.


\subsection{IMF measurement}
\label{sec:imfindexgen}
In this section we use spectral indices to estimate the IMF slope of \object\space (in Sect.~\ref{sec:pystaff}, we present our results for full spectral fitting). First, we explain our method (Sect.~\ref{sec:indexfithow}); then we explore using different sets of spectral indices,  fitting parameters (listed in Table~\ref{tab:indexset2}), and SSP models for a single stellar population (Sect.~\ref{sec:indeximf}). Then, we test the effect of taking  non-solar abundances into account, and explore which combination of indices and fitting parameters is most accurate and precise using simulated spectra (Sect.~\ref{sec:indeximfab}).  

Since the evolutionary history of a galaxy may not necessarily be described by a simple, single age, but more likely a distribution of ages, we also explore the effects of fitting combinations of two stellar populations (Sect.~\ref{sec:indeximf2}).

\subsubsection{Method}
\label{sec:indexfithow}
To fit the IMF, we 
minimized the expression 
\citep[see][]{2015MNRAS.447.1033M}  

\begin{align}
\begin{aligned}
\chi^2(\Gamma_\text{b},\text{Z},[\text{X/H}],t)= \left[\frac{t(\Gamma_\text{b})-t_\text{M}}{\sigma_\text{t}/\sqrt{N}}\right]^2\\ 
+\sum_i^N\left[\frac{\text{EW}_i-(\text{EW}_{\text{M},i}+\sum_\text{X}\Delta_{\text{X,}i}\cdot[\text{X/H}])}{\sigma_{\text{EW}_i}}\right]^2,
\label{eq:chieq}
\end{aligned}
\end{align}
where EW$_i$ denotes the data absorption-line indices,  $\sigma_\text{EW}$ the uncertainty of the measured spectral indices, EW$_{\text{M},i}$  the SSP model  indices, and $\Delta_{\text{X,}i}$ the correction of the line strength for a non-solar elemental abundance. 

The first term is used to set a prior on the age measurement. We  used  $\Gamma_\text{b}$-dependent  luminosity-weighted ages $t$($\Gamma_\text{b}$) as measured in Sect. \ref{sec:ageppxf}. We  linearly interpolated our measurements at [$\alpha$/Fe]=0.0\,dex and 0.4\,dex to   [$\alpha$/Fe]=0.2\,dex. $\sigma_\text{t}$  is the quadratic sum of the standard deviation of the two measurements (at [$\alpha$/Fe]=0.0\,dex and [$\alpha$/Fe]=0.4\,dex), and the width of a Gaussian fit to the  age distributions  (Sect. \ref{sec:ageppxf}),  
$t_\text{M}$ is the  age  of a SSP model. As in \cite{2015MNRAS.447.1033M}, we re-scale $\sigma_\text{t}$ by dividing it by the square root of the number of indices $N$. This term  biases the age estimate of the $\chi^2$ minimization to our results  from Sect.~\ref{sec:ageppxf}, and helps to break the degeneracy  when using spectral indices that are both age- and IMF-sensitive.

The second term contains the non-solar abundance corrections $\Delta_{\text{X,}i}$, which are used  when we include any elemental abundances [X/H] as fitting parameters, or fix the value of any elemental abundance. 
We derived  the non-solar abundance correction $\Delta_{\text{X,}i}$  from the model spectra of \cite{2018ApJ...854..139C} with a Kroupa IMF, and for a range of elemental abundance variations. 
For a finer sampling, we inter- and extrapolated the index measurements 
in steps of 0.1\,dex to the range $-$0.4\,dex to +0.4\,dex for [Ti/H] and [Fe/H] (original sampling was $-$0.3, 0.0,  +0.3\,dex);  $-$0.3\,dex to +0.3\,dex for [C/H] (original $-$0.15, 0.0, +0.15\,dex); 0\,dex to +1.0\,dex for [Na/H]  (original $-$0.3, 0.0, 0.3, 0.6, 0.9\,dex) and 0\,dex to +0.4\,dex for [O/H] (original 0.0, 0.3\,dex). 
We assume that the response of the indices to a single element is linear, and independent of other elemental variations, as is commonly assumed 
\citep{2013MNRAS.435..952S,2017MNRAS.468.1594A,2018MNRAS.475.1073V,2018MNRAS.477.3954P}. 
We computed the correction $\Delta_{\text{X,}i}$ of each absorption-line index to the variation of a given chemical elemental abundance
\begin{align}
\begin{aligned}
\Delta_{\text{X,}i} = \frac{\text{EW}_{\text{M,}i\text{,[X/H]}= 0}-\text{EW}_{\text{M,}i\text{[X/H]}\neq\,  0}}{\delta([\text{X/H}])}, 
\label{eq:deltael}
\end{aligned}
\end{align}
where $\text{EW}_{\text{M,}i\text{,[X/H]}= 0}$ is the index measured on the reference model spectrum with solar abundances, and $\text{EW}_{\text{M,}i\text{[X/H]}\neq\,  0}$ is the index measured on a model spectrum 
with a variation of one single element  [\text{X/H}].  

We minimized Eq.~\ref{eq:chieq} using two different sets of models for the EW$_{\text{M},i}$:  (1) \cite{2018ApJ...854..139C}  models and  (2) MILES models \citep{2015MNRAS.449.1177V}. The former Conroy  models have  ages ranging from 1-13.5\,Gyr, and \textit{Z} from $-1.0$ to 0.2\,dex, interpolated to steps of 0.1\,dex. 
As we previously derived [$\alpha$/Fe]=0.2\,dex for our data using the Mg$\textit{b}$ index, we  use the non-solar abundance correction for [Mg/H]=0.2\,dex to all Conroy model indices. 
The latter MILES models  have $t_\text{M}$ = 1--14\,Gyr and  \textit{Z} = $-$0.7\,dex to  +0.26\,dex.  We  interpolated the index measurements for a finer metallicity sampling with  $\Delta$\textit{Z} = 0.05\,dex. To account for the supersolar [$\alpha$/Fe], we tested two approaches: We interpolated the MILES model indices with [$\alpha$/Fe]=0.0 and 0.4\,dex to  [$\alpha$/Fe]=0.2\,dex, or we used the non-solar abundance correction for [Mg/H]=0.2\,dex, as for   Conroy model fits. 

In  a subset of fits, we fit two stellar populations (2SPs) instead of one single stellar population using MILES models only. In order to obtain the spectral indices of 2SPs, we combined two of the MILES models with different ages, metallicities, but the same [$\alpha$/Fe] and $\Gamma_\text{b}$. Two SSP models were combined  with weights $w_{1/2}$ = 0.0/1.0, 0.25/0.75, 0.5/0.5, or 0.75/0.25, before we measured the spectral indices on the 2SP models. We did this for [$\alpha$/Fe]=0\,dex and [$\alpha$/Fe]=0.4\,dex, and interpolated our index measurements to  [$\alpha$/Fe]=0.2\,dex. For the 2SP fit, the age prior was set such that it constrained the weighted mean age of the two stellar populations, as detailed further in Sect.~\ref{sec:indeximf2}.

We determined the best-fit parameter values as follows: We transformed our $\chi^2$ from Equ.~\ref{eq:chieq} to the likelihood $L$=$exp(-(\chi^2-min(\chi^2))/2)$, divide $L$ by the sum of all $L$, and marginalize over each fitting parameter. We calculated the  best fit value by taking the weighted mean of each parameter, weighted by the marginalized likelihood. The fitting uncertainties are determined from the range of models within the 1-$\sigma$ confidence limit, which is   calculated for the respective  degree of freedom of each fit. 

\subsubsection{Results for one single stellar population}
\label{sec:indeximf}

\begin{figure}
\gridline{\fig{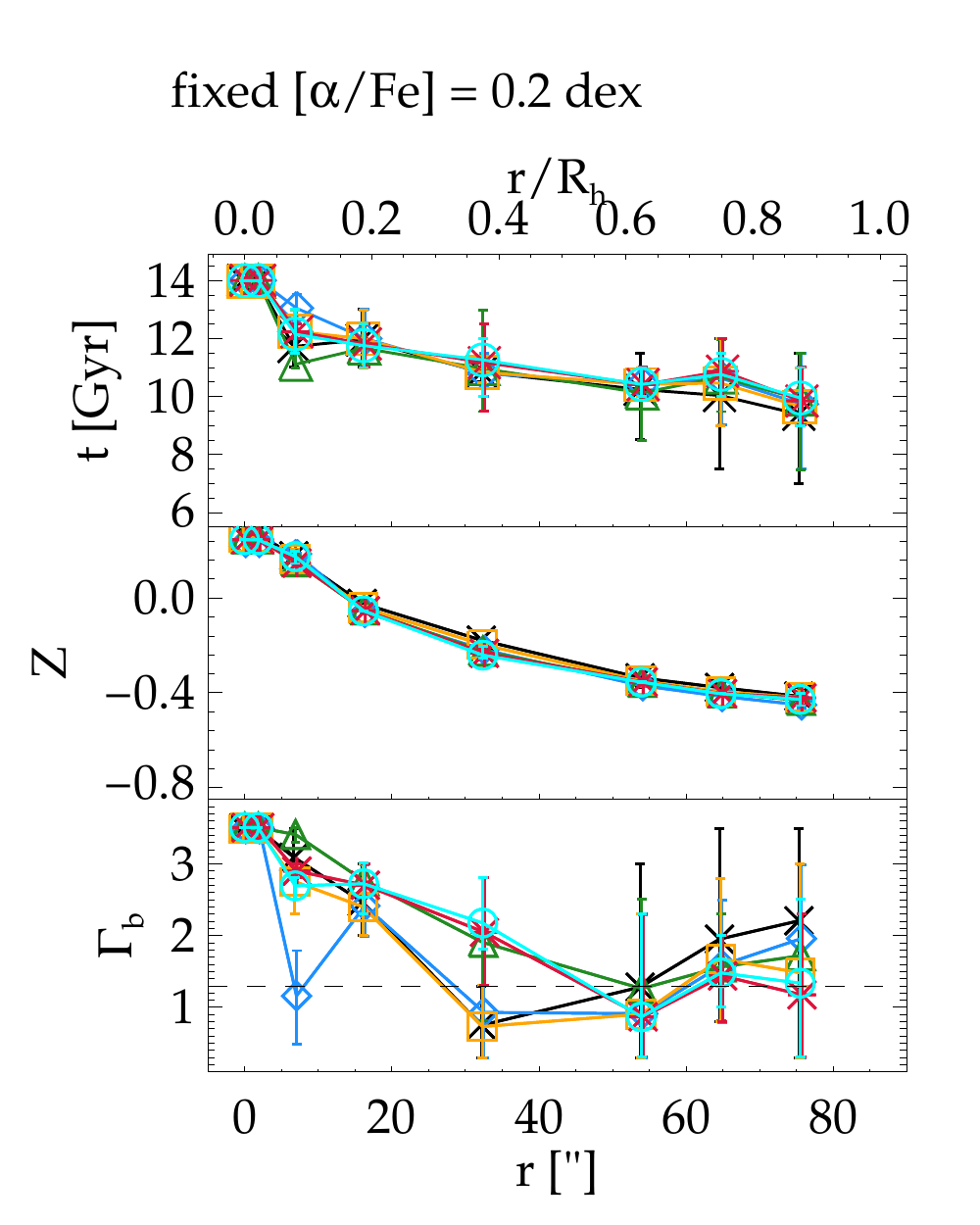}{0.4\textwidth}{}}
\caption{Results of spectral index fitting for different index subsets, fitting stellar age (top panel), metallicity \textit{Z} (center panel), and IMF slope $\Gamma_\text{b}$ (bottom panel), using MILES models interpolated to [$\alpha$/Fe]=0.2\,dex.  The horizontal dashed line denotes a Milky Way-like Kroupa IMF. Different colored symbols denote different index subsets, as listed in Table \ref{tab:indexset2}. 
 \label{fig:imf1}}
\end{figure}

\begin{figure*}
\gridline{
\fig{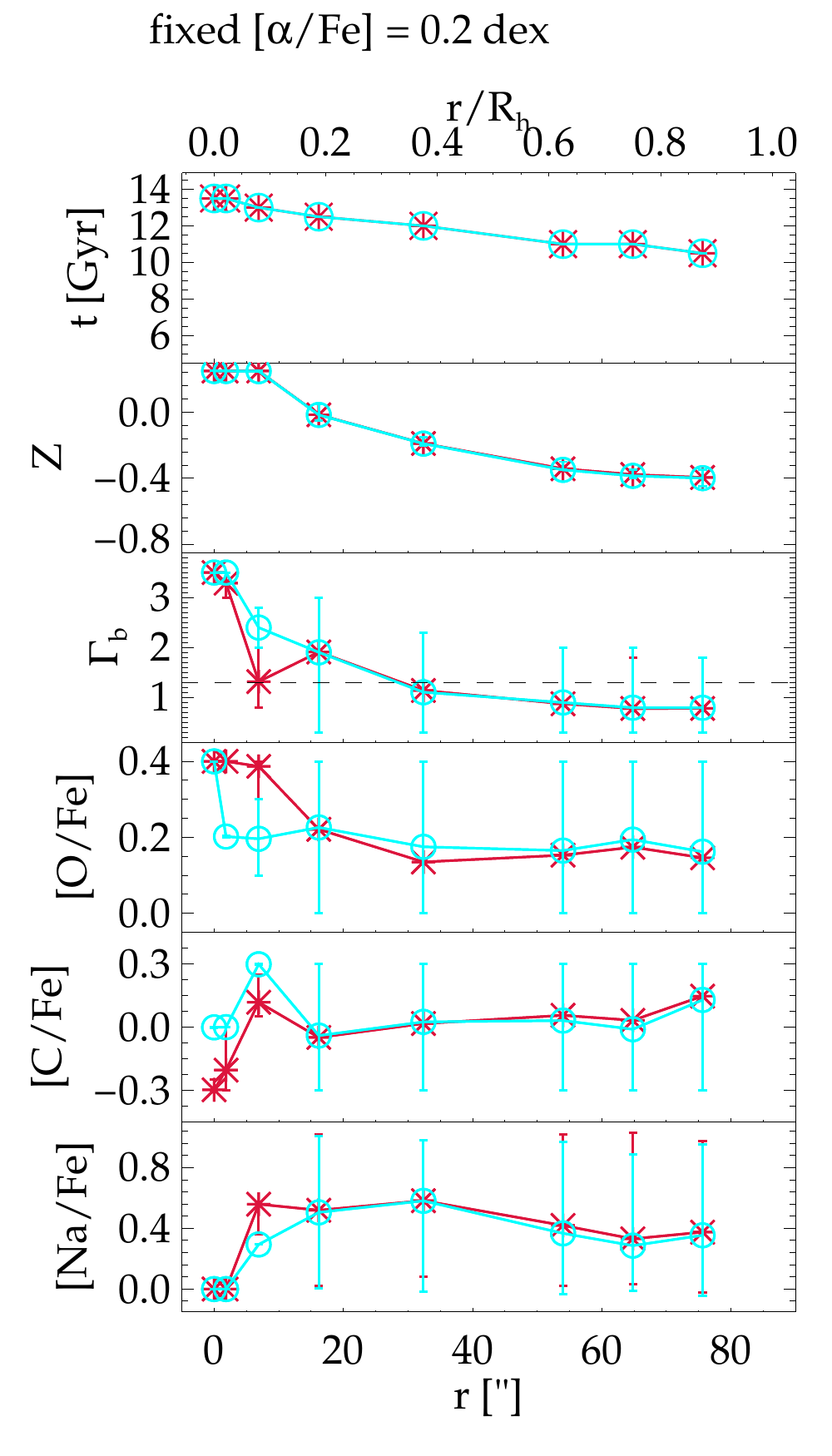}{0.3\textwidth}{}
\fig{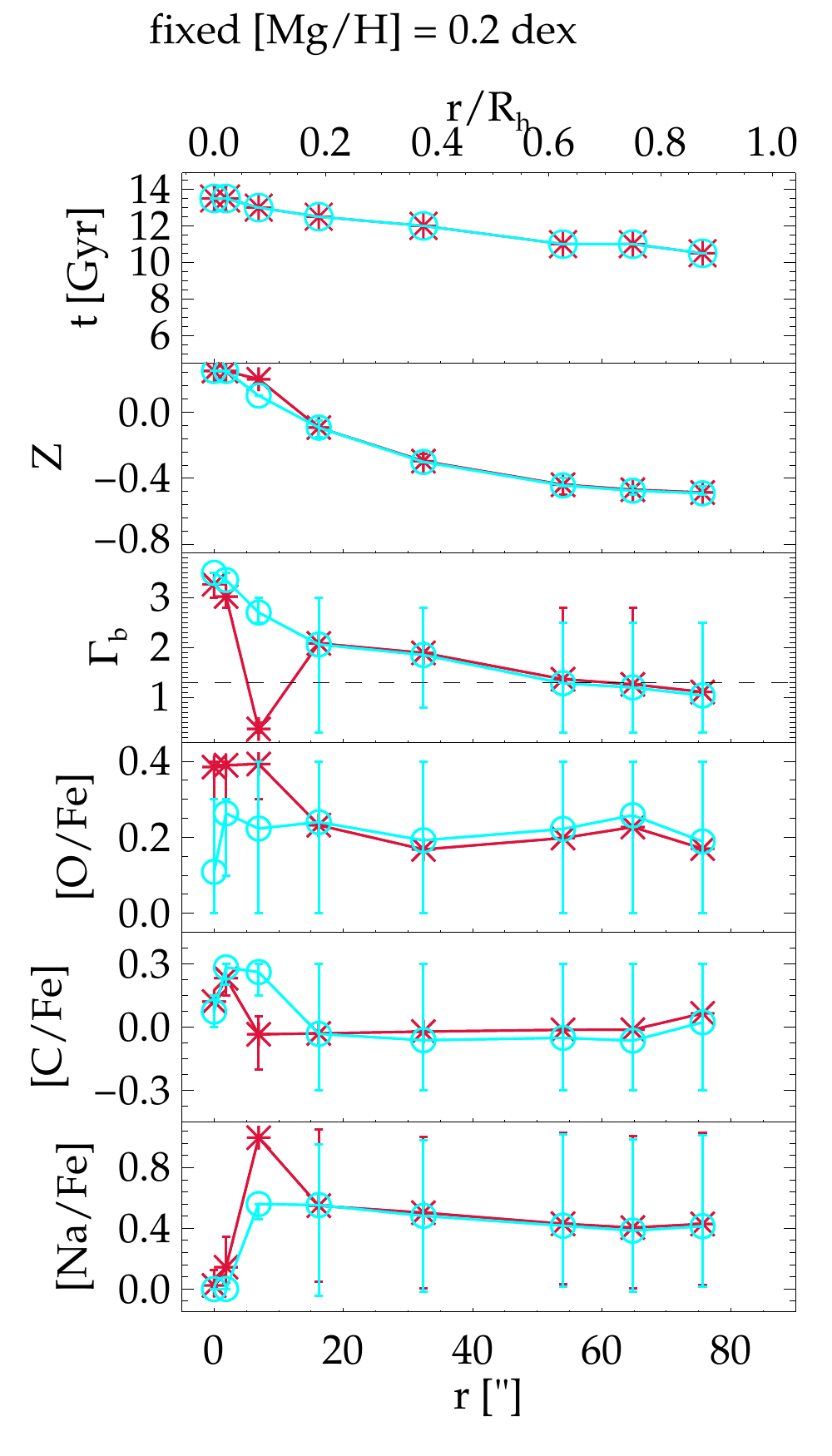}{0.3\textwidth}{}
\fig{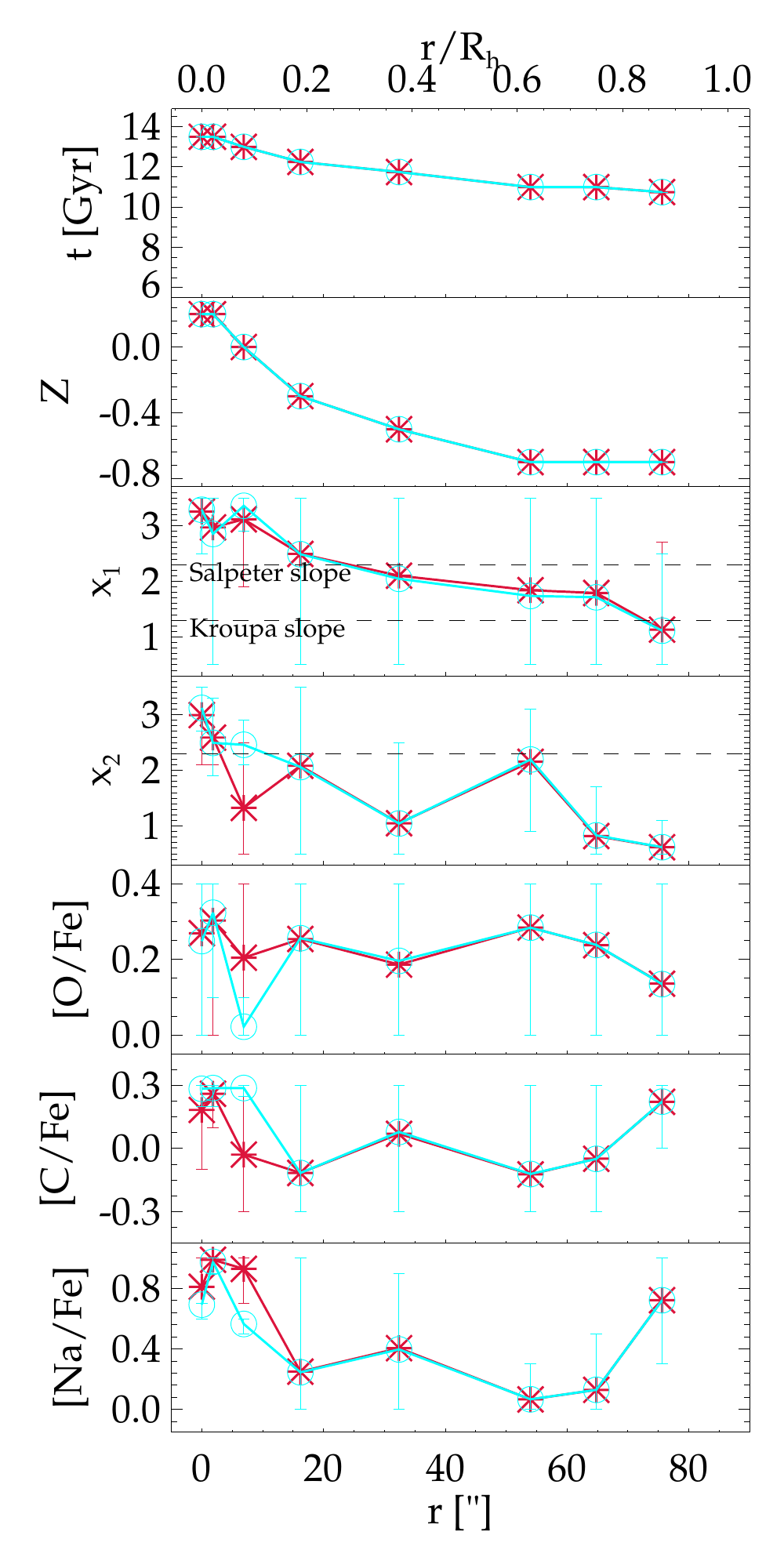}{0.3\textwidth}{}
}
\caption{
Results of spectral index fitting, including [O/H], [C/H], and [Na/H]  as fitting parameters. We used the MILES models interpolated to [$\alpha$/Fe]=0.2\,dex (left panel); the MILES models with [$\alpha$/Fe]=0\,dex, but with a fixed abundance correction  [Mg/H]=0.2\,dex (middle panel); Conroy models with a fixed abundance correction  [Mg/H]=0.2\,dex (right panel).  Note the different IMF parameterization with $\Gamma_\text{b}$ (left and central  panel with MILES models; $dN/d\log{m}\propto m^{-\Gamma_b}$); or $x_1$ and $x_2$ (right panel with Conroy models; $dN/dm\propto m^{-x_i}$), the lower horizontal dashed line denotes a Milky Way-like Kroupa IMF ($x_1$=1.3; $x_2$=2.3); the  dashed lines shown at  $x_1$=$x_2$=2.3 denote a Salpeter IMF slope. Different colored symbols denote different index subsets, as listed in Table \ref{tab:indexset2}. 
\label{fig:imf22}}
\end{figure*}

 Assuming a single stellar population, we fit  the bimodal IMF slope $\Gamma_\text{b}$, stellar metallicity \textit{Z}, and age $t$ with an age prior, which is a function of $\Gamma_\text{b}$.  
 We tried different sets of spectral indices:

Our base set of spectral indices are: MgFe', TiO$_2$, aTiO, and bTiO. This is similar to  the set  used by \cite{2016MNRAS.457.1468L}. They used Mg4780 instead of bTiO, but the index definitions are almost identical. Further, \cite{2016MNRAS.457.1468L} used  TiO$_1$, which  is unavailable to us because of IMACS chip gaps. We measured the spectral indices on the MILES models and investigated their sensitivity to different stellar population parameters.  MgFe' is a strong metallicity indicator  \citep{2003MNRAS.339..897T}, while TiO$_2$, aTiO, and bTiO are sensitive to the IMF, increasing with a larger fraction of low-mass stars (see also Fig.~\ref{fig:indexmodel} top row). This is consistent with  the results of  \cite{2014MNRAS.438.1483S},   using different stellar population models.
We decided to include also IMF-sensitive indices that decrease with a larger fraction of low-mass stars, and are sensitive to the same abundances. In particular, we tested including the indices 
Ca4592, Fe5709, and Fe5782 (Fig.~\ref{fig:indexmodel} bottom row). 
In another set we added the indices Fe4531 and Fe5406. They are mostly sensitive to age, and \textit{Z}, 
but not IMF slope.

There are other prominent indices that we were unable to use for a variety of different reasons: NaD (5900\,\AA) and the near-infrared Ca Triplet lines (8400--8800\,\AA) fall partially on  chip gaps; Na~I (8190\,\AA) was observed with a shorter exposure time and has therefore lower S/N, causing large uncertainties compared to other indices,  Na~I  therefore provides no constraining power;  the bluer Ca~H+K (3900--4000\,\AA)  region is potentially affected by H$\epsilon$ Balmer emission, and more sensitive than other spectral regions to a complex star formation history \citep{2018MNRAS.480.1973K}, which can cause  large discrepancies for  SSP models. Further, if we included the Ca~H+K index, we also would have to include  [Ca/Fe] as additional fitting parameter. The degree-of-freedom of our fit would remain the same and our ability to constrain the IMF would not improve.

Our fit results using MILES models with constant [$\alpha$/Fe] = 0.2\,dex are shown in Fig.~\ref{fig:imf1}, different symbols denote the different index combinations, as listed in Table \ref{tab:indexset2}. The result for the age (upper  panel) is rather stable  for different index combinations, due to the age prior, but for some bins the age results have a standard deviation of up to 0.7\,Gyr. 

The metallicity (middle panel) is less dependent on the index set, while the IMF slope (bottom panel) is affected by the selection of indices. The  standard deviation of $\Gamma_\text{b}$ ranges over  0--0.8 per bin. 
There are some outliers, in particular at the third  and fifth radial  bins (6.9 and 32.4\arcsec), but the overall trend is an extremely bottom-heavy IMF in the center, while the outer bins are consistent with a Kroupa IMF ($\Gamma_\text{b}$=1.3). The outliers are the fits with the  base set  (MgFe', TiO$_2$, aTiO, and bTiO, black x-symbols), the  Ca4592-set (blue diamonds), and   the Fe5782-set (orange squares). Increasing the number of indices produces a smoother IMF gradient, and decreases the uncertainties. We note that a higher age result is usually compensated by a lower IMF slope and vice versa.

For the two central bins the best-fit results of all index sets are at or close to the upper bounds of the models, for all parameters. 
This may be  related to the fact that  the MILES models are not able to match the MgFe' and several other metallicity-sensitive spectral indices in the two central bins, but have  large residuals ($\sim$8--26$\sigma$). As models with higher metallicity are not reliable to use, we had to limit  \textit{Z} to \textless 0.3\,dex, which is compensated by higher ages. We conclude that the uncertainties for our central data points, though formally low, are underestimated.

In addition, we used MILES models with [$\alpha$/Fe]=0\,dex and an elemental abundance correction for [Mg/H]=0.2\,dex instead. This results in slightly lower values for \textit{Z} (by \textless0.1\,dex), and higher values for $\Gamma_\text{b}$ (by 0.1 to 0.8\,dex), especially at large radii. Overall, the \textit{Z} gradient is steeper, and the $\Gamma_\text{b}$ gradient is shallower. The differences compared to Fig.~\ref{fig:imf1} are within the ranges of the uncertainties, but, in most cases, the fits obtain a smaller $\chi^2$ than the [$\alpha$/Fe]=0.2\,dex models.

To investigate the age-IMF dependency, we repeated our fits with fixed ages. We  fixed the stellar ages to the best-fit \textsc{pPXF} value obtained for a Milky Way-like Kroupa IMF ($\Gamma_\text{b}$=1.3). As a consequence, the scatter of $\Gamma_\text{b}$  for different index  combinations decreases, although there are still some outliers.
All changes are within or close to the uncertainties compared to the case where we fit the ages. 
To further test the influence of the stellar age on the results, we fixed the age to 8, 11, and 14\,Gyr. We found that higher ages result in  lower values of the IMF slope. The changes range from 0 to 1.3\,dex of $\Gamma_\text{b}$ per 3 Gyr change. 
 This means that  age and IMF slope are anti-correlated.

In general, when we fit the age and IMF together, we marginalize over all fitting   parameters. This takes the age-IMF degeneracy into account, and causes large uncertainties. 
When we fix the age, the IMF uncertainties are  underestimated, as they do not contain the uncertainty caused by the IMF-age degeneracy. However, as fixing the age to the $\Gamma_\text{b}$=1.3 derived value changes the IMF by \textless1$\sigma$, we fix the age in the following section, and introduce elemental abundances as fitting parameters.

\subsubsection{Results for one single stellar population with element-abundance fit}
\label{sec:indeximfab}

In this section we repeated the spectral index fits, and added elemental abundances as fitting parameters, as the selected indices are sensitive to variations of [O/H],  [C/H], [Ti/H], [Na/H], or [Fe/H]. 
We did  not fit the  base set with abundance parameters, as the degree of freedom of the fit would be zero.

 We ran simulations with mock spectra to test whether we are able to constrain the stellar populations with our sets of indices and elemental abundances. We  constructed mock spectra with  a Kroupa and a bottom-heavy IMF slope, using Conroy models. The details are described in Appendix~\ref{sec:ind-sim}. The main results are as follows: 
The low-mass IMF slope $x_1$ (0.08--0.5\,$M_\sun$) can hardly be constrained, and tends to be overestimated relative to the input values (by up to 2\,dex).  The higher mass slope $x_2$ (0.5--1.0\,$M_\sun$) is better constrained, especially with the Fe5709 and  the combined index sets, though $x_2$ tends to be underestimated (by up to 2\,dex). The most accurate results for age, \textit{Z} and $x_2$ are obtained when  the abundances [Na/H], [O/H] and [C/H] are also fit, though [O/H] and [C/H] are  underestimated, and $x_1$ overestimated (Fig.~\ref{fig:sim_results}, lower right panel).

We fit these index and abundance sets for both MILES and Conroy models. 
As the index fitting has only a small degree of freedom, we fixed the age to the values of Sect.~\ref{sec:ageppxf} for $\Gamma_\text{b}$=1.3, corresponding to Kroupa IMF. The results are shown in Fig.~\ref{fig:imf22}, left panel for MILES models with [$\alpha$/Fe]=0.2\,dex, middle panel for MILES models with [Mg/H]=0.2\,dex, and right panel for Conroy models  with [Mg/H]=0.2\,dex. For all cases the metallicity \textit{Z} has a gradient, but the values are lower by $\sim$0.3\,dex  for the Conroy models for the outer bins. 
There is  disagreement for [Na/H], [O/H] and [C/H] in the central bins between the  SSP model sets, and the abundances   have large uncertainties at large radii.  We conclude that we are not able to unambiguously constrain the considered  
elemental abundances  [Na/H], [O/H], and [C/H] with our  optical spectral index fits for \object.  
Although this fitting set-up works best for simulations, it has large uncertainties for our data, and barely constrains the IMF slope.

\begin{figure}
\gridline{\fig{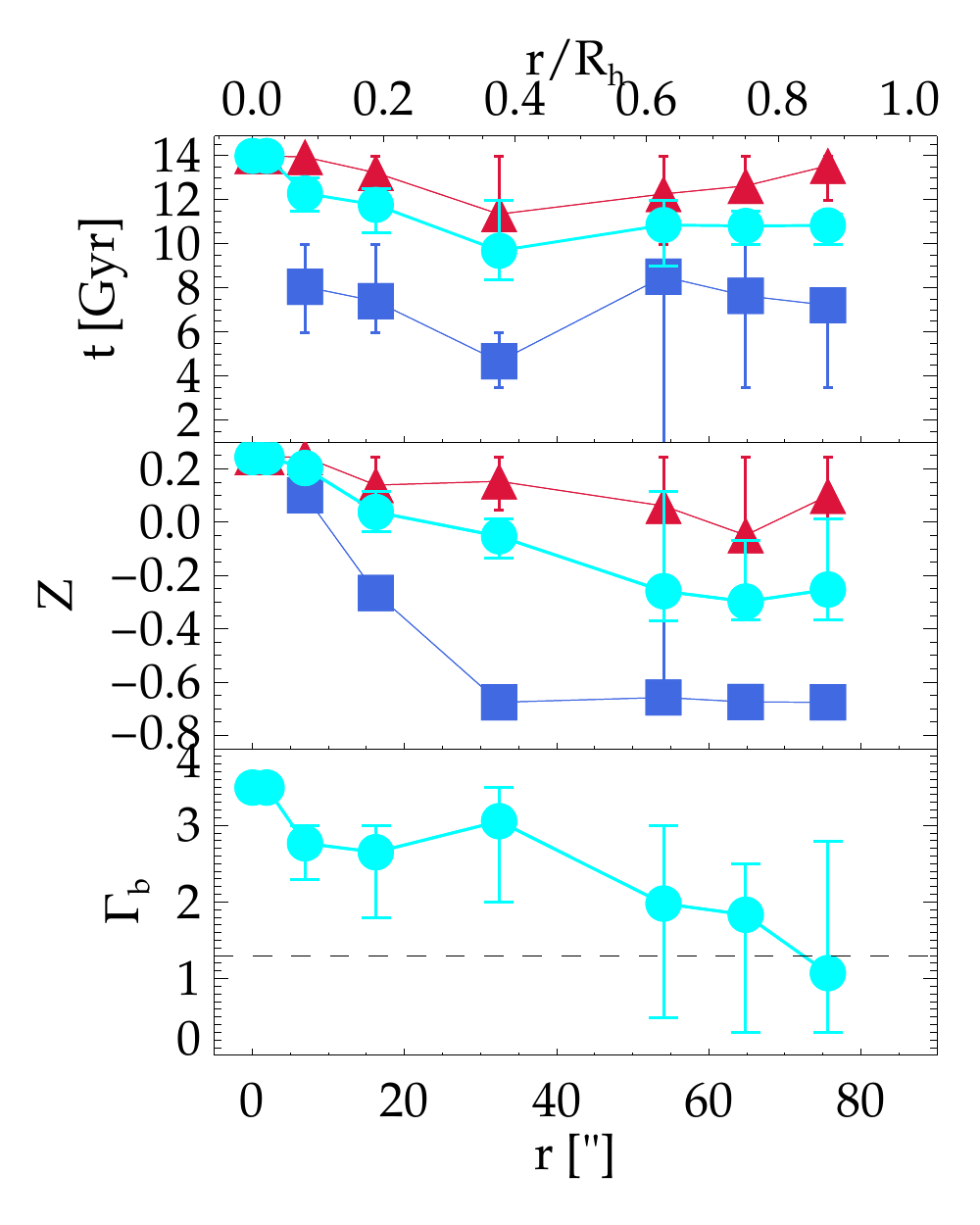}{0.4\textwidth}{}}
\caption{Results of spectral index fitting with a linear combination of two stellar populations of MILES models, fitting stellar ages (with age prior on weighted mean age), metallicities \textit{Z}, and a common IMF slope $\Gamma_\text{b}$. The horizontal dashed line denotes a Milky Way-like Kroupa IMF. Cyan circle symbols denote the weighted mean of age and metallicity, red triangles and blue squares denote the two different stellar populations.   
 \label{fig:imf2sp}}
\end{figure}

\subsubsection{Results for two stellar populations}
\label{sec:indeximf2}

We used linear combinations of two MILES SSP models to fit the IMF slope $\Gamma_\text{b}$ with two stellar populations. Because of the large number of fitting parameters ($\Gamma_\text{b}$, two ages $t_1$, $t_2$, two metallicities \textit{Z$_1$}, \textit{Z$_2$}, weights $w_{1/2}$),  we used only the  index set  with the largest number of indices (see Table~\ref{tab:indexset2}), and did not fit stellar abundances. In this toy model we assumed that both populations have the same [$\alpha$/Fe]=0.2 and $\Gamma_\text{b}$, which is not necessarily true.  We set the age prior on the weighted mean age, i.e. on $t_\text{M}=w_{1/2}\cdot t_1 + (1-w_{1/2})\cdot t_2$. Our stellar population results are shown in Fig.~\ref{fig:imf2sp}. The weighted mean ages and metallicities (cyan circles) are a linear combination of two stellar populations, shown as red triangles and blue squares. Unexpectedly, our best fit consists of  a  younger, more metal-poor population, and an older, more metal-rich population, which contributes the larger weight in most spectral bins. 
The fact that the younger population is more metal-poor than the older population suggests that either this simplified model is wrong or, perhaps that the young and old  stellar populations formed in different environments, and were mixed later. In any case, these results underscore the fact that the results in such analyses depend on initial assumptions about the (unknown) star formation history of galaxies. As \object\space is a shell galaxy, it may have an unusual star formation history.

In comparison with the fit of only one stellar population and the same index subset, the weighted mean metallicity in the outer five bins is higher by up to 0.17\,dex.   The central two bins do not require a second stellar population, and we find the same result as for the 1SP fit. 
The IMF slope is bottom-heavy reaching a maximum value of $\Gamma_\text{b}$=3.5 in the center, and  decreasing to 1.2--2.0 in the outer bins. In the outer four bins $\Gamma_\text{b}$ is higher than for the 1SP fit, though the measurements are in agreement within their uncertainties.


\section{Full spectral-fitting analysis}
\label{sec:pystaff}

Using the full information of a spectrum rather than single measurements on selected spectral regions has the advantage that all available information is used. The number of free parameters is given by the number of pixels in the spectrum rather than the number of spectral indices. This allows a  fit to a  larger number of  stellar population parameters simultaneously, and the opportunity to understand  how they influence each other. Also  \cite{2012ApJ...760...71C}, \cite{2013MNRAS.432.2632P},   \cite{2017ApJ...841...68V}, and \cite{2018MNRAS.479.2443V} used full spectral fitting to constrain the IMF. In this section we  describe our method (Sect.~\ref{sec:methodfsf}), illustrate our results (Sect.~\ref{sec:pyshort}), and discuss degenerate solutions and parameter correlations (Sect.~\ref{sec:fsfdisc}). 

\begin{deluxetable*}{ll|l|l|l}
\tablecaption{\textsc{PyStaff} fit sets \label{tab:fitset2}}
\tablecolumns{5}
\tablewidth{0pt}
\tablehead{
\colhead{Sect.} &
\colhead{Wavelength range [\AA]} &
\colhead{Free parameters} &
\colhead{Fig.} &
\colhead{Symbols\tablenotemark{a}}
}
\startdata
\ref{sec:pyshort}& 4285--4421, 4504--4629, 4744--4893, 5144--& $x_1$, $x_2$, $t$, \textit{Z}, Balmer, [Mg/H],  &  \ref{fig:pystafffit2}a&  green circles\\
& 5657,  5674--5738, 5767--5813,  6068--6417 &[Fe/H], [Ti/H], [O/H], [C/H] &\\\hline
\ref{sec:pyshort}& 4285--4421, 4504--4629, 4744--4893, 5144--& $x_1$, $x_2$, $t$, \textit{Z}, Balmer, [Mg/H],   &  \ref{fig:pystafffit2}b&  blue squares\\
& 5657,  5674--5738, 5767--5813,  6068--6417 &[Fe/H], [Ti/H], [O/H], [C/H] &\\
& &[Na/H], [Ca/H], [N/H], [Si/H] &\\\hline
\ref{sec:pyshort}\tablenotemark{b}& 4285--4421, 4504--4629, 4744--4893, 5144--& $x_1$, $x_2$, $t$, \textit{Z}, Balmer, [Mg/H],   &  \ref{fig:pystafffit}c&  orange diamonds\\
& 5657,  5674--5738, 5767--5813,  5885--6417 &[Fe/H], [Ti/H], [O/H], [C/H] &\\
& &[Na/H], [Ca/H], [N/H], [Si/H] &\\
\hline
\ref{sec:pyshort}& 4285--4421, 4504--4629, 4744--4893, 5144--& $x_1$, $x_2$, $t$, \textit{Z}, Balmer, [Mg/H],   &  \ref{fig:pystafffit}d&  red triangles\\
& 5657,  5674--5738, 5767--5813,  5885--6417,  &[Fe/H], [Ti/H], [O/H], [C/H] &\\
& 8164--8244, 8474--8524 &[Na/H], [Ca/H], [N/H], [Si/H] &\\
\hline
\hline
\enddata
\tablenotetext{a}{Fig. \ref{fig:pystaffout1}, Fig.\ref{fig:pystaffsim}}
\tablenotetext{b}{Preferred optical  fit}
\end{deluxetable*}

\subsection{Method}
\label{sec:methodfsf}
We use the \textsc{python} package \textsc{PyStaff} (Python Stellar Absorption Feature Fitting) developed by \cite{2018MNRAS.479.2443V}, implementing some of the features of \textsc{pPXF}, and  the   SSP models  of \cite{2018ApJ...854..139C}.  
 We fit stellar age [1, 14\,Gyr], \textit{Z} [-1.5, 0.3\,dex], the two IMF slopes [0.5, 3.5], and nine abundances, Na [$-0.45$, 1.0\,dex], Ca [$-0.45$, 0.45\,dex],  Fe [$-0.45$, 0.45\,dex],  C [$-0.2$, 0.2\,dex],  N [$-0.45$, 0.45\,dex],  Ti [$-0.45$, 0.45\,dex],  Mg [$-0.45$, 0.45\,dex],  Si [$-0.45$, 0.45\,dex], and O (=O, Ne, S) [0, 0.45\,dex]. We  tested  including more elemental abundances (e.g. [Cr/H], [Ba/H], [Sr/H]) to improve our fit results, but without noticeable effects.
 
 We included gas emission line templates to fit the Balmer lines, which were 
tied to the same kinematics and fixed the relative fluxes \citep[H$\gamma$=0.47$\times$H$\beta$,][]{2011ApJS..195...13O}. 
We had thus three additional fitting parameters: the gas emission line flux,  gas velocities and  gas velocity dispersion.   In agreement with the \textsc{pPXF} fits, we found that the Balmer emission line correction is negligible in the central bins, but becomes important at larger radii.  

To speed up the fit, we   resampled all spectra to a 1.25\,\AA-spaced wavelength grid. We also  converted from air to vacuum wavelengths, in which the SSP models are computed.  \textsc{PyStaff} models the spectral continuum with polynomial functions in four distinctive wavelength ranges. 
 Besides the chip gaps, we excluded bad pixel regions from the fit.

The \textsc{PyStaff} code uses the Markov-chain Monte-Carlo (MCMC) package \textsc{emcee}
 \citep{2013PASP..125..306F}.  We explored the large parameter space with 100 walkers and 8,000 steps. In addition, we did one test run with 200 walkers and 100,000 steps, 
 and obtained consistent results.

  For an easier comparison with the spectral index fitting, we fit a  wavelength region that corresponds to the spectral indices used in Sect.~\ref{sec:indeximf} with  the combined index set, and in addition the 
 H$\gamma$ and H$\beta$ regions.    The fit included 
 the spectral regions of H$\gamma_A$, Fe4531, bTiO, H$\beta$, Mg\textit{b}, Fe5270, Fe5335, Fe5406, aTiO, Fe5709, Fe5782, and TiO$_2$  indices (see Table~\ref{tab:fitset2} for exact wavelength range); results are denoted as blue square symbols in Fig.~\ref{fig:pystaffout1}. 
To improve the [Na/H] estimate we  included the red parts of the  NaD and TiO$_1$ features (denoted as orange diamond symbols in Fig.~\ref{fig:pystaffout1}.  We further added the near-infrared spectra, which include NaI (8190\,\AA) and part of the Ca1 (8500\,\AA) feature (red triangle symbols  in Fig.~\ref{fig:pystaffout1}). At  shorter wavelengths  ($\sim$3760--4190,\AA) our spectra are noisier (see also Fig.~\ref{fig:snr}), and the fit residuals are larger. It is possible that the discrepancy between models and data is caused by a complex star formation history, to which this wavelength region is more sensitive \citep{2018MNRAS.480.1973K}. To prevent potential biases, we excluded wavelengths \textless4200\,\AA\space from our fit. 
As before, we tested our fitting set-ups  in simulations as described in Appendix~\ref{sec:fsfsim}.

 We also fit  the central four spectra that are at a  P.A.=48\degr\space offset from the major axis, to test the influence of a slightly different wavelength region. 
 We obtained consistent results in most cases, with only small differences for \textit{Z}, [N/H], and [Ca/H]. 
 We conclude that the small difference in the fit wavelength range  due to the different chip gaps is negligible.

 \subsection{Stellar population results}
 \label{sec:pyshort}

 The best-fit stellar age based on \textsc{PyStaff} has a gradient of  $\sim$10 to 7\,Gyr, and is  several Gyr younger than  the ages obtained in Sect.~\ref{sec:spindex}, based on fitting  MILES models, where we found a gradient ranging from $\sim$14 to 10\,Gyr. For the metallicity we obtained a  decreasing profile, with about \textit{Z}$\lesssim$0.2\,dex in the center, decreasing to -0.4\,dex at the outermost bin, similar to our index fitting results with MILES models, but higher  than obtained with index fitting using Conroy models. 
[Mg/H] reaches  $\sim$0.22\,dex in the center, decreasing to 0.12\,dex at 54\,arcsec. We used the Mg\textit{b} index and thus Mg as  [$\alpha$/Fe] tracer  in Sect.~\ref{sec:alphaest} and \ref{sec:alphaproxy} with MILES and TMJ models, and found  values  ranging from    0.25 to 0.19\,dex. In Sect.~\ref{sec:indeximfab} we assumed that these parameters  correspond to each other, and indeed, our result for [Mg/H] is roughly consistent with  our result for [$\alpha$/Fe].

The IMF results based on \textsc{PyStaff} full spectral fitting differ from our spectral index fitting results. $x_1$ is close to a Kroupa IMF slope and constant as a function of radius; however, as for spectral index fitting, $x_1$ is not well constrained. On the other hand, we obtain a  bottom-heavy $x_2$. Though, the measurement uncertainties are large enough that a Kroupa IMF slope is only excluded by 1.3~$\sigma$ (1.8~$\sigma$ with near-infrared data) in the central two bins, but by  more than 2.5~$\sigma$ (2.7~$\sigma$) in the radial bins centered at 6.9 and 16.2\,arcsec.  There is no clear gradient for $x_2$;  at all radial bins $x_2$ is consistent with a value of 2.9 within 1~$\sigma$ (see also Table~\ref{tab:fitres}).

\begin{figure*}[ht!]
\epsscale{0.85}
\plotone{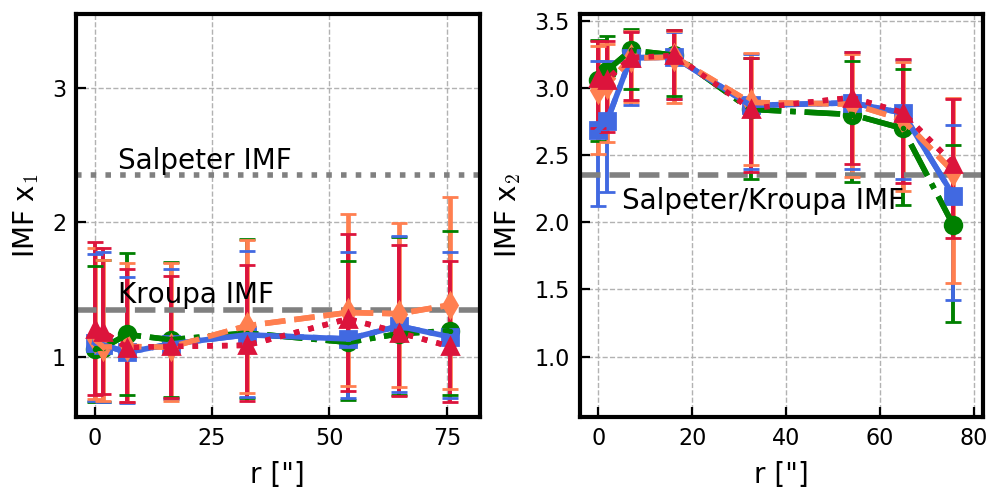}
\epsscale{1.0}
\plotone{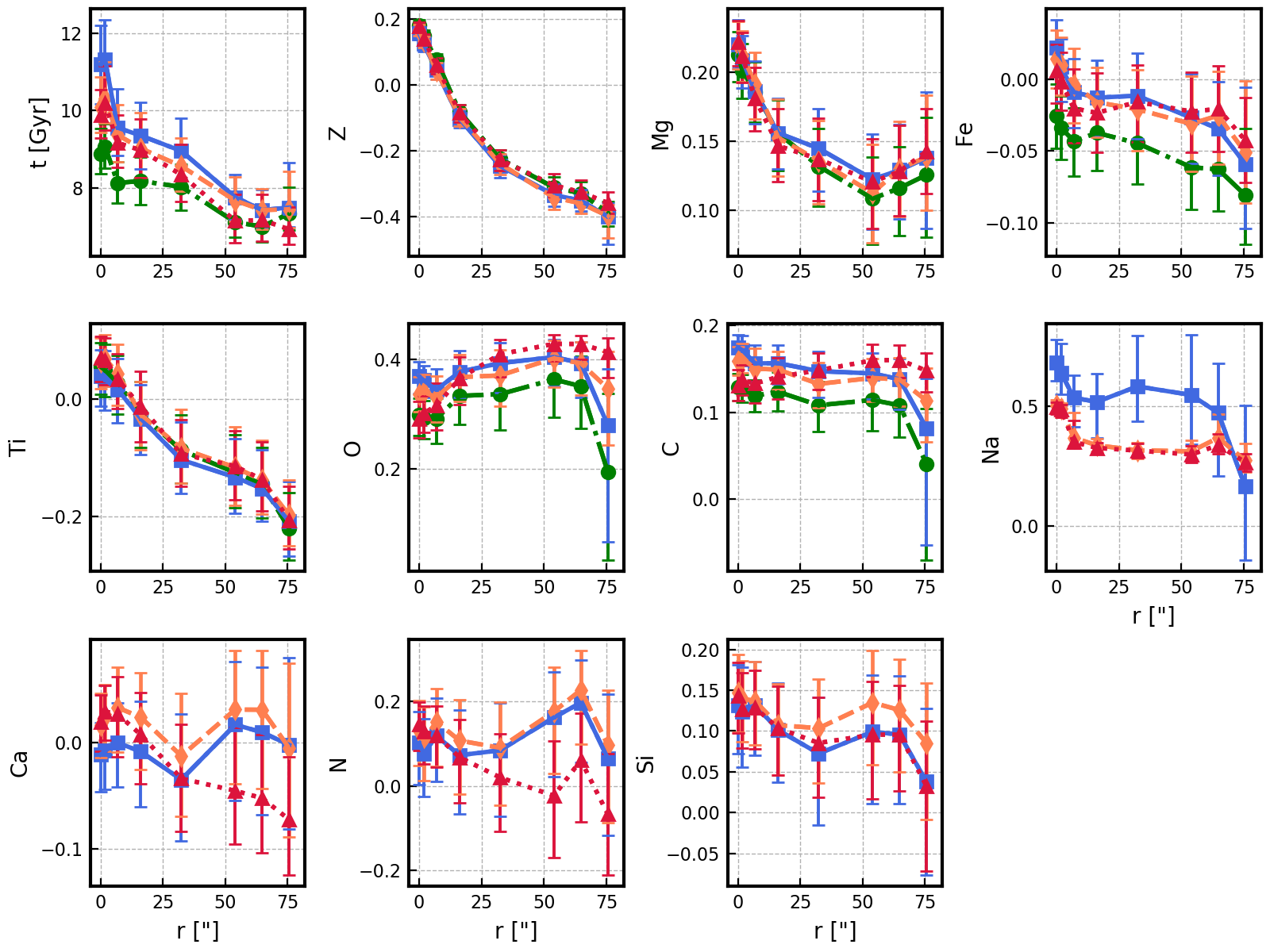}

\caption{Stellar population gradients obtained with \textsc{Pystaff} fit for \object. The different colors denote different fitting set-ups, as listed in Table \ref{tab:fitset2}. 
 The panels show as a function of radius $r$ in arcsec along the major axis, from left to right and top to bottom: IMF slopes $x_1$ (0.08-0.5\,$M_\sun$), and  $x_2$ (0.5-1.0\,$M_\sun$), where the grey dashed horizontal line marks the \cite{2001MNRAS.322..231K} IMF, the dotted line at $x_1$ marks the \cite{1955ApJ...121..161S} IMF; stellar age in Gyr; metallicity \textit{Z};  elemental abundances [Mg/H], [Fe/H], [Ti/H], [O/H], [C/H], [Na/H], [Ca/H], [N/H], [Si/H]. 
\label{fig:pystaffout1}}
\end{figure*}

As in Sect.~\ref{sec:indeximfab}, we tested how the results change when considering  different abundances in the fit. For example, we fit our data with only five  abundances  (green circle symbols in Fig.~\ref{fig:pystaffout1}, details listed in Table~\ref{tab:fitset2}) instead of nine abundance parameters. This decreases the age in the central bins by 1 Gyr. 
Also the results for several elemental abundances are influenced.
We compared the best-fit model spectra to our data  and  found that the 5-abundance best-fit model has larger fit residuals than the  9-abundance best-fit model  in the masked wavelength region 4629--4744\,\AA\space  (see also Fig. \ref{fig:pystafffit2}). 
We  also ran fits of simulated mock spectra (see Appendix~\ref{sec:fsfsim} for details), and found that, indeed, ignoring the abundances [Na/H], [N/H], [Ca/H], and [Si/H] in a 5-abundance fit leads to less accurate results for   age, \textit{Z}, [Fe/H],   and [O/H].
We conclude that it is important to add all nine tested  elemental abundances in the full spectral fit in order  to constrain the stellar populations of \object.

\subsection{Degeneracies,  correlations, and biases}
\label{sec:fsfdisc}

As is well known, several fitting parameters are correlated. We inspected the probability distribution functions (PDF) of our fits and found the following correlations: Age-metallicity anti-correlation,   $x_2$-age anti-correlation, $x_2$-O anti-correlation, O-C correlation, C-Fe correlation, and  Mg-Fe correlations. 
The $x_1$ PDF is peaked at the lower bound at most bins, the $x_2$ PDF at the upper bound in the third and fourth bin. Also [O/H] peaks at the upper bound in some fits. Overall, the PDFs in the fits with and without the near-infrared  wavelength range are very similar, except that the uncertainties are smaller when we include the near-infrared data (see also Table~\ref{tab:fitres}). 

We  used simulations of mock spectra to discover potential biases on our fitting parameters, and evaluate the accuracy and precision of our results (Appendix~\ref{sec:fsfsim}).  
Using our preferred  optical fitting set-up, with 9 abundance parameters  and including the NaD absorption line region, our simulations found accurate results (meaning in agreement with the input value within the uncertainties)  for age, \textit{Z}, $x_1$, $x_2$, [Fe/H], [Ti/H], [O/H], [Ca/H],  [N/H], and [Si/H]. [Mg/Fe] and [Na/Fe]   are slightly overestimated (by $\textless$ 0.01\,dex), [C/Fe] can be  slightly underestimated.  
Interestingly, our simulation of optical spectra show that a bottom-heavy IMF slope can be measured with a higher precision than a Kroupa IMF. The uncertainty for a Kroupa IMF  compared to a bottom-heavy IMF is greater by a factor 2.6 for $x_1$, and by a factor 2.2 for $x_2$.   For galaxies with a bottom-heavy IMF in the center and a Kroupa-like IMF at larger radii, this  complicates the IMF gradient measurement.

To test possible improvements to our analysis, we  added the near-infrared spectral features NaI (8200\,\AA) and the first Ca Triplet line ($\sim$8500\,\AA).   The near-infrared data  does not influence our results of \object\space significantly, but  the uncertainties of the IMF measurements are decreased (see  Table~\ref{tab:fitres}).  Also our simulations show that near-infrared data significantly improve the precision of $x_1$ and $x_2$.

\begin{deluxetable*}{lrrrrrrrr}
\tablecaption{\textsc{PyStaff} fit results \label{tab:fitres}}
\tablecolumns{9}
\tablewidth{0pt}
\tablehead{
\colhead{r [arcsec]} &
\colhead{0.0} &
\colhead{1.847} &
\colhead{6.9} &
\colhead{16.2}&
\colhead{32.4}&
\colhead{54.0}&
\colhead{64.8}&
\colhead{75.6}
}
\startdata
r/$R_h$ & \multicolumn{1}{c}{0.0}& \multicolumn{1}{c}{0.02}& \multicolumn{1}{c}{0.08}&\multicolumn{1}{c}{0.1875}&\multicolumn{1}{c}{0.375}&\multicolumn{1}{c}{0.625}&\multicolumn{1}{c}{0.75}&\multicolumn{1}{c}{0.875}\\
\hline
\hline
\multicolumn{3}{l}{preferred optical fit} \\
$x_1$&$ 1.1^{+0.7}_{-0.5}$&$ 1.1^{+0.6}_{-0.4}$&$ 1.1^{+0.6}_{-0.4}$&$ 1.1^{+0.6}_{-0.4}$&$ 1.2^{+0.6}_{-0.5}$&$ 1.3^{+0.7}_{-0.6}$&$ 1.3^{+0.7}_{-0.5}$&$ 1.4^{+0.8}_{-0.6}$\\
$x_2$&$ 3.0^{+0.3}_{-0.5}$&$ 3.0^{+0.3}_{-0.4}$&$ 3.2^{+0.2}_{-0.3}$&$ 3.2^{+0.2}_{-0.3}$&$ 2.9^{+0.4}_{-0.5}$&$ 2.9^{+0.4}_{-0.5}$&$ 2.8^{+0.4}_{-0.5}$&$ 2.4^{+0.6}_{-0.8}$\\
t [Gyr]&$10.1^{+0.8}_{-0.7}$&$10.3^{+0.9}_{-0.6}$&$ 9.4^{+0.8}_{-0.7}$&$ 9.0^{+0.9}_{-0.8}$&$ 8.6^{+0.7}_{-0.7}$&$ 7.7^{+0.6}_{-0.5}$&$ 7.4^{+0.5}_{-0.5}$&$ 7.4^{+1.0}_{-0.5}$\\
\textit{Z}&$ 0.17^{+0.02}_{-0.02}$&$ 0.13^{+0.02}_{-0.02}$&$ 0.04^{+0.02}_{-0.03}$&$-0.10^{+0.03}_{-0.03}$&$-0.24^{+0.03}_{-0.03}$&$-0.34^{+0.03}_{-0.04}$&$-0.36^{+0.04}_{-0.03}$&$-0.40^{+0.03}_{-0.07}$\\
Mg&$ 0.22^{+0.02}_{-0.02}$&$ 0.21^{+0.02}_{-0.02}$&$ 0.19^{+0.02}_{-0.03}$&$ 0.15^{+0.03}_{-0.03}$&$ 0.14^{+0.03}_{-0.03}$&$ 0.11^{+0.04}_{-0.04}$&$ 0.13^{+0.03}_{-0.03}$&$ 0.14^{+0.04}_{-0.04}$\\
Fe&$ 0.01^{+0.02}_{-0.02}$&$ 0.01^{+0.02}_{-0.02}$&$-0.00^{+0.03}_{-0.03}$&$-0.02^{+0.02}_{-0.02}$&$-0.02^{+0.03}_{-0.03}$&$-0.03^{+0.03}_{-0.03}$&$-0.03^{+0.03}_{-0.03}$&$-0.05^{+0.05}_{-0.04}$\\
Ti&$ 0.06^{+0.04}_{-0.05}$&$ 0.07^{+0.04}_{-0.05}$&$ 0.04^{+0.05}_{-0.05}$&$-0.03^{+0.06}_{-0.06}$&$-0.08^{+0.07}_{-0.06}$&$-0.12^{+0.07}_{-0.06}$&$-0.14^{+0.07}_{-0.06}$&$-0.20^{+0.06}_{-0.05}$\\
O&$ 0.34^{+0.03}_{-0.03}$&$ 0.34^{+0.03}_{-0.03}$&$ 0.33^{+0.04}_{-0.04}$&$ 0.37^{+0.04}_{-0.04}$&$ 0.37^{+0.05}_{-0.06}$&$ 0.40^{+0.03}_{-0.05}$&$ 0.39^{+0.04}_{-0.06}$&$ 0.35^{+0.07}_{-0.10}$\\
C&$ 0.16^{+0.02}_{-0.02}$&$ 0.16^{+0.02}_{-0.02}$&$ 0.15^{+0.02}_{-0.02}$&$ 0.15^{+0.02}_{-0.02}$&$ 0.13^{+0.02}_{-0.03}$&$ 0.14^{+0.02}_{-0.03}$&$ 0.14^{+0.02}_{-0.03}$&$ 0.11^{+0.04}_{-0.05}$\\
Na&$ 0.50^{+0.02}_{-0.03}$&$ 0.49^{+0.02}_{-0.04}$&$ 0.37^{+0.10}_{-0.03}$&$ 0.34^{+0.03}_{-0.02}$&$ 0.32^{+0.03}_{-0.03}$&$ 0.31^{+0.05}_{-0.04}$&$ 0.37^{+0.10}_{-0.05}$&$ 0.27^{+0.07}_{-0.05}$\\
Ca&$ 0.02^{+0.03}_{-0.03}$&$ 0.02^{+0.03}_{-0.03}$&$ 0.03^{+0.04}_{-0.04}$&$ 0.02^{+0.04}_{-0.05}$&$-0.01^{+0.06}_{-0.06}$&$ 0.03^{+0.06}_{-0.07}$&$ 0.03^{+0.06}_{-0.07}$&$-0.00^{+0.08}_{-0.08}$\\
N&$ 0.13^{+0.07}_{-0.08}$&$ 0.11^{+0.08}_{-0.10}$&$ 0.15^{+0.08}_{-0.11}$&$ 0.11^{+0.10}_{-0.13}$&$ 0.09^{+0.11}_{-0.14}$&$ 0.18^{+0.10}_{-0.15}$&$ 0.23^{+0.09}_{-0.13}$&$ 0.10^{+0.13}_{-0.18}$\\
Si&$ 0.15^{+0.05}_{-0.05}$&$ 0.14^{+0.05}_{-0.05}$&$ 0.14^{+0.05}_{-0.05}$&$ 0.11^{+0.05}_{-0.06}$&$ 0.10^{+0.06}_{-0.07}$&$ 0.13^{+0.06}_{-0.08}$&$ 0.13^{+0.06}_{-0.08}$&$ 0.09^{+0.07}_{-0.09}$\\
\hline
\multicolumn{3}{l}{near-infrared fit} \\
$x_1$&$ 1.2^{+0.7}_{-0.5}$&$ 1.2^{+0.6}_{-0.5}$&$ 1.1^{+0.6}_{-0.4}$&$ 1.1^{+0.5}_{-0.4}$&$ 1.1^{+0.6}_{-0.4}$&$ 1.3^{+0.6}_{-0.5}$&$ 1.2^{+0.7}_{-0.5}$&$ 1.1^{+0.6}_{-0.4}$\\
$x_2$&$ 3.1^{+0.3}_{-0.4}$&$ 3.1^{+0.3}_{-0.4}$&$ 3.2^{+0.1}_{-0.3}$&$ 3.2^{+0.2}_{-0.3}$&$ 2.9^{+0.4}_{-0.5}$&$ 2.9^{+0.3}_{-0.5}$&$ 2.8^{+0.4}_{-0.5}$&$ 2.4^{+0.5}_{-0.6}$\\
t [Gyr]&$ 9.9^{+0.7}_{-0.6}$&$10.2^{+1.0}_{-0.7}$&$ 9.2^{+0.7}_{-0.6}$&$ 9.0^{+0.8}_{-0.7}$&$ 8.3^{+0.8}_{-0.7}$&$ 7.1^{+0.7}_{-0.6}$&$ 7.2^{+0.7}_{-0.6}$&$ 6.9^{+0.6}_{-0.4}$\\
\textit{Z}&$ 0.18^{+0.01}_{-0.02}$&$ 0.14^{+0.02}_{-0.02}$&$ 0.06^{+0.02}_{-0.03}$&$-0.09^{+0.02}_{-0.03}$&$-0.23^{+0.03}_{-0.03}$&$-0.31^{+0.03}_{-0.03}$&$-0.33^{+0.04}_{-0.03}$&$-0.36^{+0.03}_{-0.03}$\\
Mg&$ 0.22^{+0.02}_{-0.02}$&$ 0.21^{+0.02}_{-0.02}$&$ 0.18^{+0.02}_{-0.02}$&$ 0.15^{+0.03}_{-0.03}$&$ 0.14^{+0.03}_{-0.03}$&$ 0.12^{+0.03}_{-0.03}$&$ 0.13^{+0.03}_{-0.03}$&$ 0.14^{+0.03}_{-0.03}$\\
Fe&$ 0.01^{+0.02}_{-0.02}$&$-0.00^{+0.02}_{-0.02}$&$-0.02^{+0.03}_{-0.02}$&$-0.02^{+0.03}_{-0.03}$&$-0.02^{+0.03}_{-0.03}$&$-0.02^{+0.03}_{-0.03}$&$-0.02^{+0.03}_{-0.03}$&$-0.04^{+0.03}_{-0.03}$\\
Ti&$ 0.07^{+0.04}_{-0.04}$&$ 0.07^{+0.04}_{-0.04}$&$ 0.03^{+0.04}_{-0.05}$&$-0.01^{+0.06}_{-0.06}$&$-0.09^{+0.06}_{-0.06}$&$-0.11^{+0.06}_{-0.06}$&$-0.14^{+0.06}_{-0.05}$&$-0.21^{+0.06}_{-0.05}$\\
O&$ 0.29^{+0.03}_{-0.03}$&$ 0.30^{+0.03}_{-0.04}$&$ 0.32^{+0.04}_{-0.04}$&$ 0.37^{+0.04}_{-0.05}$&$ 0.41^{+0.03}_{-0.04}$&$ 0.43^{+0.02}_{-0.03}$&$ 0.43^{+0.02}_{-0.03}$&$ 0.41^{+0.03}_{-0.05}$\\
C&$ 0.13^{+0.02}_{-0.02}$&$ 0.14^{+0.02}_{-0.02}$&$ 0.13^{+0.02}_{-0.02}$&$ 0.14^{+0.02}_{-0.03}$&$ 0.15^{+0.02}_{-0.02}$&$ 0.16^{+0.02}_{-0.02}$&$ 0.16^{+0.02}_{-0.02}$&$ 0.15^{+0.02}_{-0.02}$\\
Na&$ 0.50^{+0.02}_{-0.03}$&$ 0.49^{+0.02}_{-0.03}$&$ 0.35^{+0.09}_{-0.03}$&$ 0.33^{+0.02}_{-0.02}$&$ 0.32^{+0.03}_{-0.03}$&$ 0.30^{+0.04}_{-0.03}$&$ 0.34^{+0.05}_{-0.04}$&$ 0.27^{+0.04}_{-0.04}$\\
Ca&$ 0.02^{+0.03}_{-0.03}$&$ 0.03^{+0.02}_{-0.03}$&$ 0.03^{+0.04}_{-0.04}$&$ 0.01^{+0.04}_{-0.05}$&$-0.03^{+0.05}_{-0.05}$&$-0.05^{+0.06}_{-0.05}$&$-0.05^{+0.06}_{-0.05}$&$-0.07^{+0.06}_{-0.05}$\\
N&$ 0.15^{+0.05}_{-0.06}$&$ 0.13^{+0.06}_{-0.08}$&$ 0.12^{+0.07}_{-0.08}$&$ 0.07^{+0.09}_{-0.11}$&$ 0.02^{+0.10}_{-0.13}$&$-0.02^{+0.13}_{-0.15}$&$ 0.06^{+0.11}_{-0.15}$&$-0.07^{+0.14}_{-0.14}$\\
Si&$ 0.14^{+0.04}_{-0.05}$&$ 0.13^{+0.04}_{-0.05}$&$ 0.13^{+0.05}_{-0.05}$&$ 0.10^{+0.05}_{-0.06}$&$ 0.09^{+0.06}_{-0.07}$&$ 0.10^{+0.07}_{-0.08}$&$ 0.10^{+0.06}_{-0.07}$&$ 0.03^{+0.08}_{-0.10}$\\
\hline
\hline
\enddata
\end{deluxetable*}


\section{Discussion}
\label{sec:disc}


\subsection{The age result depends on the method adopted}
We have derived the stellar age under sets of  different assumptions and using a number of different published models. Here we summarize the  results, and the reasons for discrepancies.

In Sec.~\ref{sec:ageppxf} we 
  applied full spectral fitting with \textsc{pPXF} and the  \cite{2015MNRAS.449.1177V} SSP models, and we found that the age is influenced by the assumed IMF. 
 We found an age gradient ranging from $\sim$13\,Gyr in the center to 7.5--11\,Gyr at 1\,$R_h$. Using the same method, but \cite{2018ApJ...854..139C} models, the age gradient is 11--13\,Gyr in the center to 8-11\,Gyr at 1\,$R_h$.  
With \textsc{PyStaff} full-spectral fitting using Conroy models, the  age is  about 10\,Gyr in the center, and 7\,Gyr in the outer bins, and thus up to several Gyr younger.  
 As we noted earlier, the older age obtained in the central bins with \textsc{pPXF} is a consequence of the upper \textit{Z} bound to 0.26\,dex of the MILES models, and solar element abundances. 
 Because of the age-metallicity degeneracy, this lack of high-metallicity templates is  compensated by the oldest SSP model that is available. 
The \textsc{PyStaff} fit can compensate strong absorption lines by adjusting elemental abundances, and not only by older ages. 
For these reasons, we believe that the \textsc{PyStaff} age result in the center is a better estimate.
 
 At large radii, the age measurement is complicated by the possible presence of a second stellar population, which we find with \textsc{pPXF}  and a 2SP spectral index fit. 
 \textsc{PyStaff}  fits only a single stellar population best-fit spectrum,  causing an age difference of about 2.5\,Gyr.
 
 We also compared the spectral indices   H$\beta$ vs. MgFe' with model predictions. The rather high values of these indices in the center indicate a gradient opposite in sign with a younger central age. However, this method can be biased by a young sub-population in the center \citep{2007MNRAS.374..769S}, and is more sensitive to the emission line correction, and low S/N (at large radii). We therefore conclude that the full spectral fitting results are more reliable for the general population. 

Overall, the assumptions regarding [$\alpha$/Fe] or other elemental abundances, the  IMF slope,   gas emission, and one or two stellar populations,  have significant effects on the age results.  
Nevertheless, among the methods we have tested, the two most accurate methods result in an age gradient  of $\sim$3\,Gyr in the central  1\,R$_h$.


\subsection{Optical IMF measurements  in other studies}

A few  studies in the literature have relied on solely  optical spectra: 
\cite{2013MNRAS.432.2632P} used full-spectral fitting of optical spectra (3900--6800\,\AA) of ultra-compact galaxies to derive age, [Fe/H], and the low-mass IMF slope. However,  they did not consider variations of element abundances, which may introduce biases. They also studied which optical spectral regions are most sensitive to the low-mass IMF slope, and found that several IMF sensitive spectral regions were not covered by the standard index definitions.

\cite{2014MNRAS.438.1483S} applied index fitting using different sets of optical spectral indices on stacked SDSS spectra ($\sim$4700--7000\,\AA) of early-type galaxies, and introduced IMF-sensitive optical indices (bTiO, aTiO, CaH1, CaH2).  They had  additional indices in their set  compared to ours, TiO$_1$, CaH$_1$ and CaH$_2$. Assuming   solar metallicity and elemental  abundances, \cite{2014MNRAS.438.1483S} found that, in such a  case, different sets of optical indices can probe the IMF slope. However, they also note that  elemental abundances are needed  to reliably constrain the IMF slope.

In addition  \cite{2016MNRAS.457.1468L} used optical features, as we did, but including  TiO$_1$.  They found that their IMF is in agreement with constraints from the near-infrared Wing-Ford band. They studied the early-type galaxy XSG1, and found that it is   well described by a single stellar population at all radii, without an age gradient. More importantly, their best-fit age from full spectral fitting (3800--6300\,\AA) was not very sensitive to  the chosen IMF slope.  
These facts are not the case for \object: We found an age gradient, an age-IMF degeneracy, and indications for at least two stellar populations at large radii.

It appears  that  \object\space has properties that may make  the IMF measurement more complex than  for other galaxies studied in the literature. 
\object\space is a shell galaxy, and the inner-most shells are located within the extent of our data (at $\sim$8, 19, 30,  46, 56, 70\,arcsec along the major axis, \citealt{1988ApJ...326..596P,2007A&A...467.1011S}). Although the shells are faint, they indicate that \object\space has experienced a  merger event in the past, and  consists of multiple stellar populations \citep{2007A&A...467.1011S}.

\subsection{Comparison of Methods}

We applied an extensive set of model grids (see Appendix~\ref{sec:modelsum}) to constrain the stellar populations of \object\space with spectral index fitting. 
The main sources of uncertainty for spectral index fitting are the choice of spectral indices, elemental  abundance parameters, and the age-metallicity degeneracy.  
We found that our optical spectral region does not provide a large enough number of  spectral indices to unambiguously measure all relevant  elemental abundances  together with \textit{Z}, age, and IMF slope. 

Full spectral fitting enables us to take a large number of elemental abundances into account. 
Since  age, kinematics, and emission line subtraction are all fit simultaneously with the stellar populations, uncertainties from correlations of the fitting parameters are propagated appropriately.  
However, with either method, we find that a Kroupa-like $x_1$ is difficult to constrain, as the  measurement uncertainties are large. 

We extended our full-spectral fitting to the near-infrared and included  the NaI 8200\,\AA\space and the first Ca Triplet feature. Our optical full-spectral fitting results for \object\space  are not significantly changed by including near-infrared features. 
Simulations indicate a higher precision of the IMF slope with near-infrared features.   Other fitting parameters that are already reasonably well-constrained from the optical wavelength range  are not significantly affected.
With near-infrared (only optical) data we obtain a bottom-heavy IMF for $x_2$; a Kroupa IMF is excluded at the 1.2-2.9~$\sigma$ (1.1--2.9~$\sigma$) level in the inner six bins. $x_1$ is consistent with a Kroupa IMF at all radii, and a Salpeter IMF is excluded at 1.6--2.3~$\sigma$  (1.1--2.0~$\sigma$).
We conclude that near-infrared data, as widely used for IMF measurements  \citep[e.g.][]{2012ApJ...760...71C,2013MNRAS.433.3017L,2015MNRAS.447.1033M,2018MNRAS.475.1073V,2018MNRAS.478.4084S,2018MNRAS.479.2443V}
are valuable to constrain the IMF slope in the low-mass range.

For \textsc{PyStaff} full-spectral fitting, we made the assumption that \object\space is dominated by a single stellar population. However, we found indications for more than one stellar population in \object.   Assuming  a single stellar population for a spectrum with complex star formation history increases the residuals in  the   Ca~H+K (3900--4000\,\AA) spectral region \citep{2018MNRAS.480.1973K}. When we included this region to our fits, we  obtained rather large residuals,  which led us to exclude wavelengths \textless4285\,\AA.  We do not know how a complicated star formation history, with the possibility of age dependent elemental abundances and IMF slope,  influences  IMF measurements.

Finally, measuring the IMF depends on the ability of SPS
models to reproduce the properties of old, metal-rich stellar
populations with abundance patterns found in elliptical
galaxies. However, SPS models are constructed using spectral
libraries of stars in the Milky Way, which are deficient in
old, metal-rich stars. Also elemental abundances are different
in elliptical galaxies from the Milky Way, and there are
no, or not enough, stars in stellar libraries with the required
abundance patterns. It is not clear whether this introduces
unknown systematic uncertainties into IMF measurements.

\section{Summary}
\label{sec:sum}
We obtained optical (3700--6600\,\AA) and near-infrared (7900--8500\,\AA) spectroscopic data  for the elliptical  galaxy \object\space with the long-slit spectrograph IMACS on Magellan. The extracted spectra extend out to 1\,R$_h$, and have high S/N\textgreater 100\,\AA$^{-1}$. 
Using a variety of methods we derived and compared stellar ages, metallicities, [$\alpha$/Fe],  elemental abundances, and the IMF slope as a function of radius. We find that \object\space has an approximately constant  [$\alpha$/Fe]=0.2\,dex, a metallicity gradient of $\sim$0.6\,dex, and an age gradient of $\sim$3\,Gyr from the center to 1\,R$_h$.   Our fits of the IMF slope mildly  prefer bottom-heavy values  (1.9--2.9~$\sigma$ within 0.2\,R$_h$). However, the absolute values of the  IMF slope and age depend on the applied method (spectral index fitting or full spectral fitting); assumptions about the  elemental abundances (solar or fit simultaneously); assuming a single or multiple stellar populations; and the choice of SSP models (\citealt{2015MNRAS.449.1177V} or \citealt{2018ApJ...854..139C}). These results underscore the challenges in accurately measuring the IMF in unresolved galaxies.


\acknowledgments
We would like to thank Barry Madore and the  Las Campanas staff who helped us to obtain  the data.  We also thank Laura Sturch and  Marja Seidel for support and advice on data reduction. We are grateful to  Sam Vaughan for sharing the code \textsc{PyStaff}, Charlie Conroy for sharing the SSP models, and Harald Kuntschner, Alina Boecker,  Ryan Leaman for advice and useful suggestions. 
A.F.K. thanks IFIC Valencia-CSIC for hospitality, where a part of this research was carried out.
We also thank the referee for useful comments and suggestions.
This paper includes data gathered with the 6.5 meter Magellan Telescopes located at Las Campanas Observatory, Chile.

\facilities{Magellan (IMACS)} 

\bibliographystyle{yahapj}
\bibliography{bibtex-records}

\begin{thebibliography}{}
\providecommand\natexlab[1]{#1}
\providecommand\JournalTitle[1]{#1}

\bibitem[{{Alton} {et~al.}(2017){Alton}, {Smith}, \&
  {Lucey}}]{2017MNRAS.468.1594A}
{Alton}, P.~D., {Smith}, R.~J., \& {Lucey}, J.~R. 2017,
  \href{http://dx.doi.org/10.1093/mnras/stx464}{\JournalTitle{\mnras}, 468,
  1594}

\bibitem[{{Alton} {et~al.}(2018){Alton}, {Smith}, \&
  {Lucey}}]{2018MNRAS.478.4464A}
---. 2018,
  \href{http://dx.doi.org/10.1093/mnras/sty1242}{\JournalTitle{\mnras}, 478,
  4464}

\bibitem[{{Auger} {et~al.}(2010){Auger}, {Treu}, {Gavazzi}, {Bolton},
  {Koopmans}, \& {Marshall}}]{2010ApJ...721L.163A}
{Auger}, M.~W., {Treu}, T., {Gavazzi}, R., {et~al.} 2010,
  \href{http://dx.doi.org/10.1088/2041-8205/721/2/L163}{\JournalTitle{\apjl},
  721, L163}

\bibitem[{{Barber} {et~al.}(2018){Barber}, {Crain}, \&
  {Schaye}}]{2018MNRAS.479.5448B}
{Barber}, C., {Crain}, R.~A., \& {Schaye}, J. 2018,
  \href{http://dx.doi.org/10.1093/mnras/sty1826}{\JournalTitle{\mnras}, 479,
  5448}

\bibitem[{{Bastian} {et~al.}(2010){Bastian}, {Covey}, \&
  {Meyer}}]{2010ARA&A..48..339B}
{Bastian}, N., {Covey}, K.~R., \& {Meyer}, M.~R. 2010,
  \href{http://dx.doi.org/10.1146/annurev-astro-082708-101642}{\JournalTitle{\araa},
  48, 339}

\bibitem[{{Bekki}(2013)}]{2013MNRAS.436.2254B}
{Bekki}, K. 2013,
  \href{http://dx.doi.org/10.1093/mnras/stt1735}{\JournalTitle{\mnras}, 436,
  2254}

\bibitem[{{Beuing} {et~al.}(2002){Beuing}, {Bender}, {Mendes de Oliveira},
  {Thomas}, \& {Maraston}}]{2002A&A...395..431B}
{Beuing}, J., {Bender}, R., {Mendes de Oliveira}, C., {Thomas}, D., \&
  {Maraston}, C. 2002,
  \href{http://dx.doi.org/10.1051/0004-6361:20021321}{\JournalTitle{\aap}, 395,
  431}

\bibitem[{{Bonnell} {et~al.}(2007){Bonnell}, {Larson}, \&
  {Zinnecker}}]{2007prpl.conf..149B}
{Bonnell}, I.~A., {Larson}, R.~B., \& {Zinnecker}, H. 2007, in Protostars and
  Planets V, ed. B.~{Reipurth}, D.~{Jewitt}, \& K.~{Keil}, 149

\bibitem[{{Burstein} {et~al.}(1984){Burstein}, {Faber}, {Gaskell}, \&
  {Krumm}}]{1984ApJ...287..586B}
{Burstein}, D., {Faber}, S.~M., {Gaskell}, C.~M., \& {Krumm}, N. 1984,
  \href{http://dx.doi.org/10.1086/162718}{\JournalTitle{\apj}, 287, 586}

\bibitem[{{Cappellari}(2017)}]{2017MNRAS.466..798C}
{Cappellari}, M. 2017,
  \href{http://dx.doi.org/10.1093/mnras/stw3020}{\JournalTitle{\mnras}, 466,
  798}

\bibitem[{{Cappellari} \& {Emsellem}(2004)}]{2004PASP..116..138C}
{Cappellari}, M., \& {Emsellem}, E. 2004,
  \href{http://dx.doi.org/10.1086/381875}{\JournalTitle{\pasp}, 116, 138}

\bibitem[{{Cappellari} {et~al.}(2012){Cappellari}, {McDermid}, {Alatalo},
  {Blitz}, {Bois}, {Bournaud}, {Bureau}, {Crocker}, {Davies}, {Davis}, {de
  Zeeuw}, {Duc}, {Emsellem}, {Khochfar}, {Krajnovi{\'c}}, {Kuntschner},
  {Lablanche}, {Morganti}, {Naab}, {Oosterloo}, {Sarzi}, {Scott}, {Serra},
  {Weijmans}, \& {Young}}]{2012Natur.484..485C}
{Cappellari}, M., {McDermid}, R.~M., {Alatalo}, K., {et~al.} 2012,
  \href{http://dx.doi.org/10.1038/nature10972}{\JournalTitle{\nat}, 484, 485}

\bibitem[{{Cappellari} {et~al.}(2013){Cappellari}, {McDermid}, {Alatalo},
  {Blitz}, {Bois}, {Bournaud}, {Bureau}, {Crocker}, {Davies}, {Davis}, {de
  Zeeuw}, {Duc}, {Emsellem}, {Khochfar}, {Krajnovi{\'c}}, {Kuntschner},
  {Morganti}, {Naab}, {Oosterloo}, {Sarzi}, {Scott}, {Serra}, {Weijmans}, \&
  {Young}}]{2013MNRAS.432.1862C}
---. 2013, \href{http://dx.doi.org/10.1093/mnras/stt644}{\JournalTitle{\mnras},
  432, 1862}

\bibitem[{{Carlsten} {et~al.}(2017){Carlsten}, {Hau}, \&
  {Zenteno}}]{2017MNRAS.472.2889C}
{Carlsten}, S.~G., {Hau}, G.~K.~T., \& {Zenteno}, A. 2017,
  \href{http://dx.doi.org/10.1093/mnras/stx2182}{\JournalTitle{\mnras}, 472,
  2889}

\bibitem[{{Cenarro} {et~al.}(2009){Cenarro}, {Cardiel}, {Vazdekis}, \&
  {Gorgas}}]{2009MNRAS.396.1895C}
{Cenarro}, A.~J., {Cardiel}, N., {Vazdekis}, A., \& {Gorgas}, J. 2009,
  \href{http://dx.doi.org/10.1111/j.1365-2966.2009.14839.x}{\JournalTitle{\mnras},
  396, 1895}

\bibitem[{{Cenarro} {et~al.}(2003){Cenarro}, {Gorgas}, {Vazdekis}, {Cardiel},
  \& {Peletier}}]{2003MNRAS.339L..12C}
{Cenarro}, A.~J., {Gorgas}, J., {Vazdekis}, A., {Cardiel}, N., \& {Peletier},
  R.~F. 2003,
  \href{http://dx.doi.org/10.1046/j.1365-8711.2003.06360.x}{\JournalTitle{\mnras},
  339, L12}

\bibitem[{{Chabrier}(2003)}]{2003PASP..115..763C}
{Chabrier}, G. 2003,
  \href{http://dx.doi.org/10.1086/376392}{\JournalTitle{\pasp}, 115, 763}

\bibitem[{{Chabrier} {et~al.}(2014){Chabrier}, {Hennebelle}, \&
  {Charlot}}]{2014ApJ...796...75C}
{Chabrier}, G., {Hennebelle}, P., \& {Charlot}, S. 2014,
  \href{http://dx.doi.org/10.1088/0004-637X/796/2/75}{\JournalTitle{\apj}, 796,
  75}

\bibitem[{{Choi} {et~al.}(2016){Choi}, {Dotter}, {Conroy}, {Cantiello},
  {Paxton}, \& {Johnson}}]{2016ApJ...823..102C}
{Choi}, J., {Dotter}, A., {Conroy}, C., {et~al.} 2016,
  \href{http://dx.doi.org/10.3847/0004-637X/823/2/102}{\JournalTitle{\apj},
  823, 102}

\bibitem[{{Conroy} \& {van Dokkum}(2012{\natexlab{a}})}]{2012ApJ...747...69C}
{Conroy}, C., \& {van Dokkum}, P. 2012{\natexlab{a}},
  \href{http://dx.doi.org/10.1088/0004-637X/747/1/69}{\JournalTitle{\apj}, 747,
  69}

\bibitem[{{Conroy} \& {van Dokkum}(2012{\natexlab{b}})}]{2012ApJ...760...71C}
{Conroy}, C., \& {van Dokkum}, P.~G. 2012{\natexlab{b}},
  \href{http://dx.doi.org/10.1088/0004-637X/760/1/71}{\JournalTitle{\apj}, 760,
  71}

\bibitem[{{Conroy} {et~al.}(2018){Conroy}, {Villaume}, {van Dokkum}, \&
  {Lind}}]{2018ApJ...854..139C}
{Conroy}, C., {Villaume}, A., {van Dokkum}, P.~G., \& {Lind}, K. 2018,
  \href{http://dx.doi.org/10.3847/1538-4357/aaab49}{\JournalTitle{\apj}, 854,
  139}

\bibitem[{{Da Rio} {et~al.}(2009){Da Rio}, {Gouliermis}, \&
  {Henning}}]{2009ApJ...696..528D}
{Da Rio}, N., {Gouliermis}, D.~A., \& {Henning}, T. 2009,
  \href{http://dx.doi.org/10.1088/0004-637X/696/1/528}{\JournalTitle{\apj},
  696, 528}

\bibitem[{{Denicol{\'o}} {et~al.}(2005){Denicol{\'o}}, {Terlevich},
  {Terlevich}, {Forbes}, {Terlevich}, \& {Carrasco}}]{2005MNRAS.356.1440D}
{Denicol{\'o}}, G., {Terlevich}, R., {Terlevich}, E., {et~al.} 2005,
  \href{http://dx.doi.org/10.1111/j.1365-2966.2005.08583.x}{\JournalTitle{\mnras},
  356, 1440}

\bibitem[{{Dotter}(2016)}]{2016ApJS..222....8D}
{Dotter}, A. 2016,
  \href{http://dx.doi.org/10.3847/0067-0049/222/1/8}{\JournalTitle{\apjs}, 222,
  8}

\bibitem[{{Dressler} {et~al.}(2006){Dressler}, {Hare}, {Bigelow}, \&
  {Osip}}]{2006SPIE.6269E..0FD}
{Dressler}, A., {Hare}, T., {Bigelow}, B.~C., \& {Osip}, D.~J. 2006,
  \href{http://dx.doi.org/10.1117/12.670573}{in Society of Photo-Optical
  Instrumentation Engineers (SPIE) Conference Series, Vol. 6269, Society of
  Photo-Optical Instrumentation Engineers (SPIE) Conference Series}, 62690F

\bibitem[{{Ferreras} {et~al.}(2013){Ferreras}, {La Barbera}, {de La Rosa},
  {Vazdekis}, {de Carvalho}, {Falcon-Barroso}, \&
  {Ricciardelli}}]{2013MNRAS.429L..15F}
{Ferreras}, I., {La Barbera}, F., {de La Rosa}, I.~G., {et~al.} 2013,
  \href{http://dx.doi.org/10.1093/mnrasl/sls014}{\JournalTitle{\mnras}, 429,
  L15}

\bibitem[{{Ferreras} {et~al.}(2008){Ferreras}, {Saha}, \&
  {Burles}}]{2008MNRAS.383..857F}
{Ferreras}, I., {Saha}, P., \& {Burles}, S. 2008,
  \href{http://dx.doi.org/10.1111/j.1365-2966.2007.12606.x}{\JournalTitle{\mnras},
  383, 857}

\bibitem[{{Ferreras} {et~al.}(2010){Ferreras}, {Saha}, {Leier}, {Courbin}, \&
  {Falco}}]{2010MNRAS.409L..30F}
{Ferreras}, I., {Saha}, P., {Leier}, D., {Courbin}, F., \& {Falco}, E.~E. 2010,
  \href{http://dx.doi.org/10.1111/j.1745-3933.2010.00941.x}{\JournalTitle{\mnras},
  409, L30}

\bibitem[{{Fitzpatrick}(1999)}]{1999PASP..111...63F}
{Fitzpatrick}, E.~L. 1999,
  \href{http://dx.doi.org/10.1086/316293}{\JournalTitle{\pasp}, 111, 63}

\bibitem[{{Foreman-Mackey} {et~al.}(2013){Foreman-Mackey}, {Hogg}, {Lang}, \&
  {Goodman}}]{2013PASP..125..306F}
{Foreman-Mackey}, D., {Hogg}, D.~W., {Lang}, D., \& {Goodman}, J. 2013,
  \href{http://dx.doi.org/10.1086/670067}{\JournalTitle{\pasp}, 125, 306}

\bibitem[{{Garcia}(1993)}]{1993A&AS..100...47G}
{Garcia}, A.~M. 1993, \JournalTitle{\aaps}, 100, 47

\bibitem[{{Girardi} {et~al.}(2000){Girardi}, {Bressan}, {Bertelli}, \&
  {Chiosi}}]{2000A&AS..141..371G}
{Girardi}, L., {Bressan}, A., {Bertelli}, G., \& {Chiosi}, C. 2000,
  \href{http://dx.doi.org/10.1051/aas:2000126}{\JournalTitle{\aaps}, 141, 371}

\bibitem[{{Gregg}(1994)}]{1994AJ....108.2164G}
{Gregg}, M.~D. 1994,
  \href{http://dx.doi.org/10.1086/117228}{\JournalTitle{\aj}, 108, 2164}

\bibitem[{{Ho} {et~al.}(2011){Ho}, {Li}, {Barth}, {Seigar}, \&
  {Peng}}]{2011ApJS..197...21H}
{Ho}, L.~C., {Li}, Z.-Y., {Barth}, A.~J., {Seigar}, M.~S., \& {Peng}, C.~Y.
  2011,
  \href{http://dx.doi.org/10.1088/0067-0049/197/2/21}{\JournalTitle{\apjs},
  197, 21}

\bibitem[{{Hopkins}(2013)}]{2013MNRAS.433..170H}
{Hopkins}, P.~F. 2013,
  \href{http://dx.doi.org/10.1093/mnras/stt713}{\JournalTitle{\mnras}, 433,
  170}

\bibitem[{{Johansson} {et~al.}(2012){Johansson}, {Thomas}, \&
  {Maraston}}]{2012MNRAS.421.1908J}
{Johansson}, J., {Thomas}, D., \& {Maraston}, C. 2012,
  \href{http://dx.doi.org/10.1111/j.1365-2966.2011.20316.x}{\JournalTitle{\mnras},
  421, 1908}

\bibitem[{{Kacharov} {et~al.}(2018){Kacharov}, {Neumayer}, {Seth},
  {Cappellari}, {McDermid}, {Walcher}, \& {B{\"o}ker}}]{2018MNRAS.480.1973K}
{Kacharov}, N., {Neumayer}, N., {Seth}, A.~C., {et~al.} 2018,
  \href{http://dx.doi.org/10.1093/mnras/sty1985}{\JournalTitle{\mnras}, 480,
  1973}

\bibitem[{{Kausch} {et~al.}(2015){Kausch}, {Noll}, {Smette}, {Kimeswenger},
  {Barden}, {Szyszka}, {Jones}, {Sana}, {Horst}, \&
  {Kerber}}]{2015A&A...576A..78K}
{Kausch}, W., {Noll}, S., {Smette}, A., {et~al.} 2015,
  \href{http://dx.doi.org/10.1051/0004-6361/201423909}{\JournalTitle{\aap},
  576, A78}

\bibitem[{{Kroupa}(2001)}]{2001MNRAS.322..231K}
{Kroupa}, P. 2001,
  \href{http://dx.doi.org/10.1046/j.1365-8711.2001.04022.x}{\JournalTitle{\mnras},
  322, 231}

\bibitem[{{Krumholz}(2014)}]{2014PhR...539...49K}
{Krumholz}, M.~R. 2014,
  \href{http://dx.doi.org/10.1016/j.physrep.2014.02.001}{\JournalTitle{\physrep},
  539, 49}

\bibitem[{{Kuntschner}(2004)}]{2004A&A...426..737K}
{Kuntschner}, H. 2004,
  \href{http://dx.doi.org/10.1051/0004-6361:20041414}{\JournalTitle{\aap}, 426,
  737}

\bibitem[{{Kuntschner} {et~al.}(2001){Kuntschner}, {Lucey}, {Smith}, {Hudson},
  \& {Davies}}]{2001MNRAS.323..615K}
{Kuntschner}, H., {Lucey}, J.~R., {Smith}, R.~J., {Hudson}, M.~J., \& {Davies},
  R.~L. 2001,
  \href{http://dx.doi.org/10.1046/j.1365-8711.2001.04263.x}{\JournalTitle{\mnras},
  323, 615}

\bibitem[{{La Barbera} {et~al.}(2013){La Barbera}, {Ferreras}, {Vazdekis}, {de
  la Rosa}, {de Carvalho}, {Trevisan}, {Falc{\'o}n-Barroso}, \&
  {Ricciardelli}}]{2013MNRAS.433.3017L}
{La Barbera}, F., {Ferreras}, I., {Vazdekis}, A., {et~al.} 2013,
  \href{http://dx.doi.org/10.1093/mnras/stt943}{\JournalTitle{\mnras}, 433,
  3017}

\bibitem[{{La Barbera} {et~al.}(2017){La Barbera}, {Vazdekis}, {Ferreras},
  {Pasquali}, {Allende Prieto}, {R{\"o}ck}, {Aguado}, \&
  {Peletier}}]{2017MNRAS.464.3597L}
{La Barbera}, F., {Vazdekis}, A., {Ferreras}, I., {et~al.} 2017,
  \href{http://dx.doi.org/10.1093/mnras/stw2407}{\JournalTitle{\mnras}, 464,
  3597}

\bibitem[{{La Barbera} {et~al.}(2016){La Barbera}, {Vazdekis}, {Ferreras},
  {Pasquali}, {Cappellari}, {Mart{\'\i}n-Navarro}, {Sch{\"o}nebeck}, \&
  {Falc{\'o}n-Barroso}}]{2016MNRAS.457.1468L}
---. 2016,
  \href{http://dx.doi.org/10.1093/mnras/stv2996}{\JournalTitle{\mnras}, 457,
  1468}

\bibitem[{{Lagattuta} {et~al.}(2017){Lagattuta}, {Mould}, {Forbes}, {Monson},
  {Pastorello}, \& {Persson}}]{2017ApJ...846..166L}
{Lagattuta}, D.~J., {Mould}, J.~R., {Forbes}, D.~A., {et~al.} 2017,
  \href{http://dx.doi.org/10.3847/1538-4357/aa8563}{\JournalTitle{\apj}, 846,
  166}

\bibitem[{{Le Borgne} {et~al.}(2004){Le Borgne}, {Rocca-Volmerange},
  {Prugniel}, {Lan{\c{c}}on}, {Fioc}, \& {Soubiran}}]{2004A&A...425..881L}
{Le Borgne}, D., {Rocca-Volmerange}, B., {Prugniel}, P., {et~al.} 2004,
  \href{http://dx.doi.org/10.1051/0004-6361:200400044}{\JournalTitle{\aap},
  425, 881}

\bibitem[{{Leier} {et~al.}(2016){Leier}, {Ferreras}, {Saha}, {Charlot},
  {Bruzual}, \& {La Barbera}}]{2016MNRAS.459.3677L}
{Leier}, D., {Ferreras}, I., {Saha}, P., {et~al.} 2016,
  \href{http://dx.doi.org/10.1093/mnras/stw885}{\JournalTitle{\mnras}, 459,
  3677}

\bibitem[{{Li} {et~al.}(2010){Li}, {Wang}, {Abel}, \&
  {Nakamura}}]{2010ApJ...720L..26L}
{Li}, Z.-Y., {Wang}, P., {Abel}, T., \& {Nakamura}, F. 2010,
  \href{http://dx.doi.org/10.1088/2041-8205/720/1/L26}{\JournalTitle{\apjl},
  720, L26}

\bibitem[{{Lyubenova} {et~al.}(2016){Lyubenova}, {Mart{\'\i}n-Navarro}, {van de
  Ven}, {Falc{\'o}n-Barroso}, {Galbany}, {Gallazzi}, {Garc{\'\i}a-Benito},
  {Gonz{\'a}lez Delgado}, {Husemann}, {La Barbera}, {Marino}, {Mast},
  {Mendez-Abreu}, {Peletier}, {S{\'a}nchez-Bl{\'a}zquez}, {S{\'a}nchez},
  {Trager}, {van den Bosch}, {Vazdekis}, {Walcher}, {Zhu}, {Zibetti},
  {Ziegler}, {Bland-Hawthorn}, \& {CALIFA Collaboration}}]{2016MNRAS.463.3220L}
{Lyubenova}, M., {Mart{\'\i}n-Navarro}, I., {van de Ven}, G., {et~al.} 2016,
  \href{http://dx.doi.org/10.1093/mnras/stw2434}{\JournalTitle{\mnras}, 463,
  3220}

\bibitem[{{Mart{\'\i}n-Navarro}
  {et~al.}(2015{\natexlab{a}}){Mart{\'\i}n-Navarro}, {La Barbera}, {Vazdekis},
  {Falc{\'o}n-Barroso}, \& {Ferreras}}]{2015MNRAS.447.1033M}
{Mart{\'\i}n-Navarro}, I., {La Barbera}, F., {Vazdekis}, A.,
  {Falc{\'o}n-Barroso}, J., \& {Ferreras}, I. 2015{\natexlab{a}},
  \href{http://dx.doi.org/10.1093/mnras/stu2480}{\JournalTitle{\mnras}, 447,
  1033}

\bibitem[{{Mart{\'\i}n-Navarro}
  {et~al.}(2015{\natexlab{b}}){Mart{\'\i}n-Navarro}, {La Barbera}, {Vazdekis},
  {Ferr{\'e}-Mateu}, {Trujillo}, \& {Beasley}}]{2015MNRAS.451.1081M}
{Mart{\'\i}n-Navarro}, I., {La Barbera}, F., {Vazdekis}, A., {et~al.}
  2015{\natexlab{b}},
  \href{http://dx.doi.org/10.1093/mnras/stv1022}{\JournalTitle{\mnras}, 451,
  1081}

\bibitem[{{Mart{\'\i}n-Navarro}
  {et~al.}(2015{\natexlab{c}}){Mart{\'\i}n-Navarro}, {Vazdekis}, {La Barbera},
  {Falc{\'o}n-Barroso}, {Lyubenova}, {van de Ven}, {Ferreras}, {S{\'a}nchez},
  {Trager}, {Garc{\'\i}a-Benito}, {Mast}, {Mendoza},
  {S{\'a}nchez-Bl{\'a}zquez}, {Gonz{\'a}lez Delgado}, {Walcher}, \& {CALIFA
  Team}}]{2015ApJ...806L..31M}
{Mart{\'\i}n-Navarro}, I., {Vazdekis}, A., {La Barbera}, F., {et~al.}
  2015{\natexlab{c}},
  \href{http://dx.doi.org/10.1088/2041-8205/806/2/L31}{\JournalTitle{\apjl},
  806, L31}

\bibitem[{{Mart{\'\i}n-Navarro} {et~al.}(2019){Mart{\'\i}n-Navarro},
  {Lyubenova}, {van de Ven}, {Falc{\'o}n-Barroso}, {Coccato}, {Corsini},
  {Gadotti}, {Iodice}, {La Barbera}, {McDermid}, {Pinna}, {Sarzi}, {Viaene},
  {de Zeeuw}, \& {Zhu}}]{2019A&A...626A.124M}
{Mart{\'\i}n-Navarro}, I., {Lyubenova}, M., {van de Ven}, G., {et~al.} 2019,
  \href{http://dx.doi.org/10.1051/0004-6361/201935360}{\JournalTitle{\aap},
  626, A124}

\bibitem[{{Massey}(2003)}]{2003ARA&A..41...15M}
{Massey}, P. 2003,
  \href{http://dx.doi.org/10.1146/annurev.astro.41.071601.170033}{\JournalTitle{\araa},
  41, 15}

\bibitem[{{McDermid} {et~al.}(2006){McDermid}, {Emsellem}, {Shapiro}, {Bacon},
  {Bureau}, {Cappellari}, {Davies}, {de Zeeuw}, {Falc{\'o}n-Barroso},
  {Krajnovi{\'c}}, {Kuntschner}, {Peletier}, \& {Sarzi}}]{2006MNRAS.373..906M}
{McDermid}, R.~M., {Emsellem}, E., {Shapiro}, K.~L., {et~al.} 2006,
  \href{http://dx.doi.org/10.1111/j.1365-2966.2006.11065.x}{\JournalTitle{\mnras},
  373, 906}

\bibitem[{{Mulchaey} {et~al.}(2003){Mulchaey}, {Davis}, {Mushotzky}, \&
  {Burstein}}]{2003ApJS..145...39M}
{Mulchaey}, J.~S., {Davis}, D.~S., {Mushotzky}, R.~F., \& {Burstein}, D. 2003,
  \href{http://dx.doi.org/10.1086/345736}{\JournalTitle{\apjs}, 145, 39}

\bibitem[{{Newman} {et~al.}(2017){Newman}, {Smith}, {Conroy}, {Villaume}, \&
  {van Dokkum}}]{2017ApJ...845..157N}
{Newman}, A.~B., {Smith}, R.~J., {Conroy}, C., {Villaume}, A., \& {van Dokkum},
  P. 2017,
  \href{http://dx.doi.org/10.3847/1538-4357/aa816d}{\JournalTitle{\apj}, 845,
  157}

\bibitem[{{Oh} {et~al.}(2011){Oh}, {Sarzi}, {Schawinski}, \&
  {Yi}}]{2011ApJS..195...13O}
{Oh}, K., {Sarzi}, M., {Schawinski}, K., \& {Yi}, S.~K. 2011,
  \href{http://dx.doi.org/10.1088/0067-0049/195/2/13}{\JournalTitle{\apjs},
  195, 13}

\bibitem[{{Parikh} {et~al.}(2018){Parikh}, {Thomas}, {Maraston}, {Westfall},
  {Goddard}, {Lian}, {Meneses-Goytia}, {Jones}, {Vaughan}, {Andrews},
  {Bershady}, {Bizyaev}, {Brinkmann}, {Brownstein}, {Bundy}, {Drory},
  {Emsellem}, {Law}, {Newman}, {Roman-Lopes}, {Wake}, {Yan}, \&
  {Zheng}}]{2018MNRAS.477.3954P}
{Parikh}, T., {Thomas}, D., {Maraston}, C., {et~al.} 2018,
  \href{http://dx.doi.org/10.1093/mnras/sty785}{\JournalTitle{\mnras}, 477,
  3954}

\bibitem[{{Pietrinferni} {et~al.}(2004){Pietrinferni}, {Cassisi}, {Salaris}, \&
  {Castelli}}]{2004ApJ...612..168P}
{Pietrinferni}, A., {Cassisi}, S., {Salaris}, M., \& {Castelli}, F. 2004,
  \href{http://dx.doi.org/10.1086/422498}{\JournalTitle{\apj}, 612, 168}

\bibitem[{{Pietrinferni} {et~al.}(2006){Pietrinferni}, {Cassisi}, {Salaris}, \&
  {Castelli}}]{2006ApJ...642..797P}
---. 2006, \href{http://dx.doi.org/10.1086/501344}{\JournalTitle{\apj}, 642,
  797}

\bibitem[{{Podorvanyuk} {et~al.}(2013){Podorvanyuk}, {Chilingarian}, \&
  {Katkov}}]{2013MNRAS.432.2632P}
{Podorvanyuk}, N.~Y., {Chilingarian}, I.~V., \& {Katkov}, I.~Y. 2013,
  \href{http://dx.doi.org/10.1093/mnras/stt419}{\JournalTitle{\mnras}, 432,
  2632}

\bibitem[{{Prieur}(1988)}]{1988ApJ...326..596P}
{Prieur}, J.~L. 1988,
  \href{http://dx.doi.org/10.1086/166120}{\JournalTitle{\apj}, 326, 596}

\bibitem[{{Sabbi} {et~al.}(2008){Sabbi}, {Sirianni}, {Nota}, {Tosi},
  {Gallagher}, {Smith}, {Angeretti}, {Meixner}, {Oey}, {Walterbos}, \&
  {Pasquali}}]{2008AJ....135..173S}
{Sabbi}, E., {Sirianni}, M., {Nota}, A., {et~al.} 2008,
  \href{http://dx.doi.org/10.1088/0004-6256/135/1/173}{\JournalTitle{\aj}, 135,
  173}

\bibitem[{{Salpeter}(1955)}]{1955ApJ...121..161S}
{Salpeter}, E.~E. 1955,
  \href{http://dx.doi.org/10.1086/145971}{\JournalTitle{\apj}, 121, 161}

\bibitem[{{Sansom} {et~al.}(2013){Sansom}, {Milone}, {Vazdekis}, \&
  {S{\'a}nchez-Bl{\'a}zquez}}]{2013MNRAS.435..952S}
{Sansom}, A.~E., {Milone}, A. d.~C., {Vazdekis}, A., \&
  {S{\'a}nchez-Bl{\'a}zquez}, P. 2013,
  \href{http://dx.doi.org/10.1093/mnras/stt1283}{\JournalTitle{\mnras}, 435,
  952}

\bibitem[{{Sarzi} {et~al.}(2018){Sarzi}, {Spiniello}, {La Barbera},
  {Krajnovi{\'c}}, \& {van den Bosch}}]{2018MNRAS.478.4084S}
{Sarzi}, M., {Spiniello}, C., {La Barbera}, F., {Krajnovi{\'c}}, D., \& {van
  den Bosch}, R. 2018,
  \href{http://dx.doi.org/10.1093/mnras/sty1092}{\JournalTitle{\mnras}, 478,
  4084}

\bibitem[{{Sarzi} {et~al.}(2006){Sarzi}, {Falc{\'o}n-Barroso}, {Davies},
  {Bacon}, {Bureau}, {Cappellari}, {de Zeeuw}, {Emsellem}, {Fathi},
  {Krajnovi{\'c}}, {Kuntschner}, {McDermid}, \&
  {Peletier}}]{2006MNRAS.366.1151S}
{Sarzi}, M., {Falc{\'o}n-Barroso}, J., {Davies}, R.~L., {et~al.} 2006,
  \href{http://dx.doi.org/10.1111/j.1365-2966.2005.09839.x}{\JournalTitle{\mnras},
  366, 1151}

\bibitem[{{Schlafly} \& {Finkbeiner}(2011)}]{2011ApJ...737..103S}
{Schlafly}, E.~F., \& {Finkbeiner}, D.~P. 2011,
  \href{http://dx.doi.org/10.1088/0004-637X/737/2/103}{\JournalTitle{\apj},
  737, 103}

\bibitem[{{Serra} \& {Trager}(2007)}]{2007MNRAS.374..769S}
{Serra}, P., \& {Trager}, S.~C. 2007,
  \href{http://dx.doi.org/10.1111/j.1365-2966.2006.11188.x}{\JournalTitle{\mnras},
  374, 769}

\bibitem[{{Sikkema} {et~al.}(2007){Sikkema}, {Carter}, {Peletier}, {Balcells},
  {Del Burgo}, \& {Valentijn}}]{2007A&A...467.1011S}
{Sikkema}, G., {Carter}, D., {Peletier}, R.~F., {et~al.} 2007,
  \href{http://dx.doi.org/10.1051/0004-6361:20077078}{\JournalTitle{\aap}, 467,
  1011}

\bibitem[{{Smette} {et~al.}(2015){Smette}, {Sana}, {Noll}, {Horst}, {Kausch},
  {Kimeswenger}, {Barden}, {Szyszka}, {Jones}, {Gallenne}, {Vinther},
  {Ballester}, \& {Taylor}}]{2015A&A...576A..77S}
{Smette}, A., {Sana}, H., {Noll}, S., {et~al.} 2015,
  \href{http://dx.doi.org/10.1051/0004-6361/201423932}{\JournalTitle{\aap},
  576, A77}

\bibitem[{{Sonnenfeld} {et~al.}(2012){Sonnenfeld}, {Treu}, {Gavazzi},
  {Marshall}, {Auger}, {Suyu}, {Koopmans}, \& {Bolton}}]{2012ApJ...752..163S}
{Sonnenfeld}, A., {Treu}, T., {Gavazzi}, R., {et~al.} 2012,
  \href{http://dx.doi.org/10.1088/0004-637X/752/2/163}{\JournalTitle{\apj},
  752, 163}

\bibitem[{{Spiniello} {et~al.}(2011){Spiniello}, {Koopmans}, {Trager},
  {Czoske}, \& {Treu}}]{2011MNRAS.417.3000S}
{Spiniello}, C., {Koopmans}, L.~V.~E., {Trager}, S.~C., {Czoske}, O., \&
  {Treu}, T. 2011,
  \href{http://dx.doi.org/10.1111/j.1365-2966.2011.19458.x}{\JournalTitle{\mnras},
  417, 3000}

\bibitem[{{Spiniello} {et~al.}(2014){Spiniello}, {Trager}, {Koopmans}, \&
  {Conroy}}]{2014MNRAS.438.1483S}
{Spiniello}, C., {Trager}, S., {Koopmans}, L. V.~E., \& {Conroy}, C. 2014,
  \href{http://dx.doi.org/10.1093/mnras/stt2282}{\JournalTitle{\mnras}, 438,
  1483}

\bibitem[{{Spiniello} {et~al.}(2015){Spiniello}, {Trager}, \&
  {Koopmans}}]{2015ApJ...803...87S}
{Spiniello}, C., {Trager}, S.~C., \& {Koopmans}, L.~V.~E. 2015,
  \href{http://dx.doi.org/10.1088/0004-637X/803/2/87}{\JournalTitle{\apj}, 803,
  87}

\bibitem[{{Thomas} {et~al.}(2003){Thomas}, {Maraston}, \&
  {Bender}}]{2003MNRAS.339..897T}
{Thomas}, D., {Maraston}, C., \& {Bender}, R. 2003,
  \href{http://dx.doi.org/10.1046/j.1365-8711.2003.06248.x}{\JournalTitle{\mnras},
  339, 897}

\bibitem[{{Thomas} {et~al.}(2005){Thomas}, {Maraston}, {Bender}, \& {Mendes de
  Oliveira}}]{2005ApJ...621..673T}
{Thomas}, D., {Maraston}, C., {Bender}, R., \& {Mendes de Oliveira}, C. 2005,
  \href{http://dx.doi.org/10.1086/426932}{\JournalTitle{\apj}, 621, 673}

\bibitem[{{Thomas} {et~al.}(2011{\natexlab{a}}){Thomas}, {Maraston}, \&
  {Johansson}}]{2011MNRAS.412.2183T}
{Thomas}, D., {Maraston}, C., \& {Johansson}, J. 2011{\natexlab{a}},
  \href{http://dx.doi.org/10.1111/j.1365-2966.2010.18049.x}{\JournalTitle{\mnras},
  412, 2183}

\bibitem[{{Thomas} {et~al.}(2011{\natexlab{b}}){Thomas}, {Saglia}, {Bender},
  {Thomas}, {Gebhardt}, {Magorrian}, {Corsini}, {Wegner}, \&
  {Seitz}}]{2011MNRAS.415..545T}
{Thomas}, J., {Saglia}, R.~P., {Bender}, R., {et~al.} 2011{\natexlab{b}},
  \href{http://dx.doi.org/10.1111/j.1365-2966.2011.18725.x}{\JournalTitle{\mnras},
  415, 545}

\bibitem[{{Trager} {et~al.}(1998){Trager}, {Worthey}, {Faber}, {Burstein}, \&
  {Gonz{\'a}lez}}]{1998ApJS..116....1T}
{Trager}, S.~C., {Worthey}, G., {Faber}, S.~M., {Burstein}, D., \&
  {Gonz{\'a}lez}, J.~J. 1998,
  \href{http://dx.doi.org/10.1086/313099}{\JournalTitle{\apjs}, 116, 1}

\bibitem[{{Treu} {et~al.}(2010){Treu}, {Auger}, {Koopmans}, {Gavazzi},
  {Marshall}, \& {Bolton}}]{2010ApJ...709.1195T}
{Treu}, T., {Auger}, M.~W., {Koopmans}, L. V.~E., {et~al.} 2010,
  \href{http://dx.doi.org/10.1088/0004-637X/709/2/1195}{\JournalTitle{\apj},
  709, 1195}

\bibitem[{{van Dokkum} {et~al.}(2017){van Dokkum}, {Conroy}, {Villaume},
  {Brodie}, \& {Romanowsky}}]{2017ApJ...841...68V}
{van Dokkum}, P., {Conroy}, C., {Villaume}, A., {Brodie}, J., \& {Romanowsky},
  A.~J. 2017,
  \href{http://dx.doi.org/10.3847/1538-4357/aa7135}{\JournalTitle{\apj}, 841,
  68}

\bibitem[{{van Dokkum}(2001)}]{2001PASP..113.1420V}
{van Dokkum}, P.~G. 2001,
  \href{http://dx.doi.org/10.1086/323894}{\JournalTitle{\pasp}, 113, 1420}

\bibitem[{{van Dokkum} \& {Conroy}(2010)}]{2010Natur.468..940V}
{van Dokkum}, P.~G., \& {Conroy}, C. 2010,
  \href{http://dx.doi.org/10.1038/nature09578}{\JournalTitle{\nat}, 468, 940}

\bibitem[{{van Dokkum} \& {Conroy}(2011)}]{2011ApJ...735L..13V}
---. 2011,
  \href{http://dx.doi.org/10.1088/2041-8205/735/1/L13}{\JournalTitle{\apjl},
  735, L13}

\bibitem[{{van Dokkum} \& {Conroy}(2012)}]{2012ApJ...760...70V}
---. 2012,
  \href{http://dx.doi.org/10.1088/0004-637X/760/1/70}{\JournalTitle{\apj}, 760,
  70}

\bibitem[{{Vaughan} {et~al.}(2018{\natexlab{a}}){Vaughan}, {Davies},
  {Zieleniewski}, \& {Houghton}}]{2018MNRAS.475.1073V}
{Vaughan}, S.~P., {Davies}, R.~L., {Zieleniewski}, S., \& {Houghton}, R. C.~W.
  2018{\natexlab{a}},
  \href{http://dx.doi.org/10.1093/mnras/stx3199}{\JournalTitle{\mnras}, 475,
  1073}

\bibitem[{{Vaughan} {et~al.}(2018{\natexlab{b}}){Vaughan}, {Davies},
  {Zieleniewski}, \& {Houghton}}]{2018MNRAS.479.2443V}
---. 2018{\natexlab{b}},
  \href{http://dx.doi.org/10.1093/mnras/sty1434}{\JournalTitle{\mnras}, 479,
  2443}

\bibitem[{{Vazdekis} \& {Arimoto}(1999)}]{1999ApJ...525..144V}
{Vazdekis}, A., \& {Arimoto}, N. 1999,
  \href{http://dx.doi.org/10.1086/307868}{\JournalTitle{\apj}, 525, 144}

\bibitem[{{Vazdekis} {et~al.}(2016){Vazdekis}, {Koleva}, {Ricciardelli},
  {R{\"o}ck}, \& {Falc{\'o}n-Barroso}}]{2016MNRAS.463.3409V}
{Vazdekis}, A., {Koleva}, M., {Ricciardelli}, E., {R{\"o}ck}, B., \&
  {Falc{\'o}n-Barroso}, J. 2016,
  \href{http://dx.doi.org/10.1093/mnras/stw2231}{\JournalTitle{\mnras}, 463,
  3409}

\bibitem[{{Vazdekis} {et~al.}(2010){Vazdekis}, {S{\'a}nchez-Bl{\'a}zquez},
  {Falc{\'o}n-Barroso}, {Cenarro}, {Beasley}, {Cardiel}, {Gorgas}, \&
  {Peletier}}]{2010MNRAS.404.1639V}
{Vazdekis}, A., {S{\'a}nchez-Bl{\'a}zquez}, P., {Falc{\'o}n-Barroso}, J.,
  {et~al.} 2010,
  \href{http://dx.doi.org/10.1111/j.1365-2966.2010.16407.x}{\JournalTitle{\mnras},
  404, 1639}

\bibitem[{{Vazdekis} {et~al.}(2015){Vazdekis}, {Coelho}, {Cassisi},
  {Ricciardelli}, {Falc{\'o}n-Barroso}, {S{\'a}nchez-Bl{\'a}zquez}, {La
  Barbera}, {Beasley}, \& {Pietrinferni}}]{2015MNRAS.449.1177V}
{Vazdekis}, A., {Coelho}, P., {Cassisi}, S., {et~al.} 2015,
  \href{http://dx.doi.org/10.1093/mnras/stv151}{\JournalTitle{\mnras}, 449,
  1177}

\bibitem[{{Weijmans} {et~al.}(2009){Weijmans}, {Cappellari}, {Bacon}, {de
  Zeeuw}, {Emsellem}, {Falc{\'o}n-Barroso}, {Kuntschner}, {McDermid}, {van den
  Bosch}, \& {van de Ven}}]{2009MNRAS.398..561W}
{Weijmans}, A.-M., {Cappellari}, M., {Bacon}, R., {et~al.} 2009,
  \href{http://dx.doi.org/10.1111/j.1365-2966.2009.15134.x}{\JournalTitle{\mnras},
  398, 561}

\bibitem[{{Worthey} {et~al.}(1994){Worthey}, {Faber}, {Gonzalez}, \&
  {Burstein}}]{1994ApJS...94..687W}
{Worthey}, G., {Faber}, S.~M., {Gonzalez}, J.~J., \& {Burstein}, D. 1994,
  \href{http://dx.doi.org/10.1086/192087}{\JournalTitle{\apjs}, 94, 687}

\bibitem[{{Worthey} \& {Ottaviani}(1997)}]{1997ApJS..111..377W}
{Worthey}, G., \& {Ottaviani}, D.~L. 1997,
  \href{http://dx.doi.org/10.1086/313021}{\JournalTitle{\apjs}, 111, 377}

\end{thebibliography}


\begin{appendix}
\section{Technical details}

\subsection{Initial sky subtraction}
\label{sec:skyap}
We tested two different approaches for sky subtraction. In the first approach, we used the \textsc{iraf} task \textsc{background}, for which we sampled the sky at \textgreater3\,$R_h$ distance from the galaxy.   
In a second approach we extracted 1d galaxy spectra and a 1d mastersky-spectrum. The mastersky was extracted over a 500 pixel wide region, at a  distance of \textgreater3\,$R_h$ from the galaxy, applying 3$\sigma$-clipping, and  subtracted  from the extracted galaxy spectra. A comparison between the two different methods for sky subtraction shows no significant difference in  S/N obtained, and we adopted the second approach.

\subsection{Second order sky subtraction}
\label{sec:secsky}
\begin{figure}
\gridline{\fig{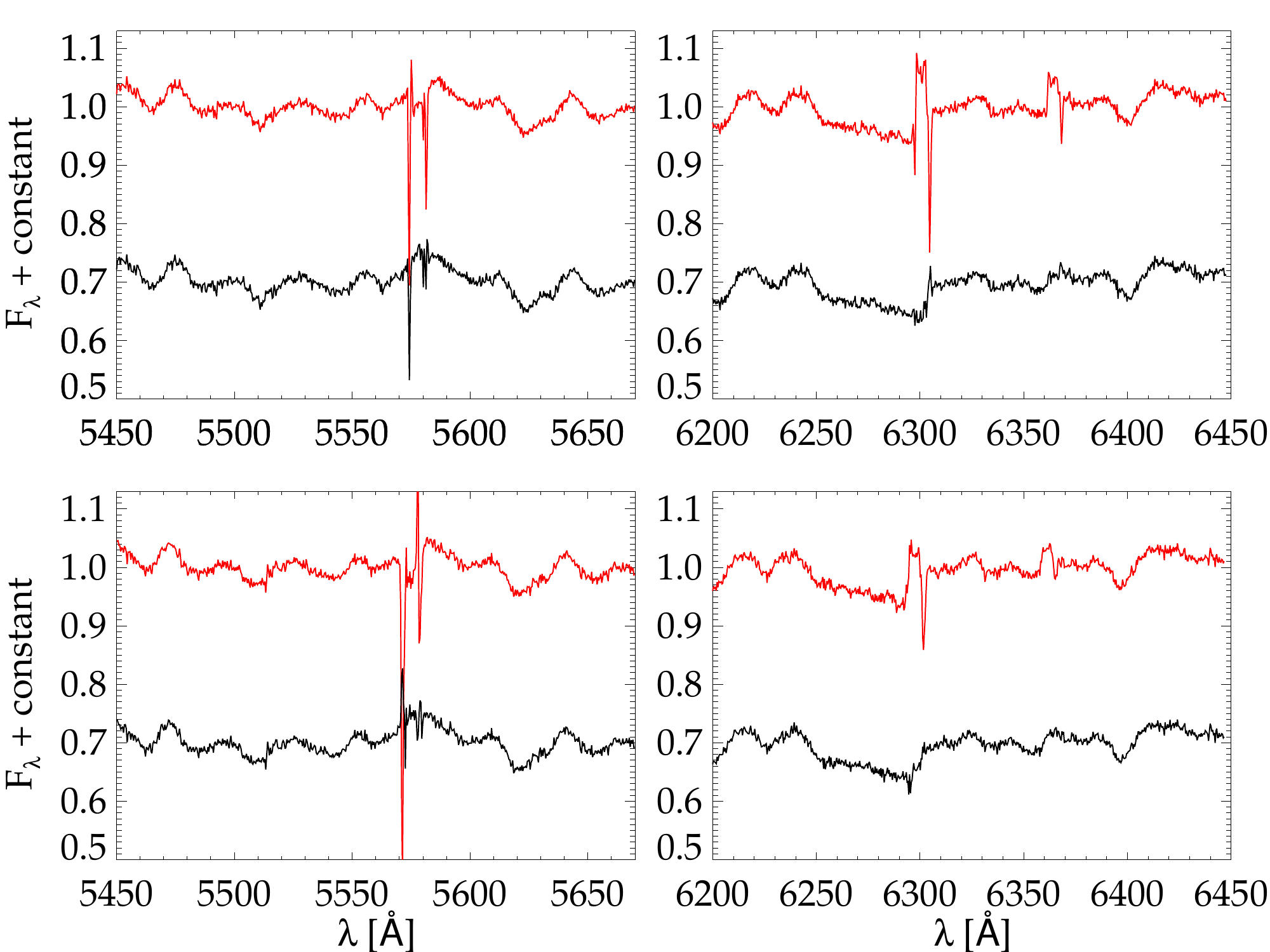}{0.45\textwidth}{}}
\caption{Spectra of \object\space before (red) and after (black) second-order sky  correction. The upper panel shows the spectra observed on 2015-5-19, and the lower panel the spectra observed on 2015-5-20, all in the extraction region 10\farcs8--21\farcs6 ($\frac{1}{8}$\,$R_h$--\onequarter\,$R_h$).   \label{fig:sky2}}
\end{figure}

As some spectra have sky subtraction residuals, we used the \textsc{idl} routine \textsc{pPXF} by \citep{2004PASP..116..138C, 2017MNRAS.466..798C}  to  fit the stellar kinematics and a sky-subtraction correction as described in \cite{2009MNRAS.398..561W}. 
The spectra were corrected for the mean Galactic extinction at the location of \object, E(B-V) = 0.0705 \citep{2011ApJ...737..103S}  before the fit, assuming the extinction curve parametrization of \cite{1999PASP..111...63F}.
We used the E-MILES library  \citep{2016MNRAS.463.3409V}  of single stellar populations (SSPs)  as templates,  
 with  four different metallicities \textit{Z} ranging from $-0.71$ to $+0.22$\,dex, 15 ages from 1 to 17.78\,Gyr,  and a \cite{2003PASP..115..763C} IMF. 
As sky templates, we used the  spectra we observed on the respective nights. For the 2015 observations, we have dedicated sky exposures as well as extracted mastersky-spectra; for the 2018 observations, we only have the extracted mastersky-spectra from the three exposures. In order to account for over-subtraction of the sky, we  included the negative of the sky spectra to the sky templates. To correct  small shifts, we have slightly shifted versions of each spectrum in the sky template library.

We found that this sky correction is only required in  wavelength regions with strong sky emission lines. 
Therefore, we subtracted the best-fit sky only for some of the chips. For the 2015 nights, we corrected two chips at the wavelength regions 5200--6700\,\AA, to correct the sky lines at 5577, 6300 and 6363\,\AA. For the 2018 night, we corrected three chips, at 4750--7100\,\AA, including the 5893\,\AA\space sky emission line, which falls on a chip gap in the 2015 data. We show an example of the second order sky subtraction in Fig.~\ref{fig:sky2}, near the sky emission-line regions.

\subsection{Spectral index uncertainties}
\label{sec:spindunc}
We obtained statistical uncertainties of the spectral indices using a set of 500  Monte Carlo runs, in which we added random noise to the spectra. The added random noise level was drawn from a Gaussian distribution with $\sigma$ given by the measured noise in the respective spectral region. We   simultaneously varied the velocity shift, by values drawn from a Gaussian distribution where   $\sigma$ was the velocity uncertainty.

We also estimated the uncertainty of the resolution and LOSVD correction  from Monte Carlo simulations. We convolved the best-fit model spectrum with a different LOSVD, according to the respective kinematic uncertainties measured with \textsc{pPXF}, in 500 runs. We included the  uncertainty for this correction to the   uncertainties of the spectral index measurements, though it was lower by a factor of 0.014 on average than the statistical uncertainty derived from the spectra.

To estimate the systematic uncertainty of the indices, we simulated the effect of the wavelength solution and illumination correction on the most central spectrum. For the wavelength calibration uncertainty, we used the root-mean-square $rms_\text{fc}$ of the wavelength calibration solution  obtained with the \textsc{iraf} task \textsc{fitcoords}, with a different value of $rms_\text{fc}$ for the different chips and thus wavelength regions. In a set of 500 Monte Carlo runs, we slightly changed the wavelength solution by values drawn from a Gaussian distribution with $\sigma$=$rms_\text{fc}$, interpolated the central spectrum to the new wavelength solution, and repeated the index measurement. The standard deviation of the 500 index measurements is used as systematic wavelength calibration uncertainty. For the spectra with high S/N, this uncertainty exceeds the statistical uncertainty of the index measurement. Depending on the index, for lower S/N spectra the wavelength calibration uncertainty becomes comparable to or smaller than the statistical uncertainty. 

To estimate the edge illumination correction uncertainty we used the two central spectra observed at the two different grating angles, say spectrum \textit{A} and \textit{B}. The chip gap of  spectrum \textit{A}  lies at the center of a chip of  spectrum \textit{B}, and vice versa. The continuum shape of spectrum \textit{A} close to the edge of a chip is affected by   edge illumination problems, and deviates from the continuum shape of spectrum \textit{B}, at the center of the chip. We used spectrum \textit{B} to estimate the deviation in spectrum \textit{A} caused by the chip edge. We defined two wavelength regions on the two sides of a chip gap, far enough from the gap to not be affected by edge illumination  problems. We  calculated the total flux in the regions, and made a linear fit to the continuum for spectra \textit{A} and \textit{B}. We divided spectra \textit{A} and \textit{B} by their respective linear continuum approximation, and computed the relative difference of the two normalized spectra. We took the standard deviation $\sigma_{\text{edge}}$  of the relative difference close to the chip edge, and consider it  as additional noise for spectrum \textit{A}, but only in the region close to the chip edge. The extent of this region differs for the different chips, but is about 80--160\,\AA\space wide.  To estimate the effect on the spectral indices, we ran another set of Monte Carlo simulations. We added random noise to the central spectrum \textit{A} in the edge region in 500 runs. The noise was drawn from a Gaussian with standard deviation $\sigma_{\text{edge}}$. We repeated the index measurements and used the standard deviation of the 500 measurements as edge illumination uncertainty. This uncertainty is zero for indices measured at the center of a chip, and in most cases less than the wavelength calibration uncertainty. Only for indices that are close to the chip edge, or even extend over the gap, the flux calibration uncertainty can exceed the wavelength calibration uncertainty.

We compared our uncertainties to values in the literature \citep{2002A&A...395..431B,2005MNRAS.356.1440D}, after transforming them to the same resolution of FWHM=14\,\AA. Our total uncertainties are smaller for  Fe5270 (by a factor of  2.5-3.7), Fe5335 (factor of 2-10), Mg\textit{b} (factor of 1.5-2.5), NaD (factor of 3); comparable for H$\delta_F$, H$\gamma_A$, Fe5406, H$\beta$, Fe5709, Fe5782, and larger for TiO$_2$ (factor  of 1.5-7).

\subsection{Regularization with \textsc{pPXF}}
\label{sec:ppxffitapp}

In this section we explain how we used \textsc{pPXF} to constrain the stellar age and metallicity. We fit the data in the wavelength region $\lambda$= 4000--5600\,\AA. \cite{2015MNRAS.447.1033M} used $\lambda$= 4600--5600\,\AA\space for the same purpose, but we extended the wavelength region to shorter $\lambda$ to make use of our longer wavelength coverage and included the  H$\delta$ and H$\gamma$ lines. The described procedure was performed separately 28 times, for MILES SSP models with [$\alpha$/Fe]=0\,dex and [$\alpha$/Fe]=0.4\,dex;  and for SSP models with  14 possible bimodal IMF slope values.

\begin{figure*}
\includegraphics[width=0.5\textwidth]{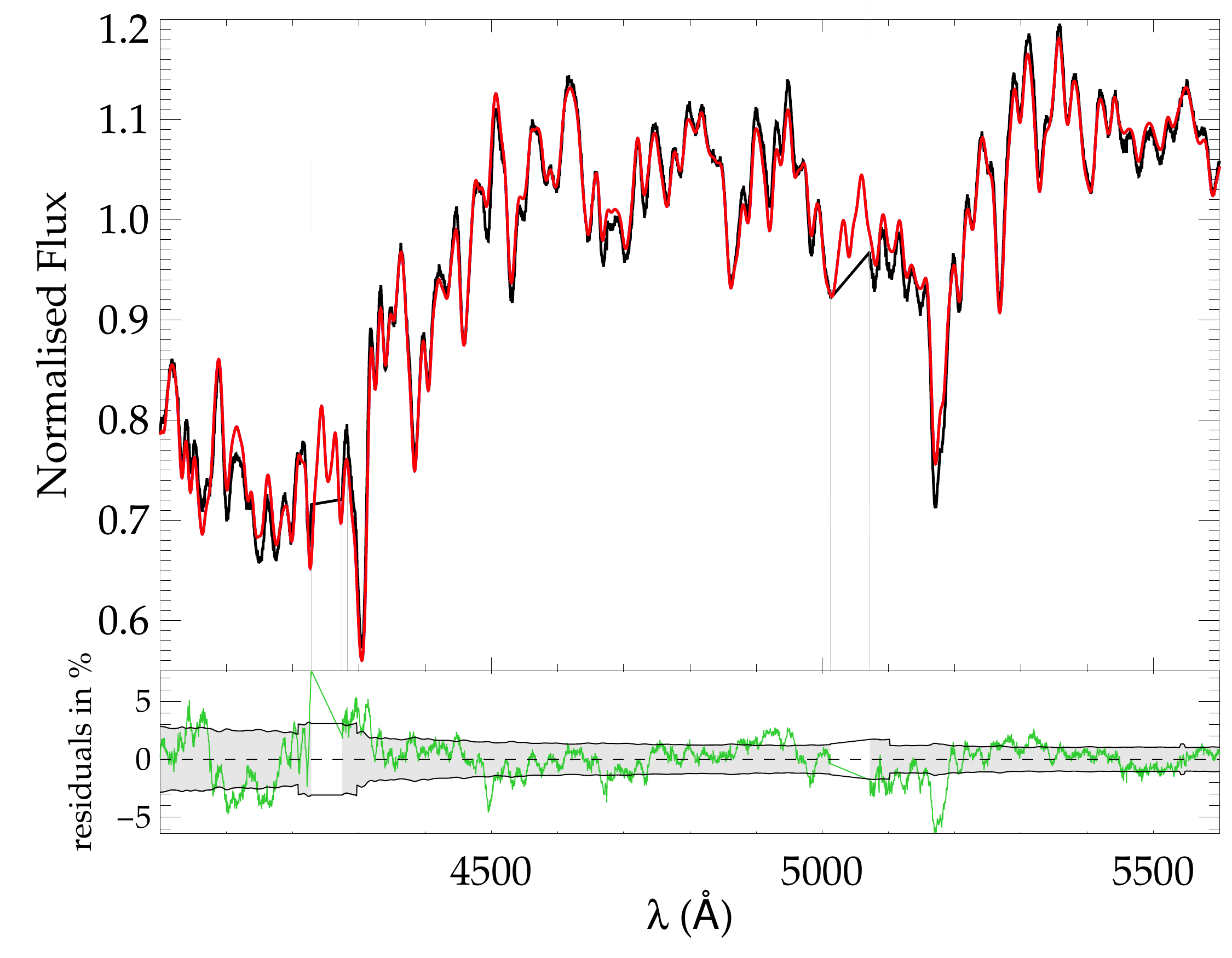}
\includegraphics[width=0.5\textwidth]{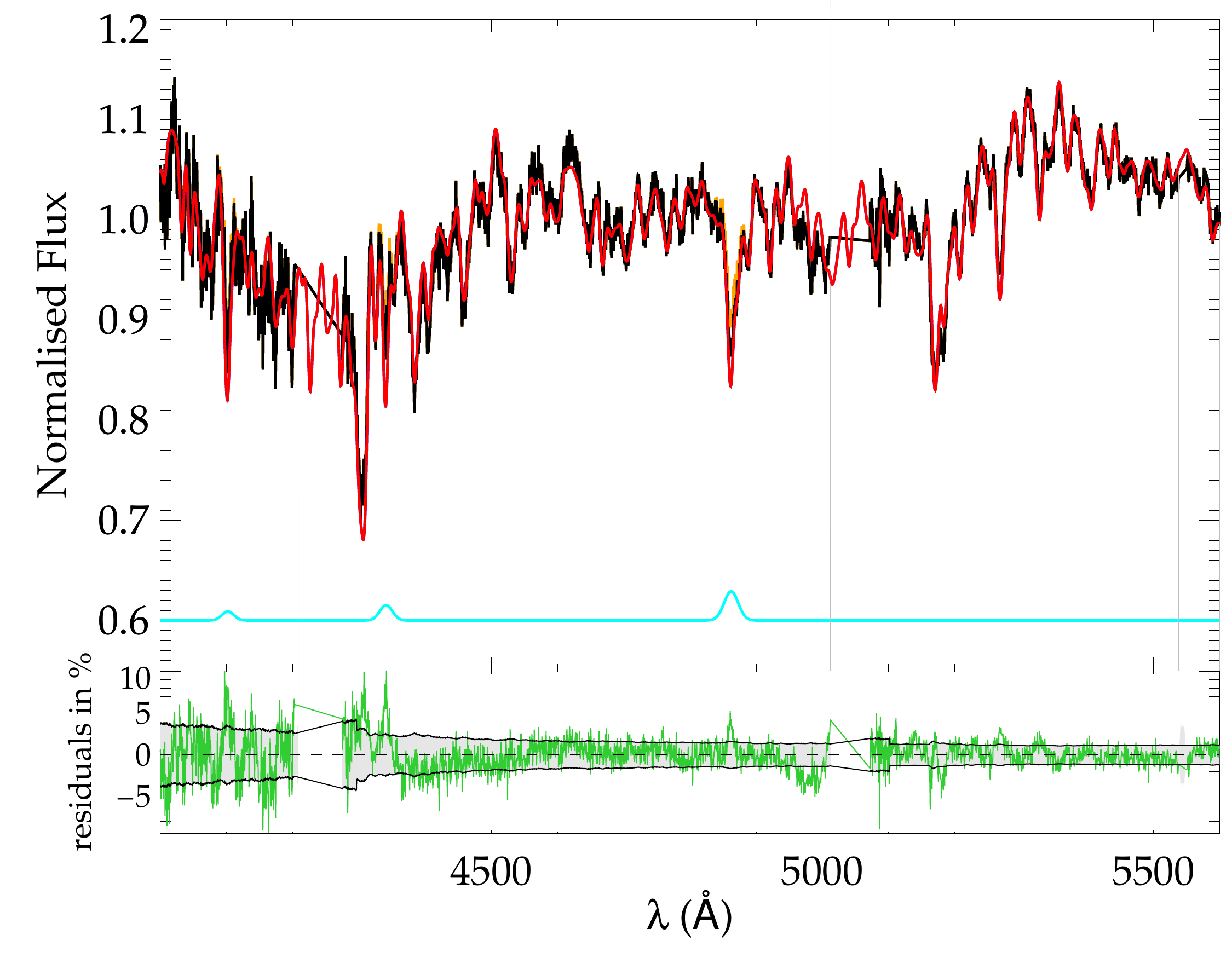}
\caption{\textsc{pPXF} fit of the most central spectrum (left panel) and outermost spectrum(right panel). Black color denotes data (after emission line correction on the right panel), red color the best-fit (using the MILES SSP models with [$\alpha$/Fe]=0.0\,dex and $\Gamma_b$=1.3), grey vertical  lines masked pixels. For the right panel, the orange colored line denotes the data before emission line correction,  cyan color the Balmer emission line flux that we subtracted. The lower panels show the residuals in per cent of the normalised flux in green, the grey shaded region denotes the error spectrum.  
}
\label{fig:ppxffit}
\end{figure*}

As a first step, we fit the stellar kinematics with  \textsc{pPXF} with additive polynomials, and  scaled the noise spectrum such that the reduced $\chi^2$=1. This is necessary for a reasonable regularization.  The next steps are similar to  Sect.~\ref{sec:kin}: we fixed the stellar kinematics,  used multiplicative polynomials,  and subtracted  the emission line flux derived with \textsc{GandALF} in Sec.~\ref{sec:kin} if necessary. We show two example fits in Fig.~\ref{fig:ppxffit}, for the central spectrum with highest S/N, and the outermost spectrum with the lowest S/N. They both have $\chi^2_{red}$=1.

Next, we undertook fits with  regularization, to   obtain a smoothly  weighted distribution for the template spectra. 
Without regularization, the optimal template may consist of several spectra with very distinct ages and metallicities, which is unphysical. Regularization produces a more realistic star formation history.  It is important to find the optimal regularization parameter for each spectrum,  to ensure that  the  weight distribution is as smooth as possible, without significantly decreasing the goodness of the fit, compared to an unregularized fit \citep{2004PASP..116..138C,2017MNRAS.466..798C}. We determined the optimal regularization by performing fits with   a range of regularization parameters, as suggested by \cite{2017MNRAS.466..798C}.  We note that our  regularization increases the $\chi^2_\text{red}$ value  of a fit usually by less than 3 per cent, to $\chi^2_{red}\lesssim$1.03.
The light-weighted mean age and metallicity are very similar between the unregularized fit, and the fit with optimal regularization.
The best-fit age distributions  of the fits   have a spread of 1--2 Gyr.  We repeated the fits with and without subtracting the gas emission obtained with \textsc{GandALF} from the Balmer line regions. The correction is only required for the outer four bins. 
As each SSP template model is normalised to 1\,M$_\sun$, we obtain mass-weights. We used the average flux of each SSP model in our wavelength region to transform the mass-weights to luminosity-weights.


\section{Stellar population models}
\label{sec:modelsum}

We used three families of stellar population models, MILES models, Conroy  models, and Thomas, Maraston \& Johansson (TMJ) models.  In this section, we briefly mention their parameter spaces and note important differences.

The MILES models are based on the code by \cite{2010MNRAS.404.1639V}; the newer E-MILES models \citep{2016MNRAS.463.3409V} extend over a larger wavelength range, from the ultra-violet to near-infrared. These models are only available as  so-called  base models, meaning that they are $\alpha$-enhanced at low metallicity, but around solar metallicity  the models are scaled-solar \citep{2010MNRAS.404.1639V}. There are two sets of  models using different isochrones, either Padova isochrones \citep{2000A&AS..141..371G} or BaSTI \citep{2004ApJ...612..168P} isochrones. 
Another set of models described in \cite{2015MNRAS.449.1177V} is available with two different values for [$\alpha$/Fe], 0.0 and 0.4\,dex, using BaSTI \citep{2004ApJ...612..168P,2006ApJ...642..797P} isochrones. Depending on the isochrones, the ages and metallicities cover different ranges: 0.063--17.8\,Gyr with seven metallicities from -2.32\,dex to +0.22 dex for Padova isochrones, and 0.03--14\,Gyr  with 12 metallicities from -2.27\,dex to 0.40\,dex for BaSTI isochrones. However,  the models with \textit{Z}=0.4\,dex  are not considered safe (\url{http://www.iac.es/proyecto/miles/pages/ssp-models/safe-ranges.php}). Unfortunately, we do not have uncertainties for the models, which would enable us to consider also models with high metallicity at 0.4\,dex. 

We used the MILES models with a bimodal IMF, which is parametrized as follows:  $dN/d \text{log} m \propto m^{-\Gamma}$, where the IMF slope $\Gamma$ is constant at  stellar masses $M \leq $0.6\,M$_\sun$, but  $\Gamma_\text{b}$ is varied  at higher stellar masses;  $\Gamma_\text{b}$ ranges  from 0.3 (extremely bottom-light, with a dearth of low-mass stars) to 3.5 (extremely bottom-heavy, with an excess of low-mass stars) for 14 possible slope values. A bimodal IMF with  $\Gamma_\text{b}$ = 1.3 is a good representation of a \cite{2001MNRAS.322..231K} like IMF shape. The upper and lower mass limits are at 100\,M$_\sun$ and 0.1\,M$_\sun$.
We used the MILES models for spectral fitting with \textsc{pPXF} in Sect.  \ref{sec:kin}, \ref{sec:ageppxf}, and Appendix  \ref{sec:secsky}, and for spectral indices in Sect.  \ref{sec:alphaest},  \ref{sec:alphaproxy}, \ref{sec:imfindexgen}, and Appendix \ref{sec:indexselect}.  This set of models is often used for spectral index fitting in the literature.

The second family of  models are from \cite{2012ApJ...747...69C}, with a newer version in \cite{2018ApJ...854..139C}. They used the MIST isochrones \citep{2016ApJS..222....8D,2016ApJ...823..102C}. The newer  models cover  five different ages (1, 3, 5, 9, 13.5\,Gyr),  five different metallicities \textit{Z} (-1.5, -1.0, -0.5, 0, 0.3\,dex), and  variations of several elemental abundances.

The \cite{2012ApJ...747...69C} models offer only Salpeter, Chabrier, and a bottom-light IMF slope. However, the newer \cite{2018ApJ...854..139C} models contain a wider range of IMF slopes, which enabled us to  use them in full-spectral fitting. The IMF is defined in a different way from the MILES SSP models. In this case the  high-mass  IMF slope is  not varied, but rather the low-mass end. Further, the slope $x$ is defined as $dN/dm\propto m^\text{-x}$.
The IMF is parametrized as follows: The high-mass slope  (1\,$M_\sun$\textless m \textless 100\,$M_\sun$) is fixed to the Salpeter value, $x_3$=2.35 (not 1.35 because of the different slope definition). For lower stellar masses, the IMF is split in two IMF slopes, $x_1$ from 0.08--0.5\,$M_\sun$, and $x_2$ from 0.5--1\,$M_\sun$. For each IMF slope $x$,  models are available for 16 slopes, ranging from 0.5 to 3.5. 

To study the elemental abundances,  \cite{2018ApJ...854..139C} published their set of   SSP models  with a \cite{2001MNRAS.322..231K} IMF  and with either solar abundances, or varying one of several elements (Na, Ca, Fe, C, N, Ti, Mg, Si, Ba) to a super- or sub-solar value, in most cases by $\pm$0.3\,dex. Some elemental abundances are only computed   with super-solar values of +0.3\,dex (O+Ne+S, Sr, Cr, Mn, Eu, K, V, Cu, Ni, Co).  O  is varied together with the elements Ne and S, for simplicity, we will simply denote [\text{O+Ne+S/H}] as  [\text{O/H}], following \cite{2012MNRAS.421.1908J}.
For the \cite{2012ApJ...747...69C} models, the elements are only varied for models with an age of 13.5\,Gyr and having  solar metallicity, adopting a   \cite{2003PASP..115..763C} IMF. We used the \cite{2018ApJ...854..139C} models to derive corrections for spectral indices with non-solar elemental abundances in Section \ref{sec:imfindexgen}.  Here  we make the assumption that a response of the spectral index to a given elemental variation does not depend on the  IMF, as we apply the same abundance correction, derived for  a SSP with a  Kroupa IMF,   for SSPs with very different IMF slopes.
We used the Conroy models in Sect. \ref{sec:imfindexgen}  and  Appendix \ref{sec:indexselect}  for spectral index response functions,  for full spectral fitting in Sect. \ref{sec:ageppxf} and \ref{sec:pystaff}, and for our simulations in Appendix~\ref{sec:simulation}.

The third family of models are from \citet[TMJ]{2011MNRAS.415..545T}. They cover ages from 0.1-15\,Gyr, metallicities from --2.25 to 0.67\,dex, [$\alpha$/Fe] = --0.3, 0.0, 0.3 and 0.5 dex. They also  include several additional elemental abundances (C, N, Ca, Na, Mg, Si, Ti) enhanced to 0.3\,dex. The TMJ models do not vary the IMF, and have only a fixed \cite{1955ApJ...121..161S} IMF. 
We have only the published spectral index measurements and  compare them to our index measurements  in Sect. \ref{sec:alphaest}, as the TMJ models have the widest range of [$\alpha$/Fe] and \textit{Z} from all model families. 

\begin{figure*}
\gridline{\fig{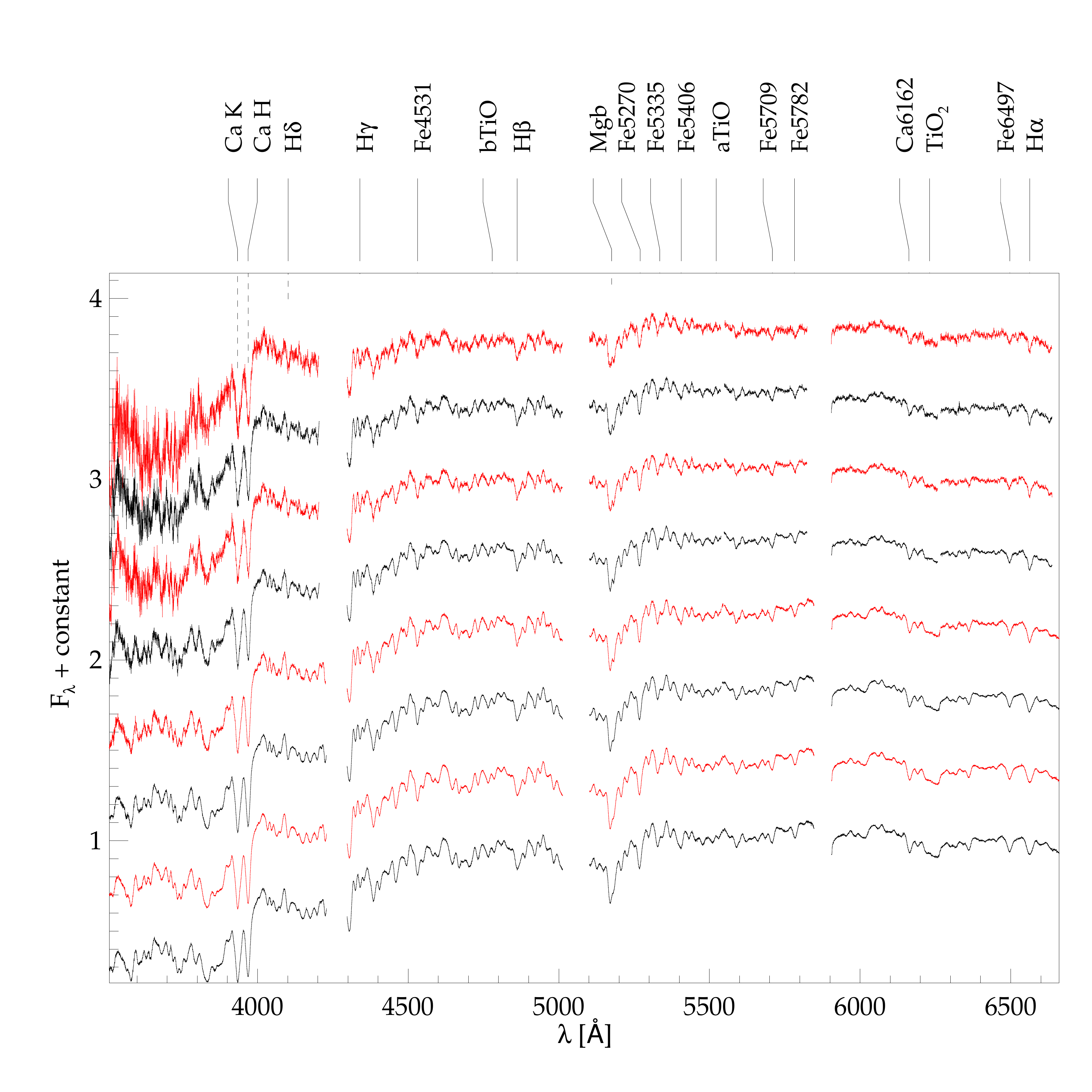}{0.96\textwidth}{}}
\caption{The final \object\space spectra  along the major axis, normalized and  shifted to rest-wavelength, from bottom to top:  central 1\farcs5, at radii 0\farcs75--3\arcsec, 3\arcsec--10\farcs8, 10\farcs8--21\farcs6,  21\farcs6--43\farcs2, 43\farcs2--64\farcs8,  43\farcs2--86\farcs4, 64\farcs8--86\farcs4. Note that the outer three spectra overlap  spatially.  Several stellar absorption lines are labeled at the top of the plot. \label{fig:spec2}}
\end{figure*}

\begin{figure*}
\gridline{\fig{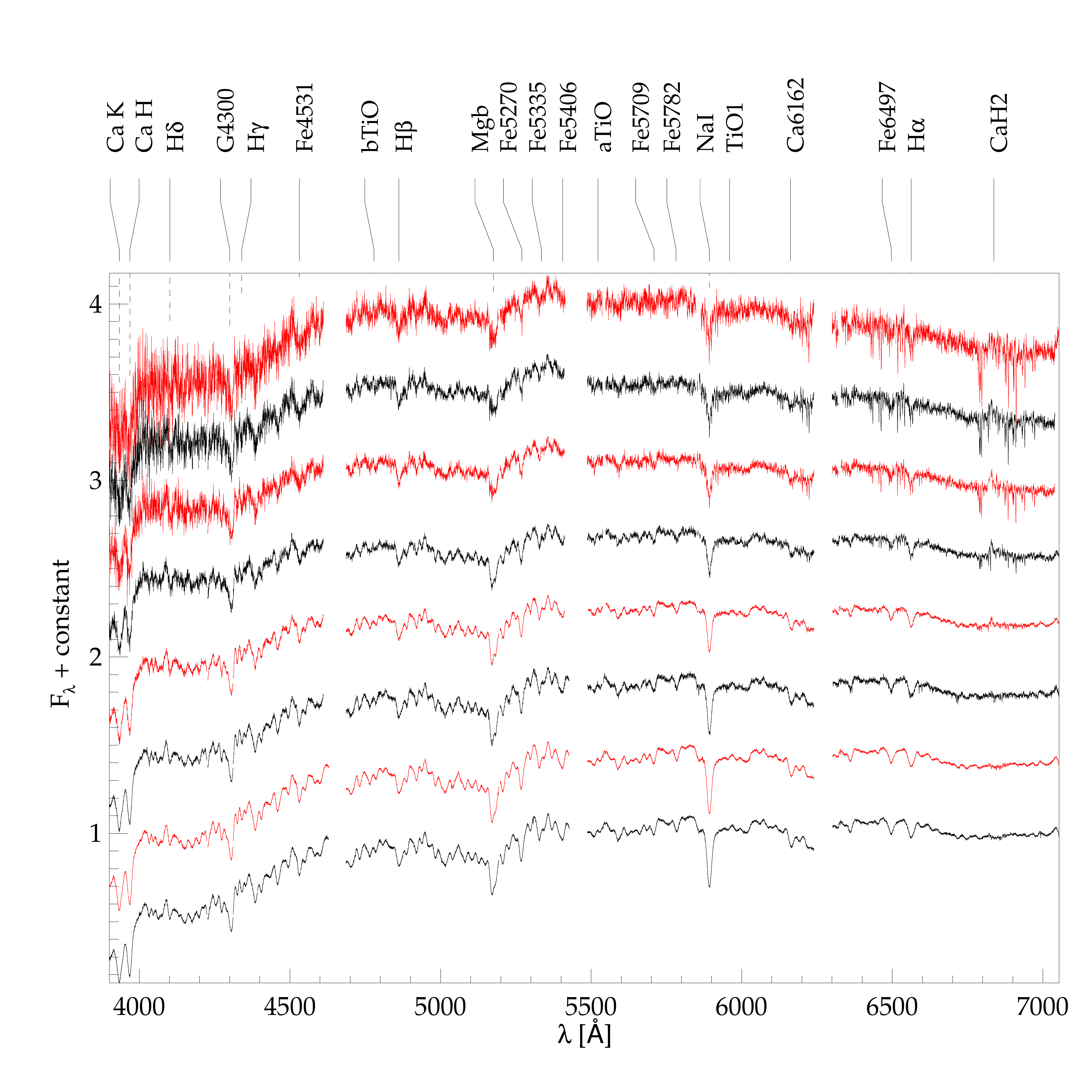}{0.96\textwidth}{}}
\caption{The final \object\space spectra at  P.A. = 48\degr\space with respect to the major axis, normalized and shifted to rest-wavelength. The extraction regions were corrected such that they cover the same isophotal regions as the spectra observed along the major axis,  from bottom to top:  central 1\farcs23, at radii 0\farcs62--2\farcs46, 2\farcs46--8\farcs86, 8\farcs86--17\farcs7,  17\farcs7--35\farcs4, 35\farcs4--53\farcs1, 35\farcs4--70\farcs9, 53\farcs1--70\farcs9. 
\label{fig:spec}}
\end{figure*}
\section{Spectral indices}

In this section we  give an overview of several absorption line indices that are in common usage, and we discuss which of them are useful for our analysis.

Spectral indices are measurements of the strength of a given absorption line. A spectral index definition usually consists of a feature band definition, i.e. a wavelength region at the location of the respective absorption line, and two continuum band definitions, i.e. wavelength regions to the red and blue of the absorption line. The continuum band definition is used to estimate the pseudo-continuum in the region of the absorption line. They can be measured in units of \AA\space or mag.

\label{sec:indexselect}

We measured spectral indices  with the index definitions listed in  \url{http://www.iac.es/proyecto/miles/programs/BANDS} in air wavelengths. From this list we selected indices with the following criteria: If there are   several index definitions for the same absorption line, but  with slightly different feature or continuum wavelength regions,  we used the index definition that was designed for galaxies rather than for globular clusters. Further, we chose  the    spectral regions that are the least affected by   emission lines and telluric absorption, and fall entirely on one  rather than two IMACS chips.

We have used the MILES SSP models
to investigate the indices' sensitivity to the IMF,  age, and overall metallicity; and the SSP models of \cite{2012ApJ...747...69C} to investigate the sensitivity to different elemental abundances. We now discuss the indices that are sensitive to age, metallicity, surface-gravity, and elemental abundances. Several indices are also marked on our spectra shown in Figs. \ref{fig:spec2} and \ref{fig:spec}. 

\textsc{Age-sensitive indices}: Many absorption-line indices are sensitive to age. Together with metallicity-sensitive  Fe indices, hydrogen lines can be used to derive the mean age of a stellar population \citep[e.g.][]{1997ApJS..111..377W,1999ApJ...525..144V,2001MNRAS.323..615K}. 
The   hydrogen line indices  for H$\alpha$, H$\beta$,  H$\gamma$, H$\delta$   can decrease  with stellar age. These indices are also sensitive to the metallicity, and H$\beta$ additionally to the IMF. They can also be  contaminated by gas emission, which makes them sensitive to   gas subtraction. 
We have all of these hydrogen lines  in at least one of the spectra (see Figs.~\ref{fig:spec2} and \ref{fig:spec}).

\textsc{Metallicity-sensitive indices}: 
The metallicity \textit{Z} influences many absorption line indices. The Fe lines  Fe5270, Fe5335, Fe5406 \citep{1998ApJS..116....1T,2006MNRAS.373..906M}, together with the above mentioned hydrogen lines, are often used to constrain the mean stellar metallicities of old stellar populations. 
These indices 
increase with increasing metallicity and age. Their sensitivity to most elemental abundances and the IMF is negligible compared to their sensitivity to the metallicity and  [Fe/H].  There are also several Fe lines at  wavelengths \textless 3800\,\AA, but since our S/N is lower in this region, and the wavelength calibration less certain, we do not consider those indices. Other Fe lines, like Fe4531, Fe5709  Fe5782, are also sensitive to some elemental abundances (Ti, Mg, O), and have a mild IMF dependence.

\begin{figure*}[ht!]
	\plotone{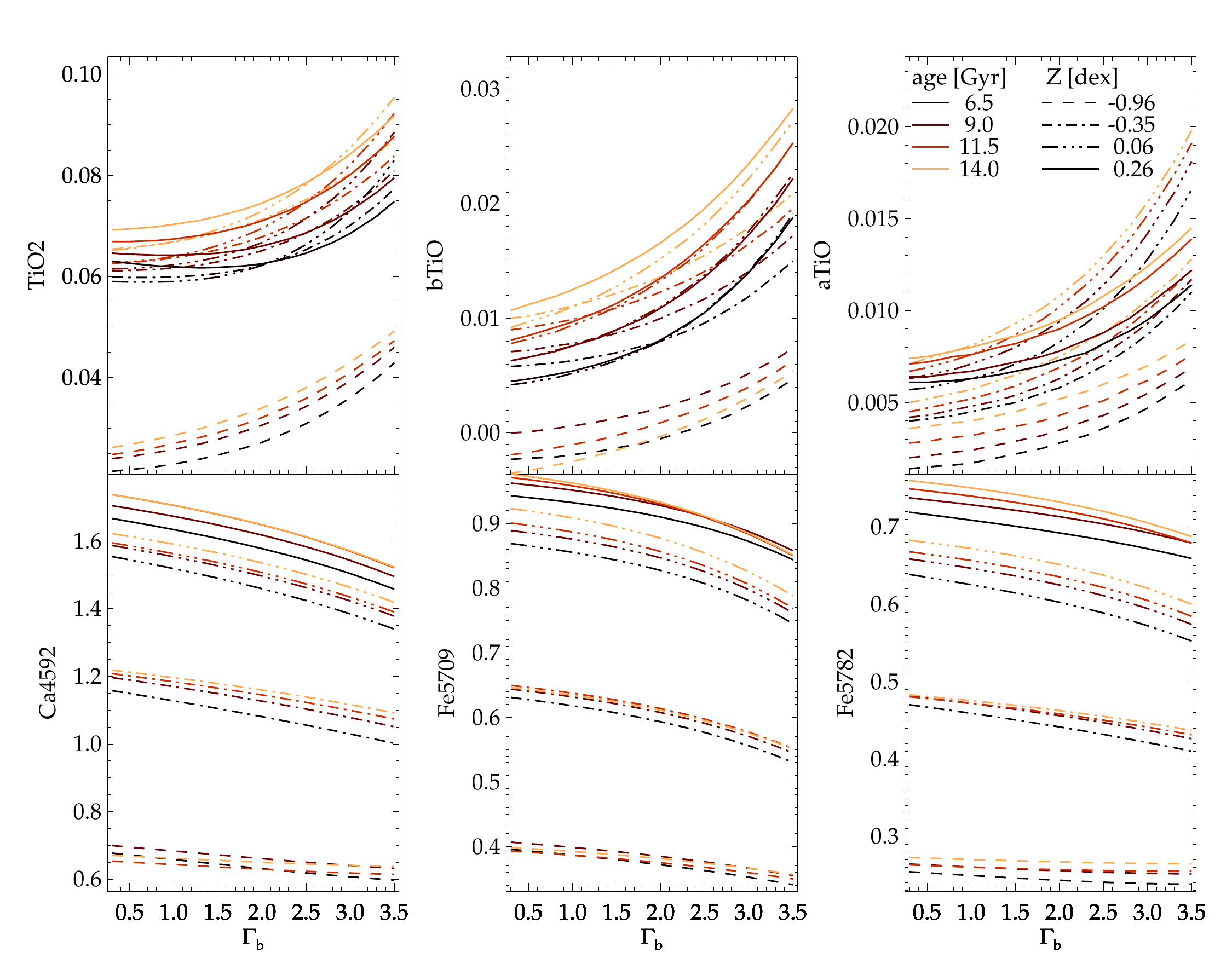}
		\caption{Dependence of different  indices on the IMF slope, upper row from left to right: TiO2, bTiO, aTiO; lower row from left to right: Ca4592, Fe5709, Fe5782. Indices were measured using the MILES models with [$\alpha$/Fe]=0\,dex, at FWHM=14\,\AA. Different colored lines denote different ages, from 6.5\,Gyr (black) to 14\,Gyr (orange) in steps of 2.5\,Gyr, different linestyles denote different metallicities, from -0.96 (dash), -0.35 (dash-dot), +0.06 (dash-dot-dot-dot), to +0.26\,dex (solid). TiO index values increase  as a function of IMF slope $\Gamma_b$, indicating a higher sensitivity to dwarf stars (i.e. to low-mass stars in old populations).  The other index values decrease as a function of $\Gamma_b$, indicating a higher sensitivity to giant stars (i.e. to higher-mass stars.)}		
	\label{fig:indexmodel}
	\end{figure*}

\textsc{Surface-gravity-sensitive indices}: 
Several indices are sensitive to the surface gravity of a star, and can provide a means to discern dwarf from giant stars in old stellar populations, and thus low-mass from intermediate-mass stars. In principle, such indices can constrain the IMF slope, assuming that the effects of age, metallicity and abundances can be accounted for. Some indices increase with a higher fraction of low-mass stars, i.e. steeper IMF slope, meaning they are stronger for dwarf stars. Other indices decrease with a higher fraction of low-mass stars; they are stronger for giant stars. 
Among the dwarf-sensitive indices are  TiO$_{2}$, bTiO,  aTiO  (see Fig.~\ref{fig:indexmodel} top row), Mg$_{2}$, NaD, TiO$_{1}$, CaH$_{1}$, CaH$_{2}$, NaI$_{8190}$,  TiO$_{0.89}$, and FeH;  
indices that are more sensitive to giant stars and increase with decreasing IMF slope are, for example, Ca4592,  Fe5709, Fe5782 (see Fig.~\ref{fig:indexmodel} bottom row), 
CaHK, H$\beta_\text{o}$, TiO$_{0.85}$ and Ca triplet lines. 
\citep[e.g.][]{2009MNRAS.396.1895C,2012ApJ...747...69C,2014MNRAS.438.1483S,2016MNRAS.457.1468L,2017MNRAS.464.3597L}.
For IMF-sensitive features in the near-infrared, further beyond the spectral range of our data, we refer the reader to \cite{2017MNRAS.468.1594A} and \cite{2017ApJ...846..166L}.

We are not able to use all of these indices, for various reasons:  
Some indices are affected  by  chip gaps, e.g.  Mg$_2$, CaH$_2$, TiO$_1$, and the Ca triplet. The NaD index  at 5890\,\AA\space is also problematic, as it  can be contaminated by interstellar absorption and  is very sensitive to [Na/H], 
which has to be considered for measuring the IMF \citep{2017MNRAS.464.3597L}. Further, NaD and Na~I are inconsistent between MILES and \cite{2012ApJ...747...69C} SSP models \citep{2015ApJ...803...87S}.
For the remaining IMF-sensitive indices, we investigated their sensitivity to abundances. For example,  CaHK, and  CaH1
are very sensitive to [Ca/H]; Ca4592 to [Fe/H]; bTiO   to [Mg/H]; H$\beta_\text{o}$ and aTiO to [Fe/H] and [Mg/H]; Fe5782 and TiO$_{2}$  to [O/H] and [Ti/H]. 

\textsc{Element-abundance-sensitive indices}: 
In order to measure the IMF slope, we have to constrain abundance variations of  elements that influence  IMF-sensitive indices. Based on our selection of IMF-sensitive indices and the SSP models of \cite{2012ApJ...747...69C}, we have to constrain the abundances of [Mg/H], [Ti/H],  [O/H], and [Fe/H]. 

For this reason,  we  searched for spectral indices that can constrain these abundances, as they are  sensitive to the given element, but less so  to other abundances and the IMF slope. For [Mg/H], this is best fulfilled for Mg\textit{b}. 
For the abundances [Ti/H] and [O/H], we did not find a spectral index that is dominated by one of them. Most  Ti- and O-sensitive indices are equally sensitive to other elemental abundances or the IMF slope. 
We did not consider spectral features with a high sensitivity to [C/H], [N/H], or [Na/H], such as CN$_{1}$, CN$_{2}$, Mg$_{1}$, or NaD. Our  list of used spectral indices used is shown in Table \ref{tab:indexset2}.


\section{Simulations}
\label{sec:simulation}
To test the reliability of our spectral index fit results and to compare different sets of spectral indices and elemental abundances, we used simulations of mock spectra having the same quality as our real data (see Lonoce  et al., in preparation).
 In particular, we investigated the level of bias and/or constraining power of the retrieved quantities when, starting from a spectrum of a certain chemical complexity, one fits only a subset of elemental abundances together with age, metallicity and IMF slopes. 
 
 For these simulations we used as input two spectra with the same chemical composition (with the exception of [C/H]), age and S/N (see Table \ref{tab:app-ind-sim}), but with different IMF slopes mimicking both a bottom-heavy  and a Kroupa-like IMF. Basic models with fixed age, metallicity and IMF slopes were taken from \cite{2018ApJ...854..139C} models and the effect of the non-solar elemental abundances on these spectra were obtained applying the respective  response functions. If certain elemental abundance values were not provided by the standard response functions, we   linearly interpolated and extrapolated them. 
The specific choice of the elemental abundance values was similar to  the results for the central radial bin of the full spectral fitting analysis (see Section \ref{sec:pystaff}), where we  took nine abundance parameters  into account in the fit.

We convolved the  Conroy models to 
a constant velocity dispersion of $284$\,km/s; this value is  similar to the  central radial bin of \object. We  then added  random noise to simulate the quality of the real data. This was done by constructing a representative S/N curve as a function of wavelength from the real data of the central radial bin and using it at each wavelength as the sigma of the random normal distribution. This S/N curve has its  peak  at around 6000\,\AA. To better delineate the biases caused by   stellar parameter degeneracies,  we did not include the effect of systematics in these simulations.

From the same input spectrum (input 1 or input 2), we constructed $1000$ noise realizations,  and considered each of them  as a single simulated object for index measurement and parameter retrieval. 
To retrieve the stellar parameters, we repeated the same fitting method as  with real data in Sect. \ref{sec:imfindexgen}, calculating the $\chi^2$ values for all the Conroy models with the same choice of elemental abundances and sets of indices (see Table \ref{tab:indexset2}), putting a prior on the age (or fixing it in the multi-abundances fits) and fixing [Mg/H]=0.2. 
We set the steps and ranges of the parameter values  as follows: 0.25 Gyr for age (ranges depending on the simulation, see Figure \ref{fig:sim_results}), 0.1 dex for \textit{Z} (from -1.0 to 0.4 dex), 0.2 for IMF slopes $x_1$ and $x_2$ (from 0.5 to 3.5), 0.05 dex for [Ti/H] and [Fe/H] (from -0.3 to 0.3 dex), 0.05 dex for [O/H] (from 0. to 0.3 dex) and 0.05 dex for [C/H] (from -0.15 to 0.3 dex).
In the case of full-spectral fitting, we used only $10$ realizations of the input spectra, and we repeated the fits as in Sec.~\ref{sec:pystaff}.

\subsection{Results of the spectral indices simulations}
\label{sec:ind-sim}

 \begin{figure*}
\begin{centering}
\includegraphics[width=5.7cm]{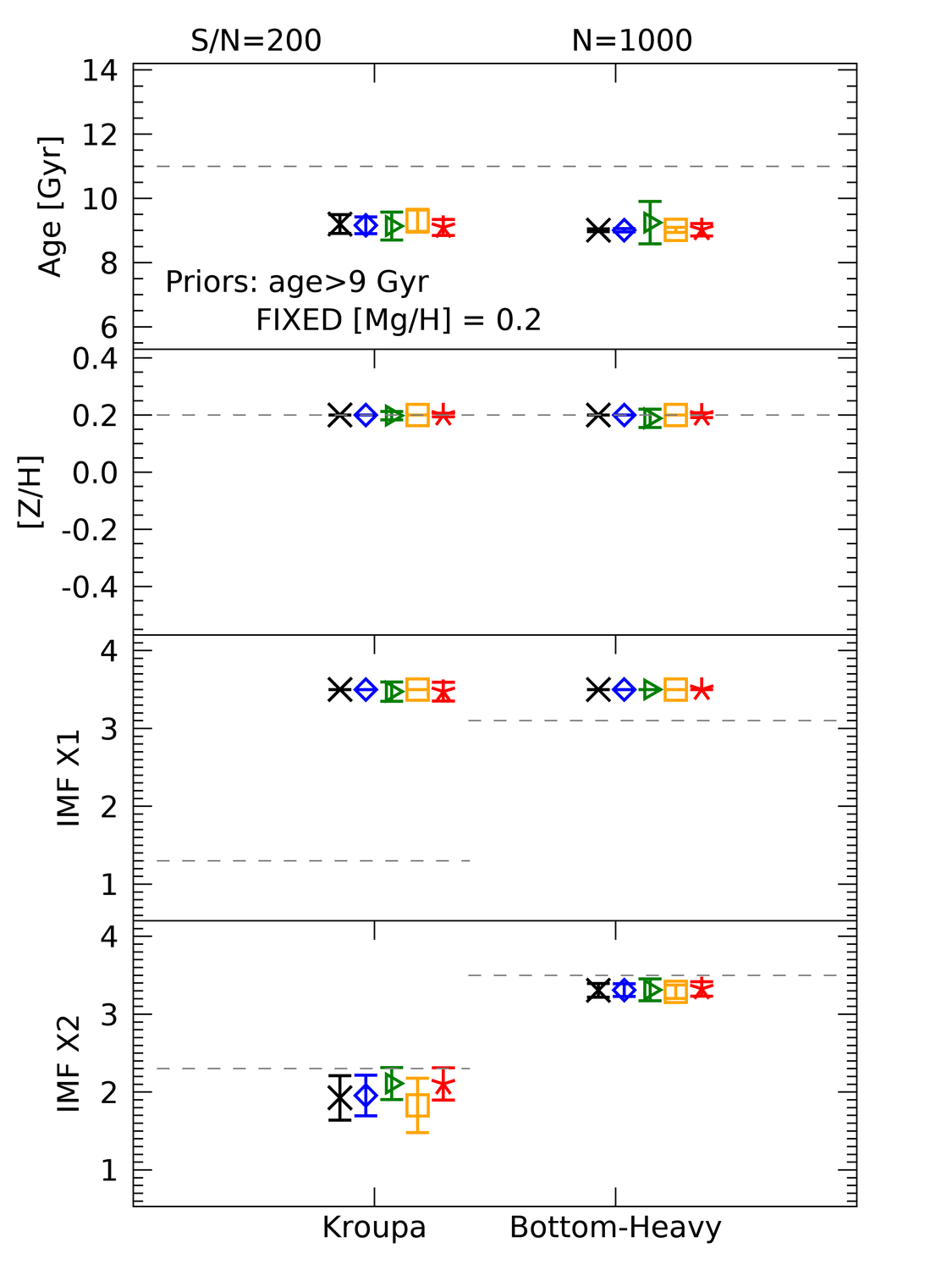}
\includegraphics[width=5.7cm]{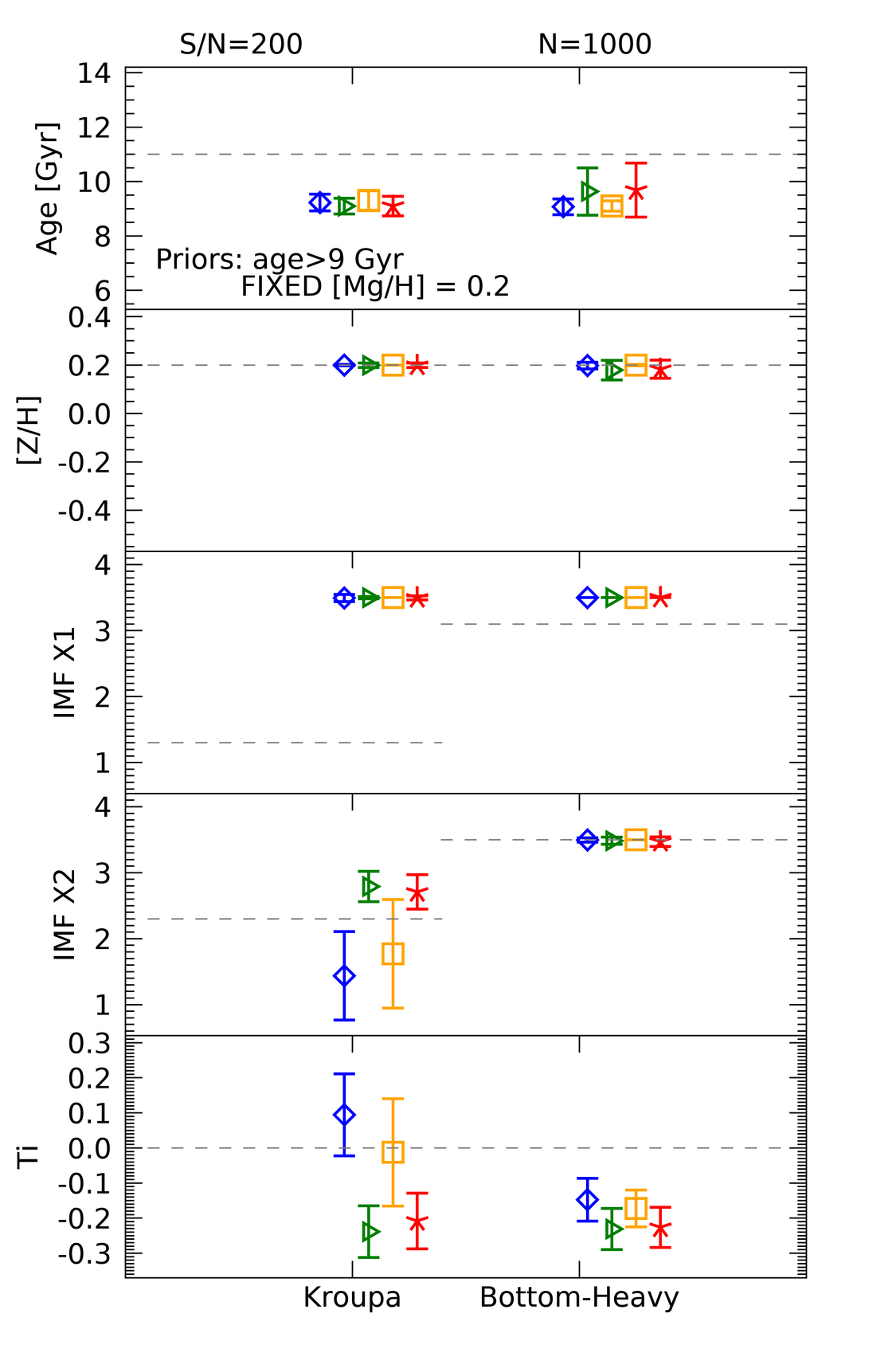}
\includegraphics[width=5.7cm]{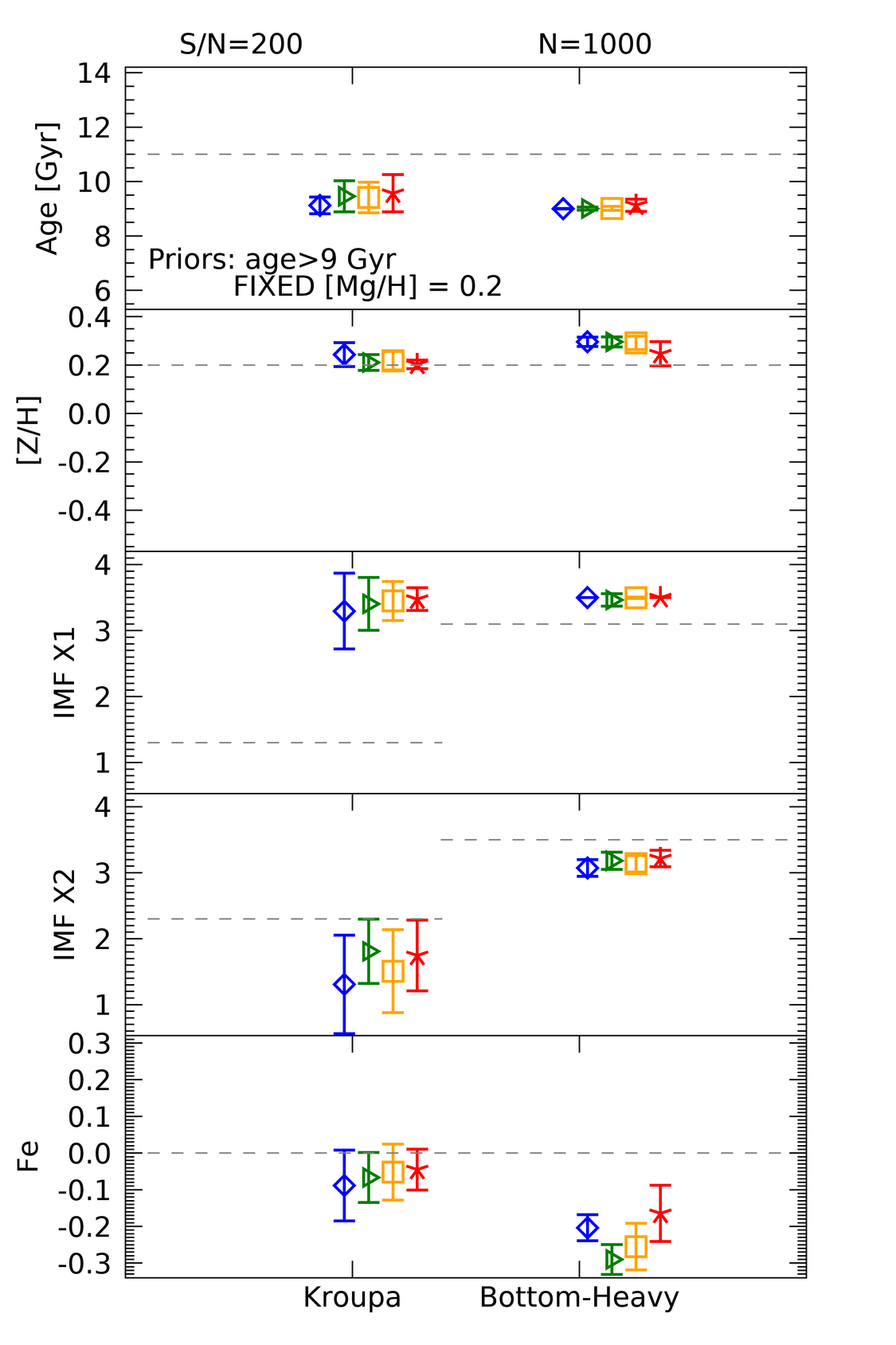}
\includegraphics[width=5.7cm]{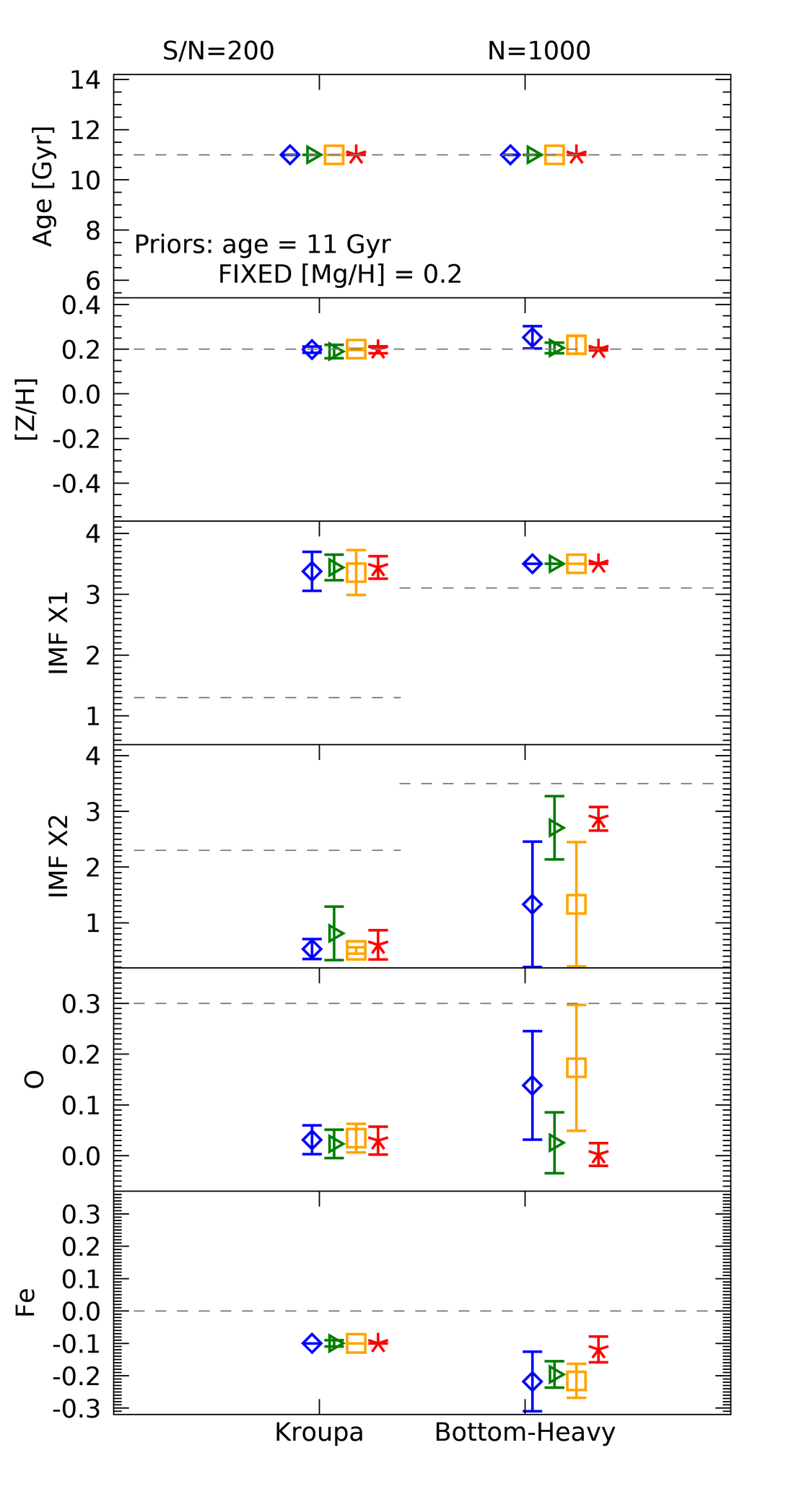}
\includegraphics[width=5.7cm]{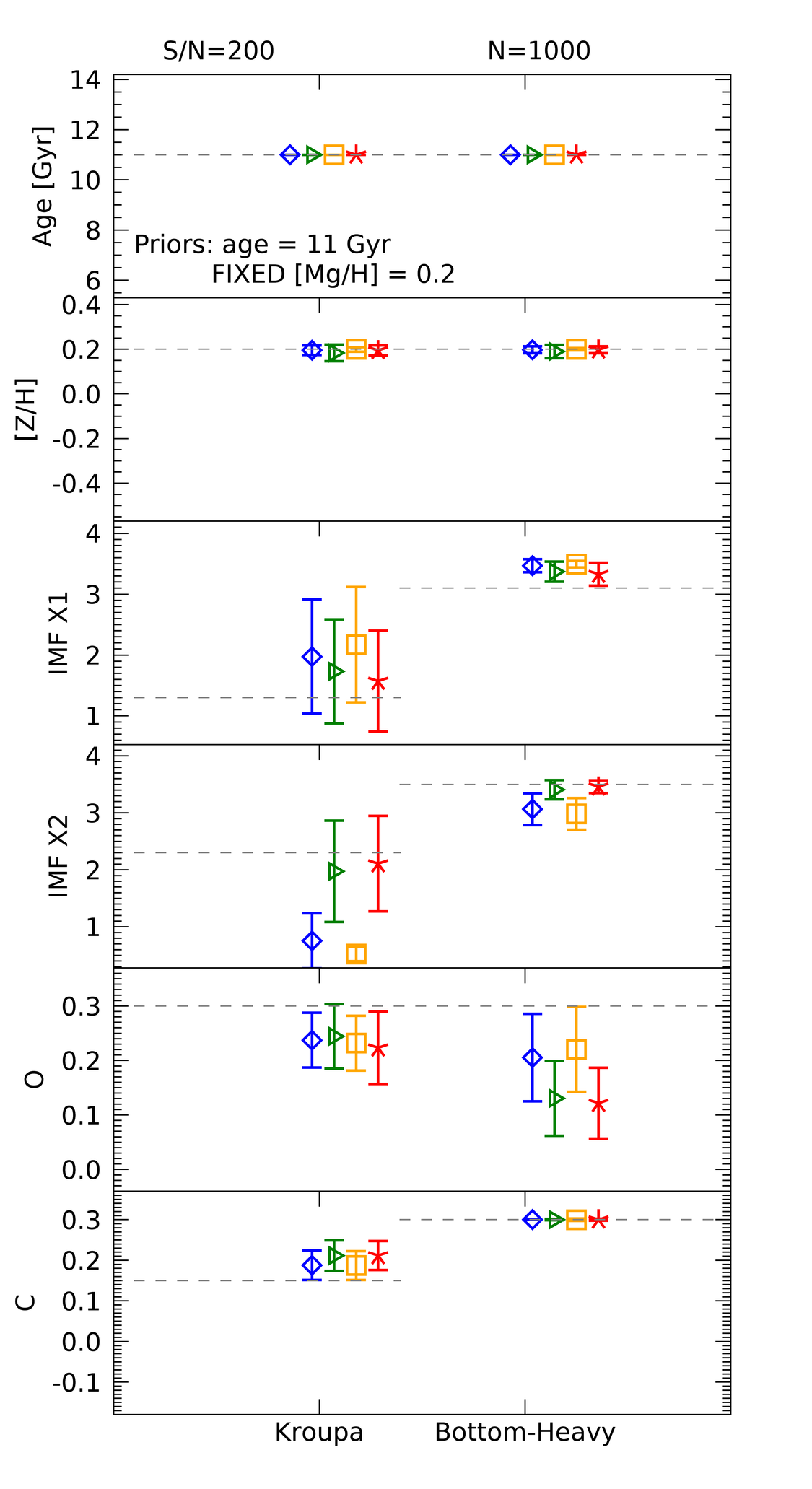}
\includegraphics[width=5.7cm]{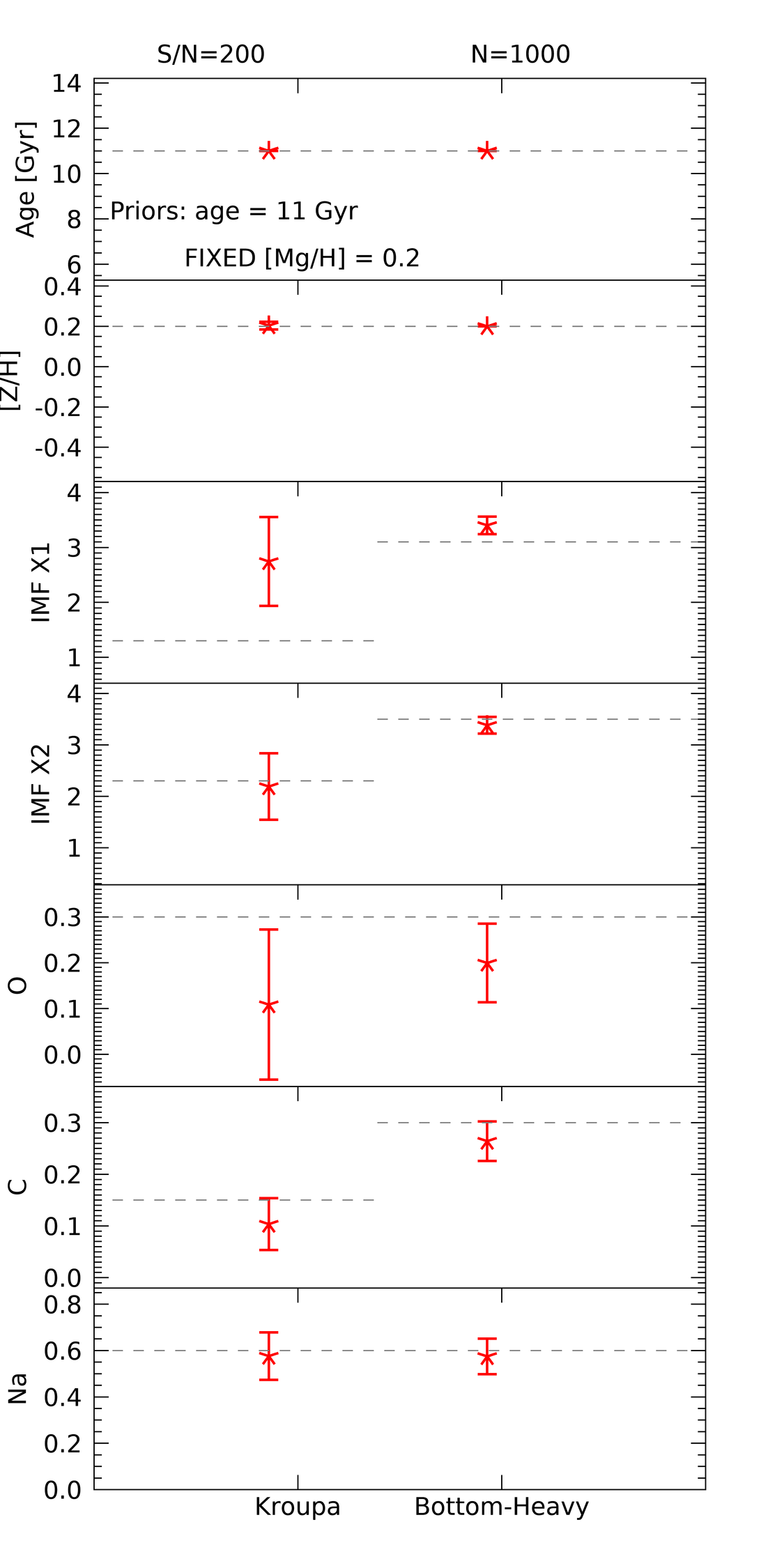}
\caption{Index simulation results: each panel shows the results of simulations for a different choice of fitted parameters. All simulations were done with S/N=200 and 1000 realizations.  Dashed horizontal grey lines show the true input spectral parameter values. Colors and symbols indicate different sets of indices as described in Table \ref{tab:indexset2}. We only show fits with a degree of freedom greater than 0. Left sets of points are the results for a Kroupa-like input IMF, right ones for a bottom-heavy input IMF. While the results for $x_1$ are often higher than  the input values, the results for $x_2$ are often lower. A bottom-heavy input IMF slope is measured with higher accuracy than a   Kroupa-like input IMF slope.}
\label{fig:sim_results}
\end{centering}
\end{figure*}

The results from our simulations are summarized in Figure \ref{fig:sim_results} for a maximum S/N of 200 per \AA.   We also simulated the case with S/N=400 per \AA, and found smaller statistical uncertainties, but similar biases. 
The $x_1$ slope, which characterizes the number of stars of  low-mass, is   poorly constrained with our available indices. Our simulations show that there is a strong bias towards  high values (at the limit of the  models), for both bottom-heavy and Kroupa input IMF.  
The least biased results are obtained  when fitting both [O/H] and [C/H], but also in this case  $x_1$ remains unconstrained with very large error bars. The choice of different index sets is irrelevant for its determination. \\

The $x_2$ slope measurements are more  encouraging, as  with some index sets the true values for $x_2$ can be retrieved.  In particular, the Fe5709  (green triangles) and combined index-sets (red asterisks)  results are close to  the input values if the IMF is bottom-heavy. However, if it is Kroupa-like,  the most accurate results for $x_2$ are obtained  when fitting simultaneously [O/H], [C/H] and [Na/H]. It is important to choose the right combination both of indices and of elemental abundances, to avoid biases.  

In general, we find that the fitted elemental abundances suffer from biases, usually independent of the choice of indices. 
Metallicity is very well constrained, with a small bias towards lower values in some cases. When the age  is not fixed (upper panels), there are   degeneracies with IMF slopes and elemental abundances, and the age is underestimated.  We thus emphasize that conclusions about the IMF are dependent on assumptions about the age of the population.

\subsection{Results of the full-spectral-fitting simulations}
\label{sec:fsfsim}
\begin{figure*}
\plotone{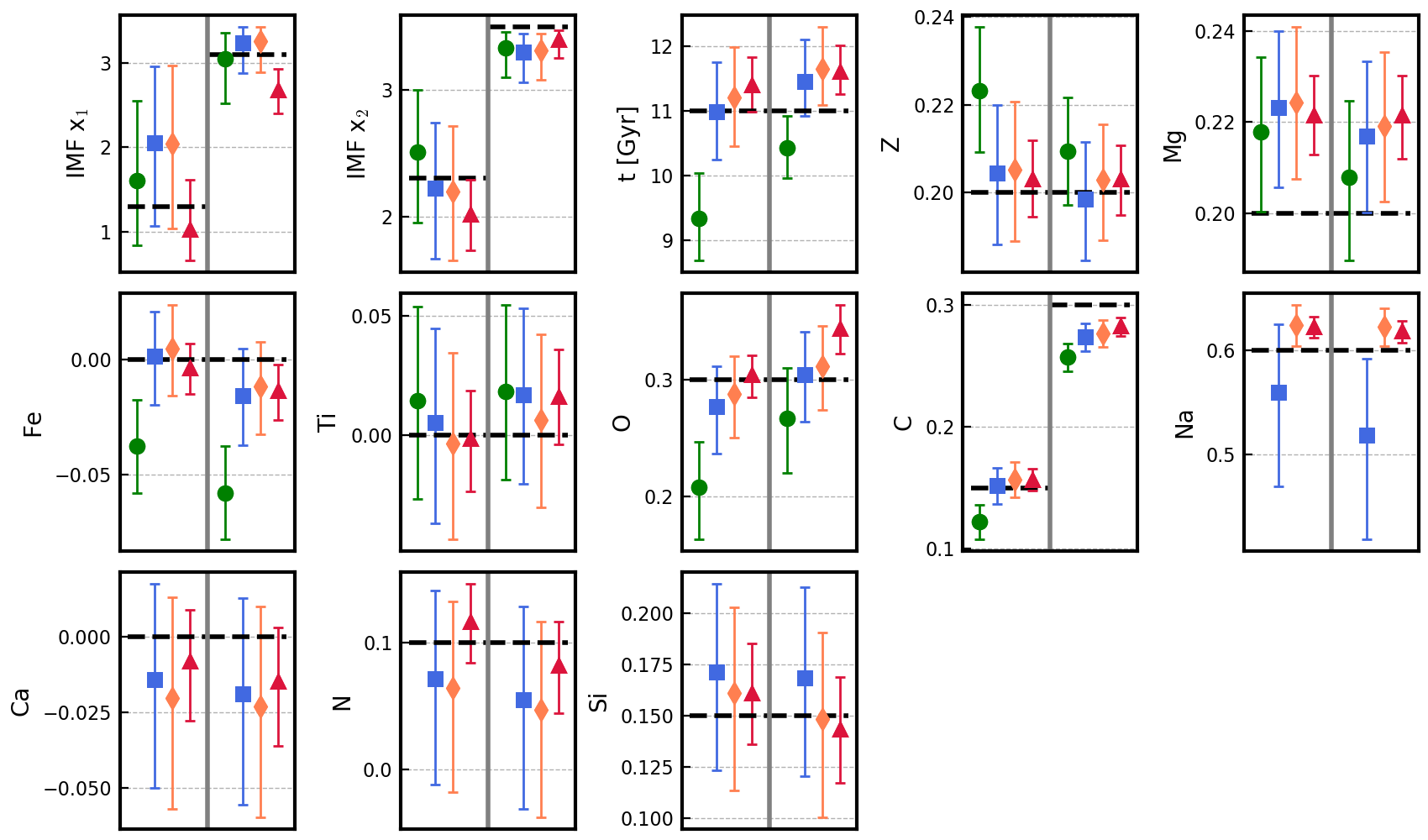}
\caption{Full-spectral fitting simulation results: The different colors represent the results of 10 realizations of mock spectra with different wavelength regions or fitting parameters. Symbols are the same as in  Figure \ref{fig:pystaffout1} and Table \ref{tab:fitset2}: Green circles denote simulations of the 5-abundance fit, blue squares the 9-abundance fit, orange diamonds the 9-abundance fit when  including the NaD absorption line region, and  red triangles when further including  the near-infrared features   NaI (8200\,\AA) and  the first CaT  line (8500\,\AA)) regions. Including more fitting parameters and spectral features improves the accuracy and precision of our fits. 
\label{fig:pystaffsim}}
\end{figure*}

We used the same spectra as for the spectral index fitting simulations, but undertook only 10 realizations. We ran our full spectral fitting (see Sect.~\ref{sec:methodfsf}) routine with 5000 steps and 100 walkers, using the same wavelength regions as for the observations. In particular, we fit the short wavelength region in one case with 5 abundances (green circles in Figure~\ref{fig:pystaffsim}), in another with 9 abundances (blue square symbols),  the longer wavelength region  including NaD with 9 abundances (orange diamonds), and including NaI and the first Ca triplet line with 9 abundances (red triangles). The exact wavelength regions of the fit and elemental abundance parameters are listed in Table~\ref{tab:fitset2}. The mean values of the 10 realisations are shown in Figure \ref{fig:pystaffsim}
for both a Kroupa IMF and a bottom-heavy IMF.

We find  that the value of $x_1$ is poorly constrained by optical fits if the IMF is Kroupa-like. Alternatively, if the IMF is bottom-heavy, the $x_1$ measurement  is more precise.  But in both cases, $x_1$  tends to be overestimated. As already found for the spectral index simulations, the $x_2$ results are more accurate and precise. Therefore it should be possible to differentiate a very bottom-heavy value of $x_2$ from a more Milky-Way like value, though  $x_2$ tends to be  slightly underestimated. 
Including   the NaI (8200\,\AA) and the first of the Ca Triplet lines (8500\,\AA), which are often used in the literature for IMF measurements,    improves the precision of  $x_1$, and $x_2$. For our data and simulations, these features have a lower S/N than the optical data. The measurements may be further improved by  spectra of the full Ca Triplet line region at a comparable S/N as the optical data.

The stellar age is best constrained when we fit all nine abundances, but it is  systematically underestimated with the 5-abundance fit, while the metallicity is slightly overestimated. These biases are less than 2\,Gyr and 0.025\,dex. Nevertheless, this shows that neglecting  relevant elemental abundances in the fit will bias the age and metallicity results to inaccurate values.

Most element abundances are always consistent with their input values, such as [Ti/H], [Ca/H], [N/H], and [Si/H]. [Mg/H] and [Na/H] are biased to slightly higher values. This bias is only small ($\textless$0.03\,dex), and may be caused by  interpolation:  \textsc{PyStaff} uses a Taylor expansion, while we used a   linear interpolation to apply the response function and  construct the simulated spectra. Usually, this causes differences on the sub-percent level. For high values of element abundances (e.g. [C/H] =0.3) we have to extrapolate, which can cause differences of few percent between the spectra in some spectral regions.  The precision of    [Na/H] improves in the  fits that include the NaD absorption lines, the uncertainties decrease by a factor of four. Several uncertainties decreases even further when including the NaI and Ca Triplet features, in particular \textit{Z}, [Mg/H], [Fe/H], [Ti/H], [Na/H], [Ca/H], [N/H], [Si/H] by factors of about 1.4 to 2. 
 
We also tested  fitting a slightly longer wavelength range that includes the Ca H+K line region and H$\alpha$, i.e. 3780\,\AA\textless$\lambda$\textless 6540\,\AA, which are covered by our observations. We expected more precise results, and  the precision  for  [Ca/H] and [N/H]  improved by about a factor of  two, but the IMF, age, and various other element abundances were unaffected.  Thus, the gain in extending our wavelength region is only small, and since our data appears to have larger systematic uncertainties in the extended region, we conclude that our results will not be improved. %

Overall, we find only small biases for element abundances ($\lesssim$0.05\,dex). While $x_1$ is slightly biased to higher values, $x_2$ is biased to lower values in the optical fits. The simulations could recover an  extremely bottom-heavy IMF, but measurements of a Kroupa IMF  with optical data alone are imprecise.  When including near-infrared wavelengths,  $x_1$ and $x_2$ are slightly biased to lower values, but overall more precise. 

\begin{deluxetable*}{lcccccccccccccc}
\tablecaption{Spectral indices simulations: input spectra parameters \label{tab:app-ind-sim}}
\tablecolumns{15}
\tablewidth{0pt}
\tablehead{
\colhead{} &
\colhead{Age [Gyr]} &
\colhead{[Z/H]} & 
\colhead{$x_1$} & 
\colhead{$x_2$}&
\colhead{[Na/H]}&
\colhead{[Ca/H]}&
\colhead{[Mg/H]}&	
\colhead{[Ti/H]}&
\colhead{[Fe/H]}&
\colhead{[C/H]}&
\colhead{[O/H]}&
\colhead{[N/H]}&
\colhead{[Si/H]}&
\colhead{S/N [\AA$^{-1}$]}
}
\startdata
Input 1  & 11. & 0.2 & 1.3 & 2.3 & 0.6 & 0.0 & 0.2 & 0.0 & 0.0 & 0.15 & 0.3 & 0.1 & 0.2 & 200 \\
Input 2  & 11. & 0.2 & 3.1 & 3.5 & 0.6 & 0.0 & 0.2 & 0.0 & 0.0 & 0.30 & 0.3 & 0.1 & 0.2 & 200 \\
\enddata
\end{deluxetable*}

\begin{figure*}
\gridline{\fig{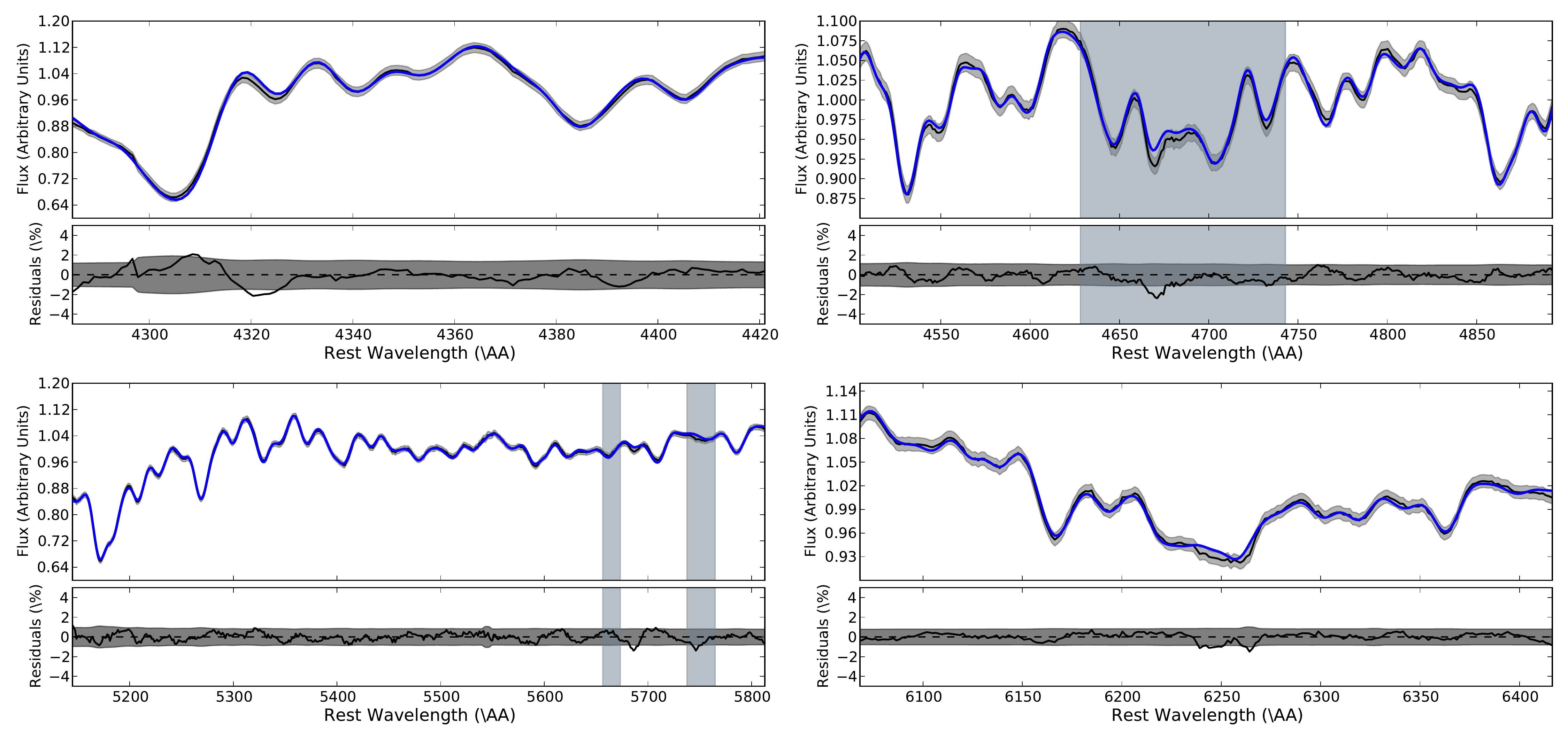}{1.0\textwidth}{(a) \textsc{PyStaff} fit with  Na, Ca, N, Si set to zero (5-abundance fit)}}
\gridline{\fig{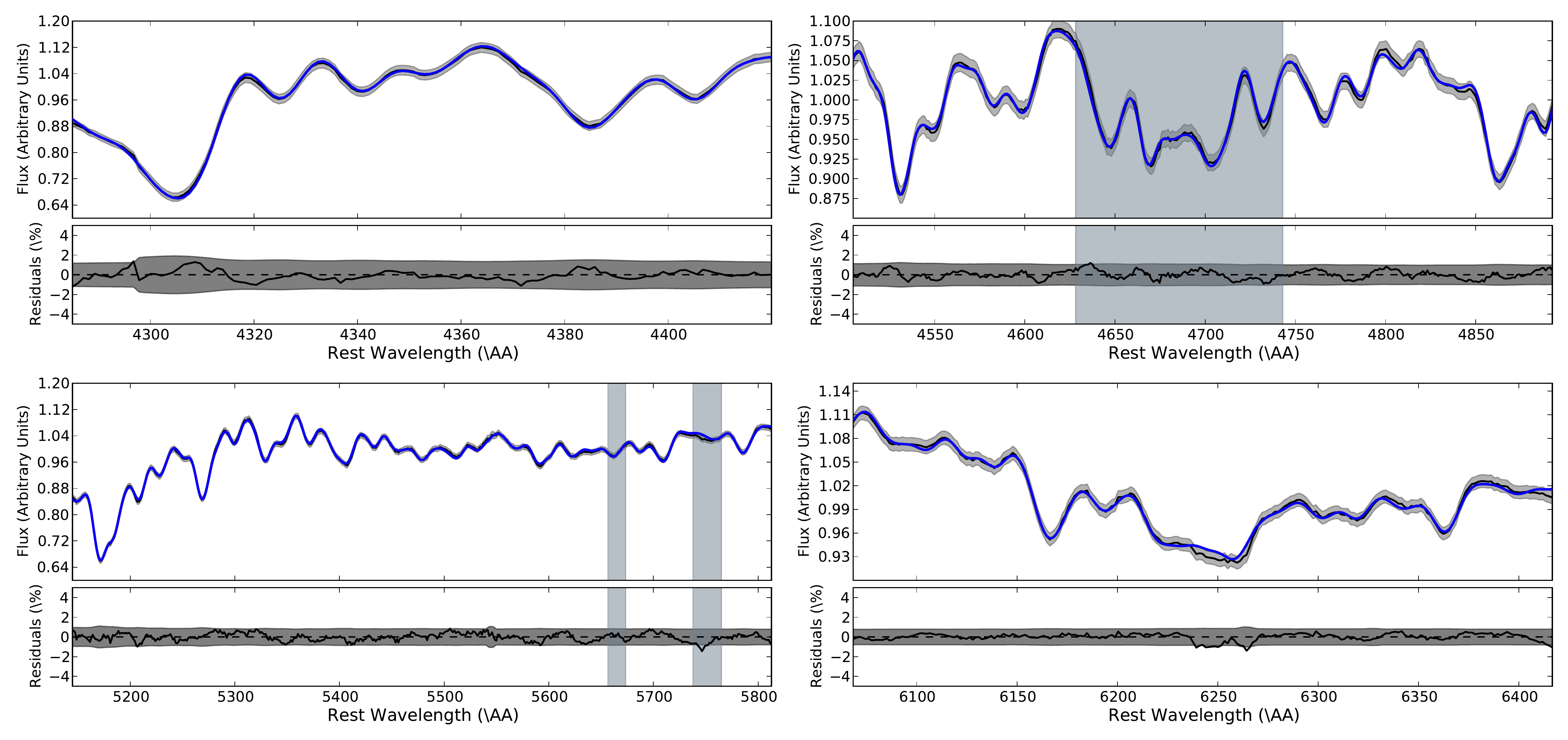}{1.0\textwidth}{(b)  \textsc{PyStaff} fit with    Na, Ca, N, Si as fitting parameters (9-abundance fit)}}
\caption{
\textsc{PyStaff} best-fit models (blue) and gas emission (red) to the central spectra (black), fitting 9 abundances. The grey color along the spectrum illustrates the noise level; vertical grey regions are excluded from the fit.  
The plot also shows the residuals in per cent as a function of wavelength. 
 The upper panel (a) is the fit with Na, Ca, N, Si fixed to zero (5-abundance fit, green circles in Figures~\ref{fig:pystaffout1} and \ref{fig:pystaffsim}), the lower panel (b) is the fit with Na, Ca, N, Si as fitting parameters (9-abundance fit, blue squares in Figures~\ref{fig:pystaffout1} and \ref{fig:pystaffsim}). 
\label{fig:pystafffit2}}
\end{figure*}  

\begin{figure*}
\gridline{\fig{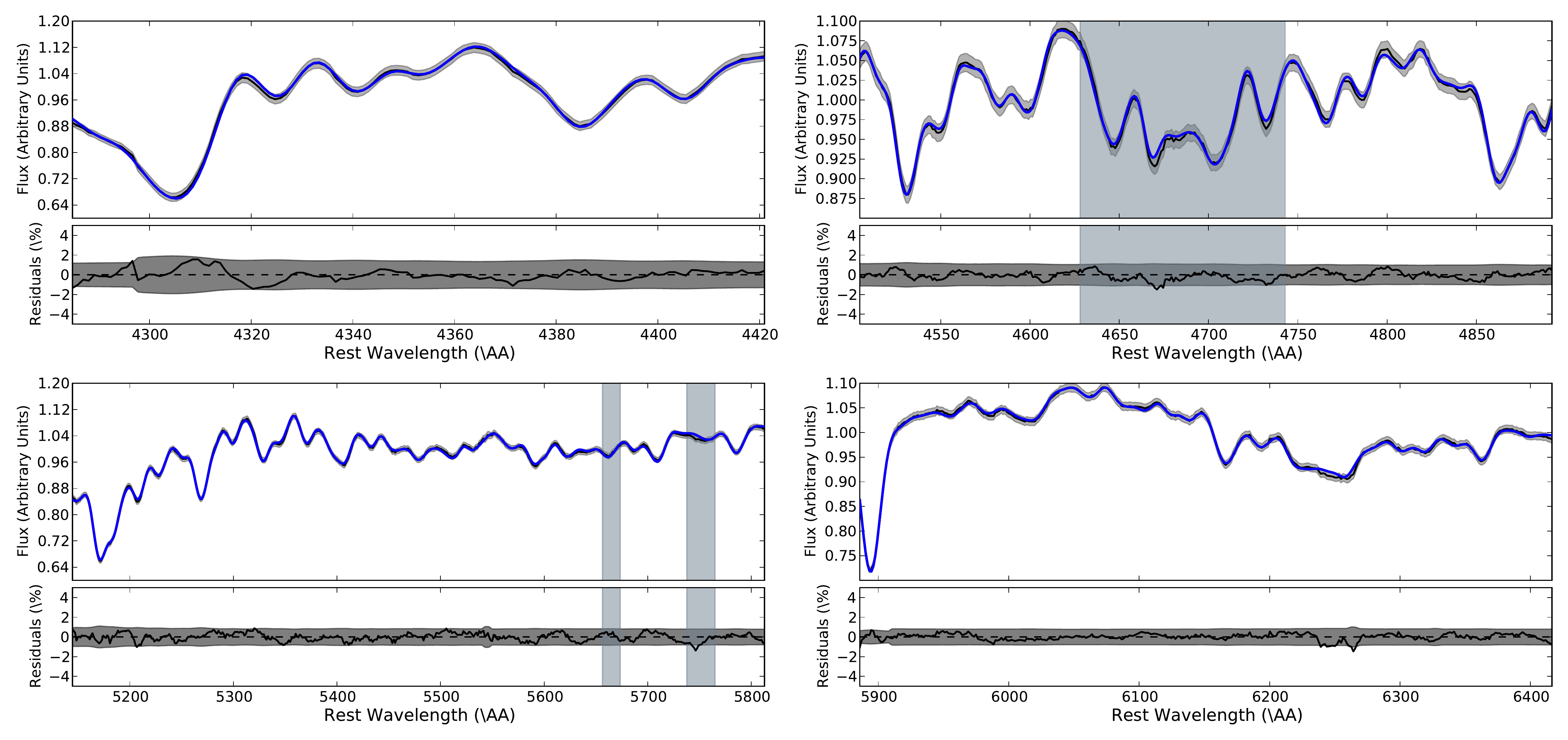}{1.0\textwidth}{(c) Preferred optical \textsc{PyStaff} fit}}
\gridline{\fig{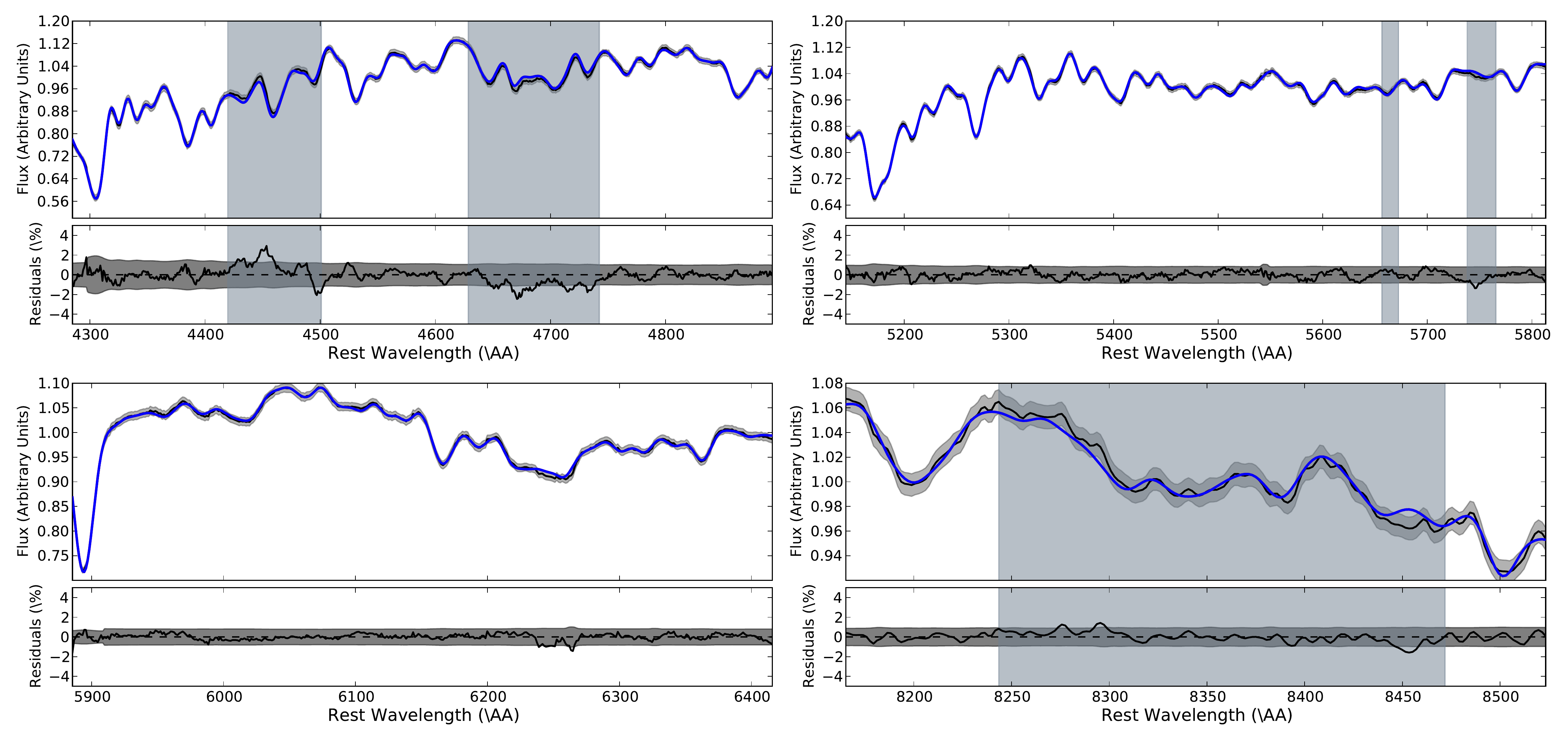}{1.0\textwidth}{(d)  \textsc{PyStaff} fit with  NaI and the first CaT line (NIR fit)}}
\caption{ 
Same as Fig. \ref{fig:pystafffit2} for different wavelength region. 
The upper panel (c) is our preferred optical fit (orange diamonds  in in Figs.~\ref{fig:pystaffout1} and \ref{fig:pystaffsim}). The lower panel (d) includes the near-infrared spectra (red triangles in  Figs.~\ref{fig:pystaffout1} and \ref{fig:pystaffsim}).
\label{fig:pystafffit}}
\end{figure*}

\end{appendix}

\end{document}